\documentclass[12pt]{article}
\usepackage{jheppub}

\usepackage{tikz}
\usepackage{subcaption}
\usetikzlibrary{shapes}
\usetikzlibrary{backgrounds}
\usetikzlibrary{decorations,decorations.pathreplacing,decorations.markings}
\usetikzlibrary{fit,calc,through}
\usetikzlibrary{external}
\usetikzlibrary{positioning}
\tikzstyle{mid>}=[decoration={markings, mark=at position 0.5 with {\arrow{>}}}, postaction={decorate}]
\tikzstyle{mid<}=[decoration={markings, mark=at position 0.5 with {\arrow{<}}}, postaction={decorate}]
\tikzstyle{upper>}=[decoration={markings, mark=at position 0.8 with {\arrow{>}}}, postaction={decorate}]
\tikzstyle{upper<}=[decoration={markings, mark=at position 0.8 with {\arrow{<}}}, postaction={decorate}]
\tikzstyle{lower<}=[decoration={markings, mark=at position 0.2 with {\arrow{<}}}, postaction={decorate}]
\usepackage{environ}
\usepackage{xargs}
\newcommandx{\NewEnvironx}[5][2,3]{%
  \expandafter\newcommandx\csname start#1\endcsname[#2][#3]{#4}%
  \NewEnviron{#1}{\csname start#1\expandafter\endcsname\BODY #5}}
\newcommand{\ladderX}{1.5}
\newcommand{\ladderY}{1.5}
\newcommand{\ladderR}{0.6}
\newcommand{\laddercoordinates}[2]{
\foreach \x in {0,...,#1} {
	\foreach \y in {0,...,#2} {
		\coordinate (l\x\y) at (\x * \ladderX, \y * \ladderY);
		\coordinate (u\x\y) at ($(l\x\y)+\ladderR*(0,\ladderY)$);
		\coordinate (d\x\y) at ($(l\x\y)+(0,\ladderY)-\ladderR*(0,\ladderY)$);
	}
}
}
\newcommand{\ladderEn}[5]{
\draw[mid>] (l#1#2) -- (d#1#2);
\draw[mid>] (d#1#2) -- ($(l#1#2)+(0,\ladderY)$) node[left] {#3};
\draw[mid>] ($(l#1#2)+(\ladderX,0)$) -- ($(u#1#2)+(\ladderX,0)$);
\draw[mid>] ($(u#1#2)+(\ladderX,0)$) -- ($(l#1#2)+(\ladderX,\ladderY)$) node[right] {#4};
\draw[mid>] (d#1#2) --node[above]{#5} ($(u#1#2)+(\ladderX,0)$);
}

\newcommand{\ladderFn}[5]{
\draw[mid>] (l#1#2) -- (u#1#2);
\draw[mid>] (u#1#2) -- ($(l#1#2)+(0,\ladderY)$) node[left] {#3};
\draw[mid>] ($(l#1#2)+(\ladderX,0)$) -- ($(d#1#2)+(\ladderX,0)$);
\draw[mid>] ($(d#1#2)+(\ladderX,0)$) -- ($(l#1#2)+(\ladderX,\ladderY)$) node[right] {#4};
\draw[mid>] ($(d#1#2)+(\ladderX,0)$) --node[above]{#5} (u#1#2);
}

\newcommand{\ladderIn}[3]{\draw[mid>] (l#1#2) -- +($#3*(0,\ladderY)$);}

\NewEnvironx{ladder}[2]{%
  \begin{tikzpicture}[baseline=13*\ladderY*#2]\laddercoordinates{#1}{#2}
  \end{tikzpicture}}

\newcommand{\fuse}[3]{\tikz[baseline=0.5cm]{
\coordinate (z1) at (0,0);
\coordinate (z2) at (1,0);
\coordinate (c) at (0.5,0.5);
\coordinate (e) at (0.5,1);
\draw[mid>] (z1) node[below] {$#1$} -- (c);
\draw[mid>] (z2) node[below] {$#2$} -- (c);
\draw[mid>] (c) -- (e) node[above] {$#3$};
}}
\newcommand{\fork}[3]{\tikz[baseline=0.5cm]{
\coordinate (z1) at (0,1);
\coordinate (z2) at (1,1);
\coordinate (c) at (0.5,0.5);
\coordinate (e) at (0.5,0);
\draw[mid<] (z1) node[above] {$#1$} -- (c);
\draw[mid<] (z2) node[above] {$#2$} -- (c);
\draw[mid<] (c) -- (e) node[below] {$#3$};
}}
\def\semicolon{;}
\def\applytolist#1{
    \expandafter\def\csname multi#1\endcsname##1{
        \def\omultiack{##1}\ifx\omultiack\semicolon
            \def\next{\relax}
        \else
            \csname #1\endcsname{##1}
            \def\next{\csname multi#1\endcsname}
        \fi
        \next}
    \csname multi#1\endcsname}
\def\calc#1{\expandafter\def\csname c#1\endcsname{{\mathcal #1}}}
\applytolist{calc}QWERTYUIOPLKJHGFDSAZXCVBNM;

\newcommand\fq{\mathfrak{q}}

\newcommand\fgl{\mathfrak{gl}}

\newcommand\bCP{\mathbb{CP}}
\newcommand\Tr{\text{Tr}}
\newcommand\bR{\mathbb{R}}
\def\bZ{{\mathbb{Z}}}
\def\bC{{\mathbb{C}}}
\def\sl{{\mathfrak{sl}}}
\def\gl{{\mathfrak{gl}}}
\newcommand\CP{\mathcal{P}}
\newcommand\CC{\mathcal{C}}
\newcommand{\qBinomial}[3][\fq]{\genfrac{[}{]}{0pt}{}{#2}{#3}_{#1}}

\newcommand{\minicircle}{%
  \tikz[baseline=-0.5ex, scale=0.15]{
    \draw[very thick] (0,0) circle (1);
  }%
}

\newcommand{\circlewithline}{%
  \tikz[baseline=-0.5ex, scale=0.15]{
    \draw[very thick] (0,0) circle (1);
    \draw[very thick] (0,-1) -- (0,1);
  }%
}

\newcommand{\neutralmodule}[2]{%
  \begin{scope}[shift={(0,#1)}]
    \def\spacing{2} 
    \foreach \lab [count=\i] in {#2} {
      \draw[very thick,postaction=decorate,
             decoration={markings, mark=at position 0.5 with {\arrow{stealth}}}]
            (\i*\spacing,0) -- (\i*\spacing,1);
      \node[anchor=east] at (\i*\spacing - 0.1,.5) {\lab};
    }
  \end{scope}%
}

\newcommand{\capmodule}[3][0.2]{%
  \begin{scope}[shift={(0,#2)}]
    \def\spacing{2} 
    \def\tagheight{#1} 
    \foreach \j in {1,...,#3} {%
      \pgfmathsetmacro{\xleft}{(2*\j - 1)*\spacing}
      \pgfmathsetmacro{\xright}{(2*\j)*\spacing}
      \draw[very thick] (\xleft,0) arc (180:0:{\spacing/2});
      \pgfmathsetmacro{\xmid}{(\xleft+\xright)/2}
      \draw[very thick] (\xmid, {\spacing/2}) -- ++(0,\tagheight);
    }%
  \end{scope}%
}

\newcommand{\cupmodule}[3][0.2]{%
  \begin{scope}[shift={(0,#2)}]
    \def\spacing{2} 
    \def\tagheight{#1} 
    \foreach \j in {1,...,#3} {%
      \pgfmathsetmacro{\xleft}{(2*\j - 1)*\spacing}
      \pgfmathsetmacro{\xright}{(2*\j)*\spacing}
      \draw[very thick] (\xleft,0) arc (180:360:{\spacing/2});
      \pgfmathsetmacro{\xmid}{(\xleft+\xright)/2}
      \draw[very thick] (\xmid, -{\spacing/2}) -- ++(0,-\tagheight);
    }%
  \end{scope}%
}

\newcommand{\Erung}[2]{%
    \def\spacing{2} 
  \pgfmathsetmacro{\rungy}{(#1 + .9)}%
  \pgfmathsetmacro{\leftx}{(#2)*\spacing}%
  \pgfmathsetmacro{\rightx}{((#2)+1)*\spacing}%
  \draw[thin,postaction={decorate},
    decoration={markings, mark=at position 0.5 with {\arrow{stealth}}}]
    (\rightx,\rungy) -- (\leftx,\rungy);
}

\newcommand{\Frung}[2]{%
    \def\spacing{2} 
  \pgfmathsetmacro{\rungy}{(#1 + .9)}%
  \pgfmathsetmacro{\leftx}{(#2)*\spacing}%
  \pgfmathsetmacro{\rightx}{((#2)+1)*\spacing}%
  \draw[thin,postaction={decorate},
    decoration={markings, mark=at position 0.5 with {\arrow{stealth}}}]
    (\leftx,\rungy) -- (\rightx,\rungy);
}

\newcommand{\braidmodule}[2]{%
  \begin{scope}[shift={(0,#1)}]
    \def\spacing{2} 
    \def\totalstrands{4} 
    \def\braidindex{#2} 
    \foreach \j in {1,...,\totalstrands} {%
      \ifnum\j=\braidindex
      \else
        \ifnum\j=\numexpr\braidindex+1\relax
        \else
          \draw[very thick] (\j*\spacing,0) -- (\j*\spacing,2);
        \fi
      \fi
    }%
    \draw[very thick] ({\braidindex*\spacing},2) ..
      controls ({\braidindex*\spacing},1.5) and ({(\braidindex+1)*\spacing},0.5) ..
      ({(\braidindex+1)*\spacing},0); 
    \pgfmathsetmacro{\xcross}{(\braidindex+0.5)*\spacing}
    \pgfmathsetmacro{\ycross}{1}
    \fill[white] (\xcross,\ycross) circle (0.1);
    
      \draw[very thick] ({(\braidindex+1)*\spacing},2) ..
      controls ({(\braidindex+1)*\spacing},1.5) and ({\braidindex*\spacing},0.5) ..
      ({\braidindex*\spacing},0);
  \end{scope}%
}

\newcommand{\braidmoduletwo}[2]{%
  \begin{scope}[shift={(0,#1)}]
    \def\spacing{2} 
    \def\totalstrands{2} 
    \def\braidindex{#2} 
    \foreach \j in {1,...,\totalstrands} {%
      \ifnum\j=\braidindex
      \else
        \ifnum\j=\numexpr\braidindex+1\relax
        \else
          \draw[very thick] (\j*\spacing,0) -- (\j*\spacing,2);
        \fi
      \fi
    }%
    \draw[very thick] ({\braidindex*\spacing},2) ..
      controls ({\braidindex*\spacing},1.5) and ({(\braidindex+1)*\spacing},0.5) ..
      ({(\braidindex+1)*\spacing},0); 
    \pgfmathsetmacro{\xcross}{(\braidindex+0.5)*\spacing}
    \pgfmathsetmacro{\ycross}{1}
    \fill[white] (\xcross,\ycross) circle (0.1);
    
      \draw[very thick] ({(\braidindex+1)*\spacing},2) ..
      controls ({(\braidindex+1)*\spacing},1.5) and ({\braidindex*\spacing},0.5) ..
      ({\braidindex*\spacing},0);
  \end{scope}%
}

\newcommand{\braidmoduletwoI}[2]{%
  \begin{scope}[shift={(0,#1)}]
    \def\spacing{2} 
    \def\totalstrands{2} 
    \def\braidindex{#2} 
    \foreach \j in {1,...,\totalstrands} {%
      \ifnum\j=\braidindex
      \else
        \ifnum\j=\numexpr\braidindex+1\relax
        \else
          \draw[very thick] (\j*\spacing,0) -- (\j*\spacing,2);
        \fi
      \fi
    }%
    \draw[very thick] ({(\braidindex+1)*\spacing},2) ..
      controls ({(\braidindex+1)*\spacing},1.5) and ({\braidindex*\spacing},0.5) ..
      ({\braidindex*\spacing},0); 
    \pgfmathsetmacro{\xcross}{(\braidindex+0.5)*\spacing}
    \pgfmathsetmacro{\ycross}{1}
    \fill[white] (\xcross,\ycross) circle (0.1);
    \draw[very thick] ({\braidindex*\spacing},2) ..
      controls ({\braidindex*\spacing},1.5) and ({(\braidindex+1)*\spacing},0.5) ..
      ({(\braidindex+1)*\spacing},0);
  \end{scope}%
}

\newcommand{\braidmoduleI}[2]{%
  \begin{scope}[shift={(0,#1)}]
    \def\spacing{2} 
    \def\totalstrands{4} 
    \def\braidindex{#2} 
    \foreach \j in {1,...,\totalstrands} {%
      \ifnum\j=\braidindex
      \else
        \ifnum\j=\numexpr\braidindex+1\relax
        \else
          \draw[very thick] (\j*\spacing,0) -- (\j*\spacing,2);
        \fi
      \fi
    }%
    \draw[very thick] ({(\braidindex+1)*\spacing},2) ..
      controls ({(\braidindex+1)*\spacing},1.5) and ({\braidindex*\spacing},0.5) ..
      ({\braidindex*\spacing},0); 
    \pgfmathsetmacro{\xcross}{(\braidindex+0.5)*\spacing}
    \pgfmathsetmacro{\ycross}{1}
    \fill[white] (\xcross,\ycross) circle (0.1);
    \draw[very thick] ({\braidindex*\spacing},2) ..
      controls ({\braidindex*\spacing},1.5) and ({(\braidindex+1)*\spacing},0.5) ..
      ({(\braidindex+1)*\spacing},0);
  \end{scope}%
}

\newcommand{\abraidmodule}[2]{%
  \begin{scope}[shift={(0,#1)}]
    \def\spacing{2} 
    \def\totalstrands{4} 
    \def\braidindex{#2} 
    \foreach \j in {1,...,\totalstrands} {%
      \ifnum\j=\braidindex
      \else
        \ifnum\j=\numexpr\braidindex+1\relax
        \else
          \draw[very thick] (\j*\spacing,0) -- (\j*\spacing,2);
        \fi
      \fi
    }%
    \draw[very thick] ({(\braidindex+1)*\spacing},2) ..
      controls ({(\braidindex+1)*\spacing},1.5) and ({\braidindex*\spacing},0.5) ..
      ({\braidindex*\spacing},0);
    \pgfmathsetmacro{\xcross}{(\braidindex+0.5)*\spacing}
    \pgfmathsetmacro{\ycross}{1}
    \fill[white] (\xcross,\ycross) circle (0.1);
    \draw[very thick] ({\braidindex*\spacing},2) ..
      controls ({\braidindex*\spacing},1.5) and ({(\braidindex+1)*\spacing},0.5) ..
      ({(\braidindex+1)*\spacing},0); 
  \end{scope}%
}

\newcommand{\Emrung}[3]{%
  \begin{scope}
    \def\spacing{2} 
    \pgfmathsetmacro{\rungy}{(#1 - 0.2)}%
    \pgfmathsetmacro{\startx}{(#2)*\spacing}%
    \pgfmathsetmacro{\endx}{(#3)*\spacing}%
    \pgfmathsetmacro{\arrowpos}{1/(2*(#3 - #2))}%
    \draw[thin,postaction={decorate},
      decoration={markings, mark=at position \arrowpos with {\arrow{stealth}}}]
      (\endx,\rungy) -- (\startx,\rungy);
    \foreach \k in {\numexpr#2+1\relax,...,\numexpr#3-1\relax} {%
      \draw[fill=white,draw=white] (\k*\spacing,\rungy) circle (0.1);
    }%
  \end{scope}%
}

\newcommand{\Fmrung}[3]{%
  \begin{scope}
    \def\spacing{2} 
    \pgfmathsetmacro{\rungy}{(#1 - 0.2)}%
    \pgfmathsetmacro{\startx}{(#2)*\spacing}%
    \pgfmathsetmacro{\endx}{(#3)*\spacing}%
    \pgfmathsetmacro{\arrowpos}{1/(2*(#3 - #2))}%
    \draw[thin,postaction={decorate},
      decoration={markings, mark=at position \arrowpos with {\arrow{stealth}}}]
      (\startx,\rungy) -- (\endx,\rungy);
    \foreach \k in {\numexpr#2+1\relax,...,\numexpr#3-1\relax} {%
      \draw[fill=white,draw=white] (\k*\spacing,\rungy) circle (0.1);
    }%
  \end{scope}%
}

\newcommand{\olambda}{\mu}
\newcommand{\omu}{\lambda}
\newcommand{\oell}{v}

\author[1]{Davide Gaiotto,}
\author[1]{Suriyah Rajalingam Kannagi,}
\author[1]{Sergio Sanjurjo}

\affiliation[1]{Perimeter Institute for Theoretical Physics, 31 Caroline Street North, Waterloo, ON N2L
2Y5, Canada}
\emailAdd{dgaiotto@perimeterinstitute.ca}
\emailAdd{srajalingamkannagi@perimeterinstitute.ca}
\emailAdd{ssanjurjo@perimeterinstitute.ca}
\title{D-branes and the planar limit of Chern-Simons theory I: Link invariants}

\abstract{We revisit the Holographic duality between $SU(N)_\kappa$ Chern-Simons theory and the A-model Topological String Theory. We develop a strategy to systematically compute the large $N$ saddles for correlation functions of Wilson lines in antisymmetric powers $\Lambda^\bullet \bC^N$ of the fundamental representation. The mathematical structures which appear in the calculation match in detail the data of dual A-model D-branes. }

\begin{document}
\maketitle

\section{Introduction}

The 't Hooft expansion of $SU(N)_\kappa$ Chern-Simons theory is expected to be holographically dual to the perturbative expansion of the A-model Topological String Theory \cite{Gopakumar:1998ki,Ooguri:1999bv}. This is a beautiful and potentially very explicit example of holography: the partition function and correlation functions of Wilson loop operators can be computed exactly with TFT techniques, and the A-model also allows for exact calculations. Ultimately, one may hope to prove the duality in full mathematical rigor. 

This goal is still out of reach for many observables, even in the planar limit. For example, we expect the partition function
$Z_{M_3}[SU(N)_\kappa]$ of the CS theory on a closed three-manifold $M_3$ to admit large $N$ saddles, which should be matched to A-model backgrounds which approach $T^* M_3$ near the boundary at infinity. For generic $M_3$, though, we do not know how to characterize the large $N$ saddles or the A-model backgrounds, nor how to match the two. The best understood example is $M_3=S^3$, believed to be dual to the A-model on the resolved conifold geometry. Quotients of $S^3$ and Lens Spaces are also well-studied, sometimes with the help of Mirror Symmetry \cite{Halmagyi:2003ze,Marino:2009dp,Borot:2015fxa}. 

The simplest observables in Chern-Simons theory are Wilson loop operators $W_{\ell,R}$, labeled by a closed framed path $\ell$ and an $SU(N)$ representation.\footnote{The height and width of the Young Tableau labeling $R$ are bounded respectively by $N$ and $\kappa$, the latter due to quantum effects.} Their correlation functions are of great mathematical interest as a source of knot and link invariants. ``Compact'' $SU(N)_\kappa$ Wilson loops supported within a 3-ball, for example, give rise to the colored HOMFLY polynomials. Correlation functions of Wilson loops along non-trivial loops in $M_3$, possibly enriched by gauge-invariant three-way junctions, can be organized into the ``Skein module'' $\mathrm{Sk}_{M_3}[SU(N)_\kappa]$ of $M_3$ \cite{przytycki2006skeinmodules,przytycki2006skeinmodules3manifolds}.

The role of Wilson loops in the 't Hooft expansion depends sensitively on the choice of representation. Schematically, specific tensor products of fundamental and/or anti-fundamental representations are associated to closed strings, while representations with order $N$ indices are associated to D-branes. Representations with order $N^2$ indices could potentially modify the asymptotic shape of the holographic dual geometry. 

In this paper we will focus on Wilson loops dual to strings and D-branes: our objective is to give a one-to-one map between the large $N$ saddles of certain Wilson loop correlation functions and an appropriate collection of A-model D-branes.   

\subsection{Closed Strings and large \texorpdfstring{$N$}{N} saddles}
Wilson loop operators $W_{\ell,\Box}$ in the fundamental representation of $SU(N)_\kappa$ have a well-studied holographic interpretation in terms of extended string worldsheets \cite{Rey:1998ik}. In a large $N$ analysis, vevs $w_\ell$ of the $(\fq - \fq^{-1})W_{\ell,\Box}$ operators are constrained by ``loop equations'', which are the planar limit of the framed HOMFLY skein relations:\footnote{Here we are not being mindful of the (minor) differences between skein relations for $U(N)_\kappa$ and $SU(N)_\kappa$ Wilson loops, which anyway drop away in the planar limit.} 
\begin{equation}
\begin{tikzpicture}[baseline=-0.65ex, scale=0.5]
  \draw[->, >=latex] (1,-1) -- (-1,1);
  \draw[white,line width=6pt] (-1,-1) -- (1,1);
  \draw[->, >=latex] (-1,-1) -- (1,1);
\end{tikzpicture}
- \begin{tikzpicture}[baseline=-0.65ex, scale=0.5]
  \draw[->, >=latex] (-1,-1) -- (1,1);
  \draw[white,line width=6pt] (1,-1) -- (-1,1);
  \draw[->, >=latex] (1,-1) -- (-1,1);
\end{tikzpicture}
= (\fq - \fq^{-1})\,\begin{tikzpicture}[baseline=-0.65ex, scale=0.5, rotate=90]
  \draw[->, >=latex] (-1,1) .. controls (0,0.5) .. (1,1);
  \draw[->, >=latex] (-1,-1) .. controls (0,-0.5) .. (1,-1);
\end{tikzpicture}
\end{equation}
with $\fq = e^{i \pi \frac{1}{\kappa + N}}$. 

The planar limit is defined as $\fq \to 1$ with constant $g = \fq^N = e^{i \pi \frac{N}{\kappa + N}}$, the exponentiated 't Hooft coupling. In the planar limit, fundamental Wilson loops turn out to be transparent to each other and insensitive to being knotted, up to overall powers of $g$ arising from changes of framing of the loop.\footnote{Even for compact Wilson loops and the associated HOMFLY polynomials, this statement appears to be somewhat non-trivial. It follows, though, from the standard large $N$ combinatorics of multi-trace correlation functions.} In particular, compact fundamental Wilson loops are only ``interesting'' beyond the planar approximation. 

On a general manifold $M_3$, the planar limit $w_\ell$ of the vevs of fundamental Wilson loops thus only depends on the homotopy class of a framed loop $\ell$. It should be possible to define a 
planar version $\mathrm{Sk}^p_{M_3}[g]$ of the Skein module restricted to (anti)fundamental Wilson lines and finite tensor products thereof.\footnote{Perhaps in terms of a Deligne category of $U_\fq(\sl_N)$ representations \cite{etingof2020representationtheorycomplexrank}.} As correlation functions of fundamental Wilson lines factor to a product of $w_\ell$'s in the planar limit, $\mathrm{Sk}^p_{M_3}[g]$ will be a commutative algebra rather than a vector space. 

In this language, a large $N$ saddle for $Z_{M_3}[SU(N)_\kappa]$ is labeled by a consistent collection of vevs $w_\ell$, i.e. a point in the spectrum of $\mathrm{Sk}^p_{M_3}[g]$. Ideally, one would like to compute $\mathrm{Sk}^p_{M_3}[g]$ and match it to the equations of motion of the dual A-model, so that any large $N$ saddle would correspond to an A-model background and vice versa.\footnote{Although this structure is a standard expected feature of the 't Hooft expansion, we have not seen it stated in this form in the literature devoted to the large $N$ analysis of Chern-Simons theory. It seems a natural starting point for a mathematical analysis of the problem.}

As a simple example of this relation, we can take a further $g \to 1$ ``tree-level planar'' limit. This limit eliminates the framing dependence and makes the planar skein relations classical. It also eliminates the back-reaction in the holographic dual geometry. The vevs $w_\ell$ satisfy the same relations as the traces of the holonomy of a flat connection on $M_3$ along a path $\ell$. 

Accordingly, any 3d $\mathrm{GL}(M)$ flat connection on $M_3$ for any non-negative integer $M$ provides a potential tree-level planar saddle: it can be embedded in $SU(N)$ to provide a classical solution of the Chern-Simons equations of motion. These saddles have an intuitive holographic interpretation: a collection of $M$ A-model D-branes in $T^* M_3$ fibered non-trivially over the base. 

We will not pursue here a further analysis of $\mathrm{Sk}^p_{M_3}[g]$. Instead we will consider a simple choice of $M_3$ and focus on D-brane probes of the dual A-model. 

\subsection{Heavy Wilson lines and D-branes}
In this paper we would like to test a ``categorical'' approach to the 't Hooft expansion \cite{tHooft:1973alw, Gaiotto:2024dwr}, where a large $N$ QFT is probed by additional degrees of freedom transforming in the fundamental representation of the gauge group. Each addition, together with a choice of large $N$ saddle for the vev of mesonic operators, is expected to be dual to a specific D-brane in a tentative String Theory dual description of the system. 
We dub this a ``fundamental modification'' of the large $N$ QFT.

If we add multiple fundamental modifications to the large $N$ QFT, we get access to new mesonic operators which are dual to the open string sectors between pairs of D-branes. This data can be compiled into a category of D-branes. This is a physically and mathematically rich object which can be employed to identify or perhaps even define the dual String Theory background.

\begin{figure}[h]
\begin{minipage}{0.5\textwidth}
\centering
\begin{tikzpicture}[scale=2.5]
  \coordinate (A) at (0,1);
  \coordinate (B) at (0.866,-0.5);
  \coordinate (C) at (-0.866,-0.5);
  
  \draw[shorten >=2mm, shorten <=2mm, very thick, 
    decoration={
      markings,
      mark=at position 0.5 with {
        \arrow{stealth};
      }
    },
    postaction={decorate}]
    (A) to[out=45, in=15] (B) to[out=195, in=-15] (C);
    
  \draw[shorten >=2mm, shorten <=2mm, very thick, 
    decoration={
      markings,
      mark=at position 0.5 with {
        \arrow{stealth};
      }
    },
    postaction={decorate}]
    (B) to[out=-75, in=-105] (C) to[out=75, in=-135] (A);
    
  \draw[shorten >=2mm, shorten <=2mm, very thick, 
    decoration={
      markings,
      mark=at position 0.5 with {
        \arrow{stealth};
        \node[above, yshift=2mm] {$k$};
      }
    },
    postaction={decorate}]
    (C) to[out=165, in=135] (A) to[out=-45, in=105] (B);
\end{tikzpicture}
\end{minipage}
\begin{minipage}{0.5\textwidth}
\centering
\begin{tikzpicture}[scale=2.5]
  \coordinate (A) at (0,1);
  \coordinate (B) at (0.866,-0.5);
  \coordinate (C) at (-0.866,-0.5);
  
  \draw[shorten >=2mm, shorten <=2mm, very thick, 
    decoration={
      markings,
      mark=at position 0.5 with {
        \arrow{stealth};
      },
      mark=at position 0.25 with {
       \node[right, xshift=2mm] {$k$};
      }
    },
    postaction={decorate}]
    (A) to[out=45, in=15] (B) to[out=195, in=-15] (C);
    
  \draw[shorten >=2mm, shorten <=2mm, very thick, 
    decoration={
      markings,
      mark=at position 0.5 with {
        \arrow{stealth};
      },
      mark=at position 0.25 with {
       \node[below, yshift=-2mm] {$k$};
      }
    },
    postaction={decorate}]
    (B) to[out=-75, in=-105] (C)  to[out=75, in=-135] node[pos=0.4, coordinate] (M) {} node[pos=0.6, coordinate] (P) {} (A) ;
    
  \draw[shorten >=2mm, shorten <=2mm, very thick, 
    decoration={
      markings,
      mark=at position 0.5 with {
        \arrow{stealth};
        \node[above, yshift=2mm] {$k-1$};
      }
    },
    postaction={decorate}]
    (C) to[out=165, in=135] node[pos=0.5, coordinate] (O) {} (A) to[out=-45, in=105] node[pos=0.5, coordinate] (N) {} (B);

    \draw[thin, 
    decoration={
      markings,
      mark=at position 0.5 with {
        \arrow{stealth};
      }
    },
    postaction={decorate}]
    (M) -- (N);
    \draw[thin, 
    decoration={
      markings,
      mark=at position 0.5 with {
        \arrow{stealth};
      }
    },
    postaction={decorate}]
    (O) -- (P);
\end{tikzpicture}
\end{minipage}
\caption{Left: the Wilson loop in the shape of a trefoil knot, decorated by the $\Lambda^k \bC^N$ representation. Right: The same knot, decorated by additional meson operators. The meson operators consist of two 1d fermions connected by a fundamental Wilson line (thin line in the Figure). The fermion number $k$ jumps across the fermion insertions, allowing one to derive recursion relations by manipulating the mesons.}\label{trefoilone}
\end{figure}
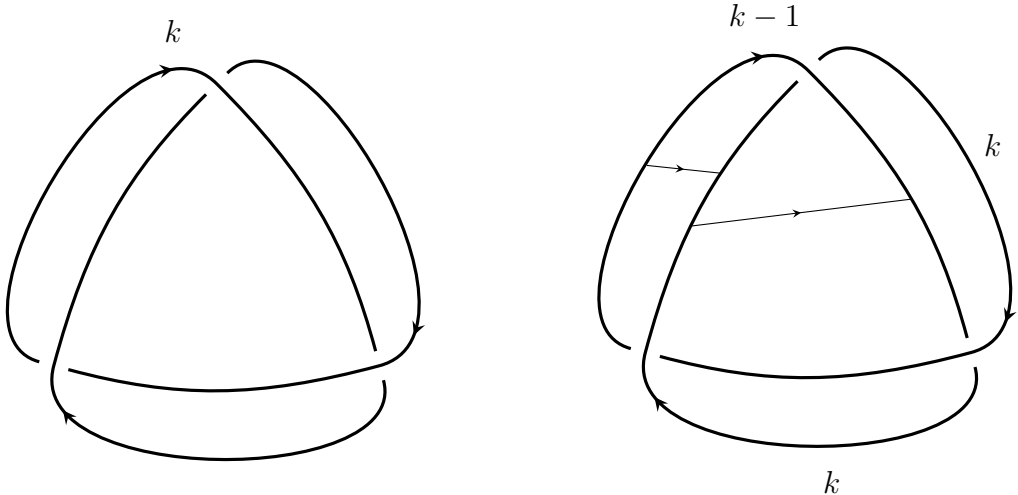

In 3d Chern-Simons theory, additional degrees of freedom can be supported in one, two or three dimensions. We will focus on the first option: an auxiliary system of $N$ complex fermions supported on a closed loop $K$ in space-time.\footnote{Non-topological modifications in two- and three- will be the subject of two separate publications, embedding large $N$ vector model holography \cite{Klebanov:2002ja,Gaberdiel:2010pz} in the A-model Topological String Theory.}  This system can be recast as a direct sum $W_{K}$ of Wilson loop operators $W_{K,k}$ labeled by the exterior powers $\Lambda^k \bC^N$ of the fundamental representation of $SU(N)$ and studied with 3d TFT methods \cite{Witten:1988hf} as colored HOMFLY knot and link invariants \cite{Morton1993}. See Figure \ref{trefoilone}.

A fundamental modification of the CS theory thus takes the form of a Wilson loop $W_{K}$ wrapping a knot $K$ (or a collection $W_L$ of Wilson loops forming a link $L$), together with a choice of large $N$ saddle for the 't Hooft expansion in the presence of the loop. 
For compact knots, correlation functions of $W_{K,k}$ satisfy difference equations in $k$ which have a non-trivial planar limit \cite{Garoufalidis:2012rt,Garoufalidis:2016zhf}, the {\it augmentation variety} of the knot/link \cite{Aganagic:2012jb, Aganagic:2013jpa},
and characterize the possible large $N$ saddles.\footnote{The augmentation variety is usually associated with the planar limit of HOMFLY polynomials colored by symmetric power representations $S^\bullet \bC^N$. We prefer to work with $\Lambda^\bullet \bC^N$, but the two options are expected to give rise to the same augmentation variety, up to a minor redefinition of variables. See Appendix \ref{app:saddle} for a detailed comparison.} 

A general expectation which has been verified for several knots in $S^3$ is that the augmentation variety defined from the difference equations matches the moduli space of A-model D-branes in the resolved conifold which are holographically dual to the knot/link \cite{Marino:2004uf,Ng_2005a,Ng_2005b,Diaconescu:2011xr, Ekholm_2012, Aganagic:2013jpa,fang2017topologicalrecursionconifoldtransition,Ekholm:2018iso,Ekholm:2019yqp}. Such calculations typically employ a geometric description of D-branes as Lagrangian submanifolds in the resolved conifold, corrected by worldsheet instantons to objects in a Fukaya category.  

Following the categorical 't Hooft expansion strategy, we will learn how to enlarge the data of the augmentation variety to include vevs of mesonic operators. We will then recognize this extra data as the sheaf-theoretic description of a dual A-brane \cite{nadler2009microlocalbranesconstructiblesheaves,nadler2008constructiblesheavesfukayacategory}, side-stepping the conventional methods of symplectic geometry. Mathematically, this sort of relation is sometimes referred to as an ``augmentation-sheaf correspondence'' \cite{gao2021augmentationssheaveslinks}. Before the current work, it was only available in a $g \to 1$ limit which we will review momentarily.

\subsection{Webs and rungs}
The references \cite{Garoufalidis:2012rt,Garoufalidis:2016zhf} propose a general strategy to compute correlation functions of $\Lambda^\bullet \bC^N$ Wilson lines and prove the existence of the associated difference equations: they consider correlation functions of more general ``webs'' consisting of $\Lambda^\bullet \bC^N$ Wilson lines connected at canonical trivalent junctions. Correlation functions of webs satisfy certain skein relations \cite{Cautis_2014} which allow one to re-organize and ultimately simplify the topology of the web. The final result is expressed as an intricate sum whose building blocks satisfy individual recursion relations, from which in principle one can derive the desired recursion relation.  

The large $N$ saddle for the 't Hooft expansion of a correlation function can be derived as a saddle evaluation of such sums. The examples in Appendix \ref{app:saddle} illustrate this general principle. In this paper we will instead employ the web manipulations to characterize the vevs of mesonic operators. The augmentation variety will emerge naturally in the process.

The meson operators available to a $\Lambda^\bullet \bC^N$ Wilson line take the form of 1d fermion bilinears connected by open Wilson lines in the $\bC^N$ fundamental representation. See Figure \ref{trefoilone}. We can denote such operators (``rungs'') as $R^{p,p'}_\ell$, with $p$ and $p'$ the starting and ending points on $K$ of the framed open path $\ell$. 
In the planar limit, the insertion of a rescaled rung operator $(\fq - \fq^{-1})R^{p,p'}_\ell$ multiplies the correlation function by a vev 
$r^{p,p'}_\ell$.  

The rung endpoints are special cases of web junctions and can be manipulated by the rules in \cite{Cautis_2014}. The resulting relations between rung operators have a nice large $N$ limit to polynomial ``open loop equations'' satisfied by the rung vevs. We expect these open loop equations to characterize the possible large $N$ saddles of the knot/link and in particular determine the augmentation variety.  

Mathematically, the open loop equations describe a planar limit  $\mathrm{Sk}^p_{M_3;K}[g;\olambda]$ of $\mathrm{Sk}_{M_3}[SU(N)_\kappa]$ where one focuses on the interaction between a $W_{K;k}$ Wilson loop with fixed $\olambda = \fq^k$ and any number of fundamental Wilson lines.\footnote{Recall that a Deligne category generalizes the category of finite-dimensional $U_\fq(\sl_N)$ representations to non-integer $N$ \cite{etingof2020representationtheorycomplexrank}. We believe it should be possible to define a module category for the Deligne which contains a generalization of $\Lambda^k \bC^N$ to non-integer $k$. It may be a useful starting point for the construction.} 

Although our main focus will be compact knots, many of the techniques we employ should apply well to more general configurations which probe the geometry of $M_3$. The loop equations for the open vevs $r^{p,p'}_\ell$  take the closed $w_\ell$ vevs as an extra input and constrain them. They should determine a dual A-model background equipped with a category of D-branes. 

We can briefly review the output of our strategy in the ``tree-level planar limit'' $g \to 1$ \cite{Ng_2012,cieliebak2017knotcontacthomologystring}. In this limit, the loop equations for $r^{p,p'}_\ell$ simplify and reduce to the classical skein relations satisfied by the transport data 
\begin{equation}
     \oell_p^R \cdot \left[\mathrm{Pexp} \oint_\ell a \right] \cdot \oell_{p'}^L \, ,
\end{equation}
for an auxiliary $\mathrm{\mathrm{GL}}(M)$ 3d flat connection $a$ on $M_3$ for some positive integer $M$. 
The connection has a ``minimal'' regular singularity wrapping the knot $K$: the monodromy around the knot has a single non-trivial eigenvalue $\olambda^2 = \fq^{2 k}$ which identifies left- and right-eigenlines $\oell^{L,R}_p$ at each point $p$ on the knot. The augmentation variety encodes the relation between $\olambda^2$ and the holonomy $\omu$ of $\oell^L_p$ along the knot.

This monodromy defect can be understood as a Gukov-Witten-like \cite{Gukov:2006jk} line defect, whose quantization reproduces the $\Lambda^\bullet \bC^N$ Wilson lines. The auxiliary 3d connection is simply a solution of the classical Chern-Simons theory equations of motion. Such 3d flat connections typically form a discrete collection. Furthermore, they are a well-known way to describe D-branes in $T^* M_3$, the expected dual geometry when the 't Hooft coupling is turned off, which have the expected asymptotic shape determined by the knot \cite{nadler2008constructiblesheavesfukayacategory,nadler2009microlocalbranesconstructiblesheaves}.

As we turn on $g$, the skein relations are deformed and the meson vevs acquire an additional dependence on the framing of the open Wilson line. Abstractly, this gives a specific deformation of the category of D-branes in $T^* M_3$, which we take as the definition of the category of D-branes in the unknown back-reacted geometry. 

We will now specialize to some convenient $M_3$'s for which we can identify the back-reacted A-model geometry and give an explicit sheaf-theoretic description of the D-brane category, making the large $N$ duality for generic $\Lambda^\bullet \bC^N$ knots, links or webs manifest. We will also find tantalizing hints of an exact match beyond the planar limit.

\subsection{Elongated knots and the phase space of D-branes}
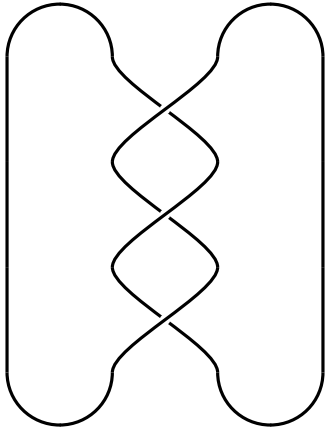
\begin{figure}[h]
    \centering
    \begin{tikzpicture}[baseline=(current bounding box.center),scale=0.7]
        \cupmodule[0]{0}{2}
        \braidmodule{0}{2}
        \braidmodule{2}{2}
        \braidmodule{4}{2}
        \capmodule[0]{6}{2}
    \end{tikzpicture}
    \caption{The trefoil knot in Schubert presentation.}\label{fig:Sch-three}
\end{figure}

We can give a rather explicit analysis for compact knots which only explore a local $\bR^3$ geometry. The basic idea is to start from a Schubert presentation of a generic knot, i.e. elongate it to a slowly evolving braid closed by cups and caps at the ends. See Figure \ref{fig:Sch-three}.

As an intermediate step in the analysis, we study the properties of rungs attached to $m=2n$ parallel $\Lambda^\bullet \bC^N$ Wilson lines with the help of ``quantum skew Howe duality'' \cite{Cautis_2014,Garoufalidis:2012rt,Garoufalidis:2016zhf}. We find it useful to compactify the transverse directions to an $S^2$, i.e. work in $M_3 = S^2 \times \bR$. The planar open loop equations satisfied by the rung vevs then define a complex symplectic ``phase space'' $\CP_{m,n}$. 

We analyze $\CP_{m,n}$ in depth and identify it with a moduli space of 2d $\mathrm{\mathrm{GL}}(n)$ flat connections with minimal regular singularities at the locations of the $m$ strands and a constant monodromy $g^2$ at infinity. This is a non-trivial deformation of the characterization we provided in the tree-level planar limit, which had the same type of minimal regular singularities but monodromy $1$ at infinity. 

We then identify $\CP_{m,n}$ as a moduli space of A-model D-branes in the deformed $A_1$ singularity $\cM_t$. Recall that $\cM_t$ is a deformation of $T^* S^2$ and is a complex symplectic manifold. We interpret it as a real symplectic manifold via the real part of the complex symplectic form and explicitly verify that the combination 
\begin{equation}
    \cM_t \times T^* \bR \, ,
\end{equation}
is a valid candidate back-reacted geometry for CS theory on $M_3 = S^2 \times \bR$. 

This identification allows us to propose a sharp holographic dictionary  
between time-independent configurations of $\Lambda^\bullet \bC^N$ Wilson lines in $S^2 \times \bR$ and time-independent A-model D-branes in $\cM_t \times T^* \bR$. The moduli space $\CP_{m,n}$ plays the role of a phase space for D-branes which asymptote to fibers of $T^* S^2$ at the locations of the strands, as expected. 

This dictionary survives when the $\Lambda^\bullet \bC^N$ strands are slowly braided, giving a slow evolution of the corresponding D-branes. In order to complete the analysis of a generic knot in the Schubert presentation, we also describe explicitly the constraints on the rung vevs enforced by the cups and caps.

With some extra work, the constraints can be formulated as a local 
``gluing'' relation between the 2d local system data associated with the vertical strands before and after the fusion of two strands. The gluing relations are a peculiar deformation of the $g\to 1$ relations which would assemble a sequence of 2d flat connections into a single 3d flat connection.\footnote{It is possible to decompose the cups and caps into a two-step process: the two strands are first fused into a single strand, which is then forced to be trivial. The first operation is the same as for $g\to 1$, but the second is deformed.}  

The sequence of 2d twisted local systems and gluing relations associated with a given open saddle defines some kind of three-dimensional twisted ``constructible sheaf'' which deforms the notion of a 3d flat connection. A 3d A-model D-brane in $\cM_t \times T^* \bR$ should admit an analogous sheaf-theoretic description, assembling a sequence of 2d slices described by 2d twisted local systems on $S^2$. 

We thus conjecture that the twisted constructible sheaves we build define D-branes in $\cM_t \times T^* \bR$ which asymptote to the conormal to the knot/link. A proof of this conjecture goes beyond our current expertise, but we will provide all the local ingredients we expect to go into such a construction. 

\subsection{Other constructions and knot presentations}
\begin{figure}[h]
\centering
\begin{tikzpicture}[thick, scale=1.5]

    \draw[rounded corners] (-0.25,-0.5) rectangle (0.75,0.5) node[midway] {Braid};

    \draw (-0.1, 0.5) 
        .. controls (-0.1, 2.5) and (3, 2.5) .. (3, 0) 
        .. controls (3, -2.5) and (-0.1, -2.5) .. (-0.1, -0.5);

    \draw (0.25, 0.5) 
        .. controls (0.25, 2) and (2.5, 2) .. (2.5, 0) 
        .. controls (2.5, -2) and (0.25, -2) .. (0.25, -0.5);

    \draw (0.6, 0.5) 
        .. controls (0.6, 1.5) and (2, 1.5) .. (2, 0) 
        .. controls (2, -1.5) and (0.6, -1.5) .. (0.6, -0.5);

\end{tikzpicture}
\caption{A schematic depiction of a knot presented as a braid closure.}\label{fig:closure}
\end{figure}
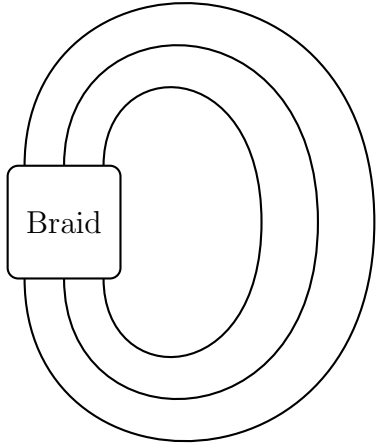

A knot or link $K$ can be represented as a braid closure, i.e. a braid with the endpoints connected together. 
This can be seen as a special case of a Schubert presentation: we add to the $r$ strands of the braid 
another set of $r$ strands with opposite labels representing the original strands looping back, and connect the original strands and new strands pairwise by cups and caps. See Figure \ref{fig:closure}.
Accordingly, the augmentation variety can be computed as described above, using $\CP_{2r,r}$.

The phase spaces $\CP_{m,n}$ can be useful beyond the study of elongated knots. If we cut any compact knot or link by a plane cutting $n$ strands, the rung vevs in the neighborhood of the plane will determine a point in $\CP_{2n,n}$. One can describe explicitly how the point varies as a function of the choice of plane. This should give some kind of ``Radon transform'' of the dual D-brane, capturing 2d sections of all possible orientations. 

Our tools can also be specialized to the study of closed braids in $M_3 = S^2 \times S^1$. Quantum mechanically, the braid operations on $m$ strands are implemented by unitary operators on the finite-dimensional Hilbert space of the CS theory on $S^2$ and the $S^2 \times S^1$ correlation function computes the trace of these operators. 

In the planar limit, the meson vevs evolve along $S^1$ by a combination of the braid group action and a certain rescaling parameterized by $\omu$. A planar saddle is thus labelled by a point in $\CP_{m,n}$ which comes back to itself under that transformation. Geometrically, we find an $S^1$ family of 2d $\mathrm{\mathrm{GL}}(n)$ flat connections on the $(m+1)$-punctured sphere, with holonomy $g^2$ at infinity and 
$m$ extra minimal regular punctures, whose holonomy remains locally constant as the position of the punctures is braided continuously along $S^1$. 

This is the same as the data of a 3d $\mathrm{\mathrm{GL}}(n)$ flat connection on $S^2 \times S^1$ with appropriate co-dimension $1$ regular singularities. It maps directly to the data of an A-model D-brane in 
\begin{equation}
    \cM_t\times T^* S^1  \, ,
\end{equation}
with asymptotic shape controlled by the braid. We thus have a fully explicit match between open planar large $N$ saddles and dual D-branes for this geometry as well.

Finally, it is interesting to observe that the phase spaces $\CP_{m,n}$ which appear in our planar analysis have a reasonably canonical quantization \cite{jordan2023quantumcharactervarieties} as skein algebras $\mathrm{Sk}[U(n)]$ on the $(m+1)$-punctured sphere, which reproduces the algebra of rung operators in the dual $SU(N)_\kappa$ CS theory. When $N$ is an integer, we expect unitarity to constrain the possible representations of this algebra to coincide with the Hilbert space of the dual $SU(N)_\kappa$ CS theory. Accordingly, it seems plausible to conjecture that our planar holographic dictionary can be uniquely extended beyond the planar approximation. An important step would be to constrain the quantization of the Lagrangian submanifolds in $\CP_{m,n}$ which encode the planar cup and cap constraints. 

\subsection{D-branes and symplectic geometry}
Although the categorical 't Hooft expansion led us to a sheaf-theoretic description of A-branes, it is still interesting to make contact with a more traditional description in terms of symplectic geometry and Lagrangian submanifolds.

The ``symplectic'' version of the augmentation variety can be expressed in terms of a Differential Graded Algebra (DGA) associated with the knot, which has a curve-counting definition but also a combinatorial presentation involving the braid closure presentation of a knot. See \cite{Ng_2014} for a review of the construction.

The ingredients of this combinatorial presentation closely resemble our computational tools, with auxiliary variables which behave under braiding in the same way as rung vevs. It should be possible to match the two presentations in detail and thus demonstrate that the planar limit of $\Lambda^\bullet \bC^N$ knot invariants matches the augmentation variety 
computed from symplectic geometry. We leave such an exercise to future work. 

Another very interesting problem is to give a precise symplectic geometry interpretation to the open loop equations for generic $M_3$. 

\subsection{Structure of the Paper}
Section \ref{sec:CStH} discusses the large $N$ expansion of $\Lambda^k \bC^N$ Wilson line correlation functions in $SU(N)_\kappa$ Chern-Simons theory, aka colored HOMFLY polynomials. Section \ref{sec:cQG} discusses the relation between elongated knots/links and $U_\fq(\gl_m)$ and its planar limit as a phase space. Section \ref{sec:char} discusses how to assemble the rung vev data into the data of a 2d flat connection. Section \ref{sec:Amodel} discusses the A-model interpretation of the $U_\fq(\gl_m)$ constructions. Section \ref{sec:conclude} reviews our conclusions and discusses some possible future research directions. 

Appendix \ref{app:classS} recalls constructions in Supersymmetric Quantum Field Theory with eight supercharges which help characterize $\CP_{m,n}$ and its quantization. Appendix \ref{app:inter} discusses SQFT interfaces which help characterize Lagrangian correspondences between $\CP_{m,n}$'s. Appendix \ref{app:rung_algebra} discusses the relation between the algebra of mesonic operators and quantum groups. Appendix \ref{app:braiding_algebra}
discusses the planar limit of the algebraic results derived in Appendix \ref{app:rung_algebra}. Appendix \ref{app:rules} reviews some results of \cite{Cautis_2014} and their planar limit. Appendix \ref{app:knots} lists the augmentation varieties we computed for several knots. Appendix \ref{app:saddle} compares our results with the recursion relations of HOMFLY polynomials and previous literature.

\section{Chern-Simons theory and the 't Hooft expansion} \label{sec:CStH}
Calculations in $SU(N)_\kappa$ Chern-Simons theory which employ TFT technology, such as partition functions on non-trivial manifolds and correlation functions of Wilson loop operators, are naturally expressed in terms of the parameter 
\begin{equation}
    \fq = e^{\frac{i \pi}{\kappa + N}} \, ,
\end{equation}
giving a useful re-summation of the standard perturbative expansion in $\kappa^{-1}$.\footnote{In some situations, the fractional power $\fq^{\frac{1}{N}}$ will appear. This can be avoided by working with a $U(N)_\kappa$ gauge group, but we prefer not to do so.}

We will often encounter the combination 
\begin{equation}
    \fq^N = e^{i \pi\frac{N}{\kappa + N}} \equiv g  \, .
\end{equation}
This is a convenient repackaging of the 't Hooft coupling. In the planar limit, we send $\fq \to 1$ while keeping $g = \fq^N$ fixed. 

Wilson loop operators $W_{\ell,R}$ compute the trace in some representation $R$ of the holonomy of the gauge connection along a framed closed path $\ell$:
\begin{equation}
    W_{\ell,R} = \Tr_R\,\mathrm{Pexp} \oint_\ell A \, .
\end{equation}
An important property of CS theory is that a collection of loop operators supported in a ball (whose boundary is not intersecting other defects) will contribute to a correlation function in a manner independent of the rest of the setup: 
\begin{equation}
    \langle W_{\ell,R} \rangle \equiv \frac{\langle W_{\ell,R} \rangle_{M_3}}{\langle 1 \rangle_{M_3}} 
\end{equation}
does not depend on the choice of space-time three-manifold $M_3$ and can be computed e.g. on $S^3$. These correlation functions for a knotted Wilson loop or multiple linked loops will be our main topic of interest. 

\subsection{Fundamental Wilson loops}
As an instructive example, consider a fundamental Wilson loop in the shape of an unknot (i.e. a circle):
\begin{equation}
    \langle W_{\minicircle,\bC^N} \rangle= [N]_\fq = \frac{\fq^N -\fq^{-N}}{\fq-\fq^{-1}} \, .
\end{equation}
This has a reasonable behavior in the planar limit: a fundamental Wilson line is a (somewhat peculiar) single-trace operator. In the 't Hooft expansion, such an operator scales as $N$. The rescaled operator $(\fq- \fq^{-1})W_{\minicircle,\bC^N}$ has a finite limit \begin{equation}
    w_{\minicircle} = g - g^{-1} \, .
\end{equation}

Although the expectation value of a fundamental Wilson loop depends non-trivially on its topology, giving rise to the HOMFLY knot and link invariants, much of the topology washes away in the planar limit: the expectation value of $s$ loops rescaled by $(\fq- \fq^{-1})^s$ always limits to $(g - g^{-1})^s g^f$ for some integer $f$.\footnote{It is easy to see that two loops can be unlinked, as the skein relation for passing a loop through the other gives $(\fq - \fq^{-1})$ times a term with a single loop, resulting in a total suppression of $(\fq - \fq^{-1})^2$. Verifying that the vev of individual loops only depends on the shape through a power of $g$ takes a bit more work. It must be true because of the combinatorics of the large $N$ expansion.}

In either case, the interesting topological content of the HOMFLY invariants is hidden in the subleading terms of the planar expansion. Verifying this aspect of the holographic duality for Chern-Simons theory thus requires an intricate analysis of higher genus amplitudes in the A-model topological string theory. We will not pursue that objective in this paper. 

\subsection{Heavy Wilson lines}
The analogous quantity for a higher anti-symmetric power of the fundamental representation is
\begin{equation}
    \langle W_{\minicircle,\Lambda^k \bC^N} \rangle=\qBinomial{N}{k} = \frac{[N]_\fq \cdots [N-k+1]_\fq}{[k]_\fq \cdots [1]_\fq}  \, .
\end{equation}

As mentioned in the Introduction, we are interested in Wilson loops defined via 1d free fermions, with action
\begin{equation}
    \int_\ell \left[\bar \psi \dot \psi - \bar \psi A \psi - a \bar \psi \psi\right] \, .
\end{equation}
We included a $U(1)$ background connection $a$ with holonomy $\omu$ coupling to the fermion number to keep track of the irreducible pieces in the fermionic Fock space $\Lambda^\bullet \bC^N$. 

The resulting loop operator is a trace on $\Lambda^\bullet \bC^N$, i.e. a sum of traces in $\Lambda^k \bC^N$ weighted by $(-\omu)^k$.\footnote{We use conventions where the fermions are periodic around the loop, so we include a factor of $(-1)^F$ in the trace.} This gives a generating function of antisymmetric Wilson loop correlation functions:
\begin{equation}
    \langle W_{\minicircle}(\omu) \rangle \equiv \sum_{k=0}^N (-\omu)^{k}\langle W_{\minicircle,\Lambda^k \bC^N} \rangle = \prod_{k=1}^N \left(1- \omu \, \fq^{2k-N-1}\right) \, .
\end{equation}

The definition in terms of auxiliary fundamental degrees of freedom makes it clear that $W_{\minicircle}(\omu)$ will behave in the planar limit as an open string partition function for some kind of D-brane:
\begin{equation} \label{eq:brane_action}
   \frac{i \pi}{\kappa + N} \log \langle W_{\minicircle}(\omu) \rangle \sim  S^{\mathrm{open}}_{\minicircle}(\omu) \, .
\end{equation}
Here it is not difficult to compute the planar limit
\begin{equation}
    S^{\mathrm{open}}_{\minicircle}(\omu) =\int_0^{\log g} d \sigma \log \left(1-\omu g^{-1} e^{2\sigma}\right) = \frac12\mathrm{Li}_2(\omu g^{-1}) -  \frac12\mathrm{Li}_2(\omu g) \, .
\end{equation}
The logarithmic derivative 
\begin{equation} \label{eq:mu_def}
    \log \olambda \equiv \omu \partial_\omu S^{\mathrm{open}}_{\minicircle}(\omu)
\end{equation}
is nicer:
\begin{equation}
    \olambda^2 = \frac{1 - \omu g}{1 - \omu g^{-1}} \, .
\end{equation}
The inverse relation 
\begin{equation}
    \omu = -\frac{\olambda - \olambda^{-1}}{g \olambda^{-1} - g^{-1} \olambda}
\end{equation}
reflects the recursion relation:
\begin{equation}
    \langle W_{\minicircle,\Lambda^k \bC^N} \rangle =\frac{[N-k+1]_\fq}{[k]_\fq} \langle W_{\minicircle,\Lambda^{k-1} \bC^N} \rangle  =\frac{\fq^{N-k+1}-\fq^{k-N-1}}{\fq^k-\fq^{-k}}\langle W_{\minicircle,\Lambda^{k-1} \bC^N} \rangle  \, ,
\end{equation}
whose coefficients are nice functions of $\fq$, $g= \fq^N$ and $\olambda = \fq^k$. We can also compute directly 
\begin{equation}
    \frac{i \pi}{\kappa + N}\log \langle W_{\minicircle,\Lambda^k \bC^N} \rangle \sim  \int_{0}^{\log \olambda} d \sigma \left(\log (g e^{-\sigma} - g^{-1} e^\sigma) -\log (e^\sigma- e^{-\sigma}) \right) \, .
\end{equation}
We see that both $\langle W_{\minicircle}(\omu) \rangle$ as a function of $\omu$ and $\langle W_{\minicircle,\Lambda^k \bC^N} \rangle$ as a function of $\olambda = \fq^k$ have good planar limits, related by a Legendre transform. 

A particularly useful formulation \cite{Aganagic:2013jpa} of the planar limit expresses the relation between $\omu$ and $\olambda$ as a Lagrangian submanifold $\cL_{\minicircle} \subset \bC^* \times \bC^*$, with symplectic form 
$\mathrm{d} \log\! \omu \, \mathrm{d}\log\!\olambda$:
\begin{equation}
    g^{-1} \olambda \omu -\olambda + \olambda^{-1} - g \olambda^{-1}\omu=0 \, .
\end{equation}
We will call this the {\it augmentation variety} of the unknot.\footnote{This is a slight abuse of language: strictly speaking the augmentation variety describes the result of a symplectic geometry geometric calculation which should reproduce the moduli space of A-model D-branes dual to the large $N$ saddles \cite{Aganagic:2013jpa}. What is defined here is a QFT quantity which should conjecturally match the geometric augmentation variety.} 

One expects an analogous planar limit for correlation functions  $\langle W_{K,\Lambda^k \bC^N} \rangle$ and $\langle W_{K}(\omu) \rangle$ associated with any knot $K$, leading to more intricate augmentation varieties 
$\cL_K \subset \bC^* \times \bC^*$. Experimentally, $\cL_K$ is always defined as the vanishing locus of a Laurent polynomial in $\olambda$, $\omu$ and $g$, with remarkable relations to the theory of cluster varieties and to spaces of vacua of 3d ${\cal N}=2$ gauge theories compactified on a circle \cite{Ekholm:2018eee}. 

Similarly, the correlation function 
\begin{equation}
    \langle W_{L;\Lambda^{k_1} \bC^N,\ldots, \Lambda^{k_n} \bC^N} \rangle
\end{equation} of a collection $L$ of linked Wilson lines or the dual generating function 
\begin{equation}
    \langle W_{L}(\omu_1, \ldots, \omu_n) \rangle
\end{equation}
will have a planar limit controlled by the relation between $\olambda_i = \fq^{k_i}$ and $\omu_i$, defining a complex Lagrangian augmentation variety $\cL_{L}\subset (\bC^* \times \bC^*)^n$.

A general goal is to reproduce the augmentation variety of a generic knot/link from the classical geometry of D-branes in the dual String Theory description of the system, using tools which could be generalized to Wilson loops in a general $M_3$. Ideally, both sides should be described in a purely algebraic manner, to allow for further generalizations to non-geometric examples of Holography.

\subsection{Mesonic operators and recursion relations}
We now recall an important aspect of the planar expansion, which was beautifully illustrated in the study of ``giant graviton'' D-branes \cite{Jiang:2019xdz,Budzik:2021fyh,Gaiotto:2024dwr}: the 't Hooft expansion in the presence of fundamental modifications depends on a choice of large $N$ D-brane saddle described by the expectation values of meson operators. 

In the case at hand, meson operators are bilinears of 1d fermions, possibly supported on different Wilson lines or at different locations of a single Wilson line. Gauge invariance requires the fermion insertions to be joined by open fundamental Wilson lines, supported on some path $\ell$ in the complement of the original knot/link. This defines the mesonic operators $R^{p,p'}_\ell$ mentioned in the introduction. To avoid confusion, we will refer to the original $\Lambda^\bullet \bC^N$ Wilson lines as ``strands'' and the ones appearing in mesonic operators as ``rungs''. Note that the $k_i$ labels on the strands jump by $\pm 1$ across fermion insertions. 

At the leading order in the 't Hooft expansion, each mesonic operator insertion can be dressed by a factor of $(\fq-\fq^{-1})$ and then replaced by a finite vev $r^{p,p'}_\ell$, which is independent of the presence of other mesonic insertions. The only subtlety is a ``charge conservation'' constraint: the total number of fermion and anti-fermion insertions along a strand must be $0$ for the rule to apply, otherwise the correlation function vanishes. As a consequence, the $r^{p,p'}_\ell$ rung vevs are only defined up to a collective rescaling 
\begin{equation}
    r^{p,p'}_\ell \to r^{p,p'}_\ell c_p c_{p'}^{-1}
\end{equation}
where $c_p$ is an invertible locally constant function on $K$. It is convenient to treat the $r^{p,p'}_\ell$ as sections of some flat line bundle on $K$ with holonomy $\omu$, promoting the rescaling to a local   $GL(1)$ gauge invariance along $K$.

We identify fermionic insertions (in some implicit renormalization scheme) with special cases of the 
\begin{equation}
    W_{\Lambda^{k} \bC^N} \times W_{\Lambda^{l} \bC^N} \leftrightarrow W_{\Lambda^{k+l} \bC^N}
\end{equation}
trivalent junctions defined and studied in \cite{Cautis_2014} with the help of the representation theory of the $U_\fq(\sl_N)$ quantum group. The main difference between our setup and the reference is that we will assign Grassmann parity $k\;\mathrm{mod}\;2$ to the representations $\Lambda^k \bC^N$, which results in some extra signs detailed below. 
From this point on, we will assume $N$ is even, so that the trivial $\Lambda^N \bC^N$ representation has the same Grassmann parity as $\Lambda^0 \bC^N$. 

The reference gives us a collection of local rules, which we can employ to reorganize the topology of a collection of mesonic insertions, say by moving endpoints across each other, passing a fundamental rung across other rungs or strands, and creating/removing contractible rungs. See Appendix \ref{app:rung_algebra} for a detailed discussion. 

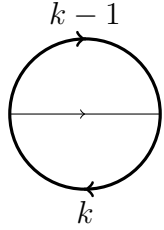
\begin{figure}[h]
\centering
\begin{tikzpicture}[baseline=-3]
\coordinate (A) at (1,0) {};
\coordinate (B) at (-1,0) {};
\coordinate (C) at (0,1) {};
\coordinate (D) at (0,-1) {};
\draw[mid>] (B) -- (A);
\draw[very thick] (0,0) circle (1);
\coordinate (A1) at (0.01,1) {};
\coordinate (B1) at (-0.01,1) {};
\draw[mid>, very thick] (B1) -- node[above] {$k-1$} (A1);
\coordinate (A2) at (-0.01,-1) {};
\coordinate (B2) at (0.01,-1) {};
\draw[mid>, very thick] (B2) -- node[below] {$k$} (A2);
\end{tikzpicture}
\caption{Unknot with a mesonic insertion.}
\label{fig:unknot}
\end{figure}

As a simple example, consider the unknot with a single contractible mesonic insertion. See Figure \ref{fig:unknot}. This can be contracted in two inequivalent manners, using either of the two ``bubble removal'' rules from Figure \ref{fig:rungs_moves}. The result is the crucial recursion relation we encountered before:
\begin{equation}
   [k]_\fq \langle W_{\minicircle,\Lambda^k \bC^N} \rangle = [N-k+1]_\fq \langle W_{\minicircle,\Lambda^{k-1} \bC^N} \rangle \, .
\end{equation}

\begin{figure}[h]
    \centering
    \begin{equation}
        \begin{tikzpicture}[baseline=20]
        \foreach \n in {0,...,3} {
        	\coordinate (z\n) at (0.4*\n, 0.8*\n);
        }
        \draw[very thick, mid>] (z0) -- node[right] {$k$} (z1);
        \draw[very thick, mid>] (z2) -- node[right] {$k$} (z3);
        \draw[very thick, mid>] (z1) to[out=150,in=-190] node[left] {$k{-}1$} (z2);
        \draw[mid>] (z1) to[out=-30,in=0] node[right] {$1$} (z2);
        \end{tikzpicture}
        = [k]_\fq
        \tikz[baseline=20]{\draw[very thick, mid>] (0,0) -- node[right] {$k$} (1,2);}\quad\quad\quad
        \begin{tikzpicture}[baseline=20]
        \foreach \n in {0,...,3} {
        	\coordinate (z\n) at (0.4*\n, 0.8*\n);
        }
        \draw[very thick,mid>] (z0) -- node[right] {$k$} (z1);
        \draw[very thick,mid>] (z2) -- node[right] {$k$} (z3);
        \draw[mid<] (z1) to[out=150,in=-190] node[left] {$1$} (z2);
        \draw[very thick,mid>] (z1) to[out=-30,in=0] node[right] {$k{+}1$} (z2);
        \end{tikzpicture} = [N-k]_\fq
        \tikz[baseline=20]{\draw[very thick,mid>] (0,0) -- node[right] {$k$} (1,2);}
        \label{eq:rung_bigon}
    \end{equation}
    \caption{The ``bubble removal'' rules.}
    \label{fig:rungs_moves}
\end{figure}

We can attempt a similar manipulation for any knot $K$, starting from a contractible meson insertion onto a loop with label $k$ and deforming it to a contractible insertion on a loop with label $k-1$. In the process, we will typically need to pass the rung across $K$ or itself multiple times, producing additional terms with multiple rung insertions. We can then attempt to contract away the extra rungs, etc. 

At finite $N$, this procedure will create an increasing collection of linear relations between correlation functions with multiple rung insertions (``open loop equations''). We will demonstrate later on that this procedure ends: all decorated correlation functions can be reduced to undecorated ones, and in the process they determine the $\langle W_{K;\Lambda^{k} \bC^N}\rangle$ correlation functions for all $k$.

We will learn how to take the planar limit of each elementary operation in the process, so that the planar open loop equations can be derived without any finite $N$ intermediate step. The result is a collection of polynomial relations between rung vevs $r^{p,p'}_\ell$, with coefficients which are Laurent polynomials in $\olambda$, $\omu$ and $g$. Eliminating the rungs, we obtain relations which we expect to reproduce the augmentation variety $\cL_K$. A similar strategy can be applied to links and even ``networks'' of $\Lambda^\bullet \bC^N$ Wilson lines joined at junctions.

\subsection{Planar tree-level}

As mentioned in the Introduction, the planar open loop equations simplify further in the $g \to 1$ limit (``tree-level planar limit''). In that limit (adjusting some signs):
\begin{enumerate}
    \item The framing of open fundamental lines is immaterial. 
    \item The difference between a rung over-crossing a strand from the right and under-crossing it is the combination of two rungs ending and starting on that strand: 
\begin{equation}
    \begin{tikzpicture}[scale=1, baseline=(current bounding box.center)]
    \node (C) at (0,1) {};
    \node (D) at (0,-1) {};
    \node (A) at (1,0) {};
    \node (B) at (-1,0) {};
    \node (i) at (intersection of A--B and D--C) {};
    \draw[mid>] (A) -- (B);
    \draw[very thick, mid>] (D) -- (i);
    \draw[very thick, mid>] (i) -- (C);
  \end{tikzpicture}
  \quad-\quad
  \begin{tikzpicture}[scale=1, baseline=(current bounding box.center)]
    \node (C) at (0,1) {};
    \node (D) at (0,-1) {};
    \node (A) at (1,0) {};
    \node (B) at (-1,0) {};
    \node (i) at (intersection of A--B and D--C) {};
    \draw[mid>] (A) -- (i);
    \draw[mid>] (i) -- (B);
    \draw[very thick, mid>] (D) -- (C);
    \end{tikzpicture}
  \quad=\quad
  \begin{tikzpicture}[scale=1, baseline=(current bounding box.center)]
    \draw[very thick, mid>] (0,-1) -- (0,1);
    \draw[mid>] (1,0.5) -- (0,0.5);
    \draw[mid>] (0,-0.5) -- (-1,-0.5);
  \end{tikzpicture}
  \quad=\quad
  \begin{tikzpicture}[scale=1, baseline=(current bounding box.center)]
    \draw[very thick, mid>] (0,-1) -- (0,1);
    \draw[mid>] (1,-0.5) -- (0,-0.5);
    \draw[mid>] (0,0.5) -- (-1,0.5);
    \end{tikzpicture}
\end{equation}
    \item Rotating the endpoint of a rung by an angle of $\pi$ around a strand costs a factor of $\olambda$ associated to that strand.
\end{enumerate}
The relations satisfied by the rung vevs precisely reproduce the relations satisfied by the transport coefficients of certain 3d flat connections defined on the complement of the knot. 

Indeed, consider a 3d flat connection whose monodromy $M_p$ around the knot at a point $p$ satisfies 
\begin{equation}
    M_p = 1+ \oell_p^L \oell_p^R \, ,
\end{equation}
for a left eigenvector $\oell_p^L$ and a right eigenvector $\oell_p^R$, normalized so that 
\begin{equation}
    \oell_p^R \cdot \oell_p^L = \olambda^2-1 \, ,
\end{equation}
where $\olambda^2$ is the non-trivial monodromy eigenvalue. This is a ``minimal regular'' defect for a 3d flat connection. 

Given such a connection, we can associate to each open path $\ell$ between points $p$ and $p'$ on the knot the inner product $t_{\ell} \equiv \oell^R_p \cdot \oell^L_{p'}$ computed along $\ell$.\footnote{In order to lighten the notation, we leave implicit the path-ordered exponential computing the holonomy from $p'$ to $p$ along $\ell$.} We can also define traces of the holonomy along closed paths $\ell$. These quantities satisfy simple relations when $\ell$ is carried across the knot. For example, $t_{\ell}$ jumps by a product of two $t$'s ending at the intersection point. They also get multiplied by $\olambda^2$ whenever the endpoints of $\ell$ rotate around the knot. 

These relations are identical to the tree-level planar skein relations, with the rung vev $r^{p,p'}_\ell$ identified with $t_{\ell}$. We thus find a one-to-one correspondence between tree-level planar saddles for $W_{K,\Lambda^\bullet \bC^N}$
and representations of the knot group, i.e. the fundamental group of the knot complement $M_3/K$, with minimal regular holonomy around the knot. The $\omu$ parameter is read off as the holonomy of $\oell^L$ along the knot. 

As we also mentioned in the Introduction and we will review further in the second half of the paper, the holographic dual description of $W_{K,\Lambda^\bullet \bC^N}$ in the tree-level planar limit involves A-model D-branes in $T^* M_3$ which go to infinity along the co-normal bundle to $K$. Such D-branes can be directly defined in terms of 3d flat connections on the knot complement, bypassing a geometric description. 

The tree-level planar skein relations are sometimes organized into the definition of a ``chord algebra'' \cite{cieliebak2017knotcontacthomologystring}, which at least for $M_3 = S^3$ is proven to match the construction of Lagrangian submanifolds in $T^* S^3$ with appropriate asymptotics. 

It is interesting to compare the considerable amount of mathematical effort required to obtain such geometric descriptions of the dual D-branes and the relatively straightforward presentation as 3d flat connections. Mathematically, the latter description uses implicitly some powerful theorems about A-branes \cite{nadler2009microlocalbranesconstructiblesheaves}. Physically, it describes complicated configurations of D-branes as a systematic deformation of simpler ones. 

As a simple example of this discussion, consider the trefoil knot from Figure \ref{trefoilone}. The fundamental group of the knot complement is well known. Define some generators $a$, $b$ and $c$ associated with loops that start above the figure and pass behind one of the three arcs in the picture, from right to left if the strand is pointing upwards. Then we have relations $a = c b c^{-1}$, $b = a c a^{-1}$, and $c = b a b^{-1}$ around the three crossings.  

If we write $M_a = 1 + \oell^L_a \oell^R_a$, etcetera, as the corresponding monodromy matrices, the relations become
\begin{equation}
    \oell^L_a \left[\oell^R_a + (\oell^R_a\cdot \oell^L_c)\oell^R_c \right] = \left[\oell^L_b + (\oell^R_c\cdot \oell^L_b)\oell^L_c \right]\oell^R_b   \, ,
\end{equation}
and cyclic rotations thereof. Contracting these relations with all possible $\oell^L$'s and $\oell^R$'s gives cubic relations among inner products, which reproduce all the non-trivial skein relations available near each of the three crossings. 

All three $M$'s have the same non-trivial eigenvalue $\olambda^2$. The monodromy $\omu$ of the eigenline can be computed by following a full loop, 
as the action of $M_c M_b M_a$ on the eigenline $\oell_b$. But that is the same as $\olambda^2 M_c^2 \oell_b$, so $\oell_b$ is an eigenvector of $M_c^2$. 
It is easy to see that unless $\olambda^2=1$, the only option is for all three vectors to be proportional to each other, and thus $\omu = \olambda^6$
is the $g \to 1$ augmentation variety. The 3d flat connection can be taken to have rank $1$.

\section{The rung algebra as a quantum group} \label{sec:cQG}
In order to systematize the process further, we now ``elongate'' the knot/link along the $x^3$ direction to reduce it to a sequence of braiding and fusion operations on collections of parallel strands. 

For most values of $x^3$, we keep the strands close to the $x^2=0$ plane, well-separated along the $x^1$ direction and pointing towards positive $x^3$. 
We braid individual pairs of strands at definite locations along $x^3$. We end consecutive pairs of strands at large positive and negative $x^3$ by smooth ``cups'' and ``caps''. See Figures \ref{fig:eloun} and \ref{fig:badun}. 

\begin{figure}[h]
    \centering
    \begin{tikzpicture}[baseline=(current bounding box.center)]
        \cupmodule{0}{1}
        \neutralmodule{0}{$k$,$N\!-\!k$}
        \capmodule{1}{1}
    \end{tikzpicture}
    \hspace{1cm}
    \begin{tikzpicture}[baseline=(current bounding box.center)]
        \cupmodule{0}{1}
        \neutralmodule{0}{$k$,$N\!-\!k$}
        \Erung{0}{1}
        \capmodule{1}{1}
    \end{tikzpicture}
    \hspace{1cm}
    \begin{tikzpicture}[baseline=(current bounding box.center)]
        \cupmodule{0}{1}
        \neutralmodule{0}{$k$,$N\!-\!k$}
        \Frung{0}{1}
        \capmodule{1}{1}
    \end{tikzpicture}
    \caption{Left: The elongated unknot. The ``tags'' on the semi-circles represent a fusion into the trivial $\Lambda^N \bC^N$ representation, in the convention of \cite{Cautis_2014}. They are useful in keeping track of some important signs. This convention produces the standard unknot times $(-1)^k$. Middle and right: the two non-trivial rungs ``$E$'' and ``$F$'' which can be placed at some $x^3$. Above the $E$ rung, the left and right labels are $k+1$ and $N-k-1$ respectively.}\label{fig:eloun}
\end{figure}
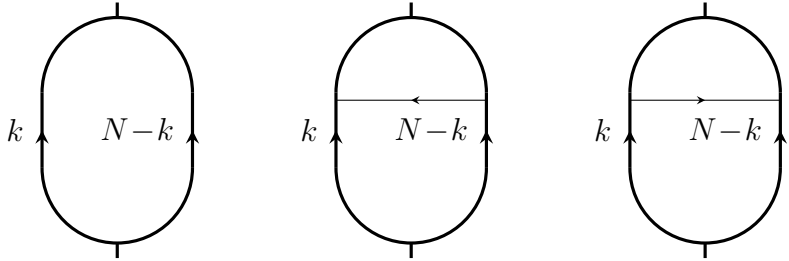
\begin{figure}[h]
    \centering
    \begin{tikzpicture}[baseline=(current bounding box.center)]
        \cupmodule{0}{2}
        \neutralmodule{0}{$k$,$N\!-\!k$,$N\!-\!k$,$k$}
        \braidmodule{1}{2}
        \capmodule{3}{2}
    \end{tikzpicture}
    \caption{Unknot with a twist. The choice of tags produces again an extra factor of $(-1)^k$.}\label{fig:badun}
\end{figure}
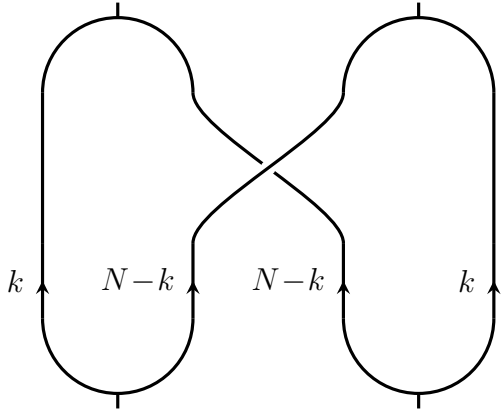

The basic idea is to first characterize the behavior of rungs localized at some fixed generic $x^3$, then relate rungs above and below braid operations, and then analyze the interaction between rungs and strands. 

Accordingly, consider a collection of $m$ parallel strands extended along the $x^3$ direction and placed at $x^2=0$ with a specific relative order along the $x^1$ direction, each defined from a set of $N$ complex fermions. We can define a ``rung algebra'' over $\bZ[\fq, \fq^{-1}]$, whose elements are collections of fundamental rungs supported within a finite $x^3$ interval, modulo local manipulations. We add to the algebra the evaluation operators $K_i^{\pm 1} = \fq^{\pm k_i}$, which read out the fermion number along each strand. The algebra operation is simply concatenation along the $x^3$ direction.

Following \cite{Cautis_2014}, we can generate the whole algebra from rungs $E_i$ and $F_i$ stretched between consecutive strands, together with the $K_i^{\pm 1}$. As reviewed in Appendix \ref{app:rung_algebra}, these generators satisfy relations independent of $N$ which coincide with the standard definition of the quantum group $U_\fq(\gl_m)$. They also satisfy some extra relations depending on $N$, characteristic of the representation of $U_\fq(\gl_m)$ on the $N$-th power of $\Lambda^\bullet \bC^m$. 

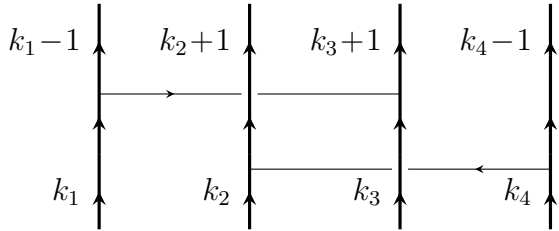
\begin{figure}[h]
    \centering
    \begin{tikzpicture}[baseline=(current bounding box.center)] 
        \Emrung{1}{2}{4}
        \neutralmodule{0}{$k_1$,$k_2$,$k_3$,$k_4$}
        \Fmrung{2}{1}{3}
        \neutralmodule{1}{,,,}
        \neutralmodule{2}{$k_1\!-\!1$,$k_2\!+\!1$,$k_3\!+\!1$,$k_4\!-\!1$}
    \end{tikzpicture}
    \caption{The $F_{-;1,3}E_{-;4,2}$ operator.
    }\label{fig:mrung}
\end{figure}

The quantum group $U_\fq(\gl_m)$ has a ``PBW'' linear basis consisting of normal-ordered products of $K_i^{\pm 1}$ as well as $m(m-1)$ generators $E_{-;i,j}$ and $F_{-;i,j}$, which we can identify with rung operators that join any pair of strands while passing behind intermediate strands. See Figure \ref{fig:mrung}. 

The identification of basis elements with combinations of rungs allows us to formulate a simple proposal for the planar limit of the rung algebra: the rescaled rungs 
\begin{equation}
    e_{-;i,j} \equiv (\fq-\fq^{-1}) E_{-;i,j} \qquad \qquad f_{-;i,j} \equiv (\fq-\fq^{-1}) F_{-;i,j} \, ,
\end{equation}
and $K_i$ all have a good planar limit. 

As commutators of the above generators are all proportional to $(\fq- \fq^{-1})$, the planar limit is a classical limit of $U_\fq(\gl_m)$ in the sense of deformation quantization: we can define a Poisson algebra consisting of polynomials in $e_{-;i,j}$, $f_{-;i,j}$ and $K_i^\pm$, with a Poisson bracket that captures 
the leading term in the commutators. 

This is interpreted as the Poisson algebra of holomorphic functions on a complex phase space $\CP_m$, which is denoted in the mathematical literature as the dual Poisson Lie group $\mathrm{\mathrm{\mathrm{GL}}}_m^*$: the space of pairs of upper triangular and lower triangular matrices with equal diagonal. See Appendix \ref{app:braiding_algebra} for details. We will come back to reality conditions on rung vevs at a later stage.

This has the following implication. Take some knot or link $L$ and slice it by a plane intersecting $m$ strands, roughly ordered along a line. If the strands point in the wrong direction, we can flip them by the duality operation $k \to N-k$ at the price of inserting ``tags'' as detailed in \cite{Cautis_2014}.  Then the information contained in the planar vevs of all the possible rungs drawn in the neighborhood of the plane can be distilled into a point in $\CP_m$. 


Using the rules of \cite{Cautis_2014}, we can compute what happens when a rung operator is transported across a braid, as well as relations between rung operators which appear when they are brought to a collection of cups or caps. We do so in Appendix \ref{app:rung_algebra}. The relations we find have a good planar limit. They give an action of the braid group on $\CP_m$ \cite{boalch2002gbundlesisomonodromyquantumweyl} and constraints on rung vevs enforced by cups and caps. The relations define certain submanifolds 
\begin{equation} 
    \cL_{\mathrm{cups}}\subset \CP_m \times (\bC^*)^{2s}  \, ,
\end{equation}
and $\cL_{\mathrm{caps}}\subset \CP_m$. The $(\bC^*)^{2s}$ factor contains the information about $\olambda$ and $\omu$ for the $s$ connected components of the link. We find the augmentation variety $\cL_K$
by intersecting these constraints. Schematically:\footnote{Strictly speaking, the intersection contains a bit more information than the augmentation variety: a given point in $\cL_K$ may lift to multiple points in $\CP_m$. We have not seen this happen in examples, but it is a possibility.}
\begin{equation}\label{eq:planarm}
   \cL_K = \cL_{\mathrm{caps}} \cap (B_{\mathrm{braid}} \circ \cL_{\mathrm{cups}}) \subset (\bC^*)^{2s} \, .
\end{equation}
The braid group action on $\CP_m$ is rather intuitive: the rung endpoints are braided in the obvious way following the strands, we apply skein relations if needed to push the rungs behind the strands, and include some extra framing factors when the horizontal orientation of a rung is flipped. The cup and cap constraints are less intuitive. We will come back to them momentarily.

At first sight, it is not obvious that this computational scheme should always produce the desired outcome. For example, one may wonder if skein relations involving non-horizontal rungs may impose further constraints. The Lagrangian nature of the augmentation variety for links is also not manifest. 

We will now address these concerns by taking into account some extra constraints on rung vevs which hold for any compact knot, link or network.

\subsection{Casimir operators and charge at infinity}
The $U_\fq(\gl_m)$ has $m$ Casimir elements which generate the center. With some work, it is possible to relate these Casimir elements to the vevs of circular Wilson loops which wrap around the whole collection of $m$ strands. These can be expressed in terms of rungs via skein relations but also obviously commute with any other rung operators. 

When the strands are part of a compact knot, these circular Wilson loops are contractible and have a vev equal to the quantum dimension of their representation label. This constrains the Casimir elements of $U_\fq(\gl_m)$ to specific values. There are also further constraints. For example, an $E_{-;m,1}$ generator could be passed  below the rest of the knot to become a rung passing in front of all strands and then pushed across the strands with skein relations, giving an extra polynomial constraint. Or a circular Wilson loop surrounding the first $s$ strands can be deformed to a loop surrounding the remaining $m-s$.

These constraints are most easily characterized by compactifying the directions transverse to the strands to an $S^2$. This operation makes no difference for compact knots, as their vev would be the same in $\bR^3$ and in $S^2\times\bR$. It allows us to introduce a useful environment for our calculations: the Hilbert space of the system of $m$ Wilson lines in $S^2$. 

The Hilbert space can be computed by borrowing another result from \cite{Cautis_2014}, as the $U_\fq(\sl_N)$-invariant part of the tensor product $\Lambda^{\bullet}\bC^{m N}$ of representations associated with individual strands. This is the quantum-corrected version of the naive classical answer in Chern-Simons gauge theory: the $\mathrm{SL}(N)$-invariant part of the Fock space for the system of 1d fermions.

Namely, ``Quantum Skew Howe Duality'' decomposes the exterior algebra $\Lambda^{\bullet}\bC^{m N}$ as a direct sum of products of irreducible representations of $U_\fq(\sl_N)$ and $U_\fq(\gl_m)$:
\begin{equation}
    \Lambda^{\bullet}\bC^{m N} = \oplus_Y R_Y[\sl_N] \otimes R_{Y^t}[\gl_m] \, 
\end{equation}
where $Y$ is a Young Tableau fitting in an $m \times N$ rectangle, 
$R_Y[\sl_N]$ is the irreducible representation of $U_\fq(\sl_N)$ labelled by $Y$ and $R_{Y^t}[\gl_m]$ is the irreducible representation of $U_\fq(\gl_m)$ labeled by the transpose Young Tableau. 

In order for $R_Y[\sl_N]$ to be the trivial representation, $Y$ must be a rectangle with height $N$. We will denote as $n$ the number or columns.
The Hilbert space is thus a direct sum of irreducible finite-dimensional unitary representations $\cH_{m,n}$ of $U_\fq(\gl_m)$, labeled by rectangular Young Tableaux with $N$ columns of height $n$ ranging from $0$ to $m$. The sum of the strands' labels in each sector is $n N$, so strands belonging to a knot or link have $n=m/2$. We will sometimes consider more general compact networks of $\Lambda^{\bullet}\bC^N$ Wilson lines, for which other values of $n$ are possible. Extreme cases $n=0$ and $n=m$ correspond to setting all labels to $0$ or $N$, which are configurations annihilated by all of the $E_i$ and $F_i$ rung operators.  

A collection of ``cups'', or any other way to end a collection of $m$ strands with total label $n N$, creates by definition a state in $\cH_{m,n}$. The braid operations also act as linear operators on $\cH_{m,n}$. The vev of an elongated knot, link or network $K$ can be computed as the expectation value of a braid operator between states built by cups and caps. Schematically:
\begin{equation}\label{eq:qm}
    \langle K\rangle = \langle \mathrm{caps}|B_{\mathrm{braid}}|\mathrm{cups}\rangle \, .
\end{equation}
As the Hilbert space $\cH_{m,n}$ is an irreducible representation of 
$U_\fq(\gl_m)$, the inner products between any two states are uniquely determined by the $U_\fq(\gl_m)$ action on the states. 

As we will illustrate in examples, this means that the norm of cup and cap states and the expectation values of braid operators between cup and cap states are uniquely determined by recursion relations produced by comparing the action of horizontal rung operators on the future and past of the knot, using the results of Appendix \ref{app:rung_algebra}.

The planar computational strategy (\ref{eq:planarm}) is thus the classical limit of the finite $N$ computation (\ref{eq:qm}). Both the finite $N$ and planar computations benefit from a further observation: the rung operators 
satisfy extra relations when acting on $\cH_{m,n}$. At finite $N$, the relations define a quotient $A_{m,n}$ of $U_\fq(\gl_m)$, a simpler algebra acting irreducibly on $\cH_{m,n}$. 

More importantly for us, the planar relations cut out a complex symplectic slice $\CP_{m,n}$ of $\CP_m$ (the planar limit of $A_{m,n}$) and all of the ingredients of (\ref{eq:planarm}) can be defined within $\CP_{m,n}$. Determining the explicit form of $A_{m,n}$ and $\CP_{m,n}$ is an interesting exercise in representation theory. We bypass the exercise by invoking some Supersymmetric Quantum Field Theory technology in  Appendix \ref{app:classS}.

Namely, we identify $A_{m,n}$ and $\CP_{m,n}$ with the fusion algebras of BPS line operators in certain theories of class S \cite{Kapustin:2007wm,Gaiotto:2010be}, which in turn gives us multiple interpretations of $\CP_{m,n}$ as moduli spaces of 2d flat connections. We also identify $\CP_{m,n}$ as the moduli space of certain A-model D-branes with a specific asymptotic shape in the deformed $A_1$ singularity $\cM_t$, which is perfectly consistent with the holographic duality we are trying to establish. 

We will not attempt to prove that $\CP_{m,n}$ precisely captures the constraints on $\CP_m$ arising in our context. Indeed, we do not need to do so: the cups and caps constraints on $\CP_{2n}$ land automatically in $\CP_{2n,n}$ and the braid group action on $\CP_{2n}$
preserves $\CP_{2n,n}$. Accordingly, we do not lose any generality by restricting to $\CP_{m,n}$, and we gain a conjectural holographic interpretation of our calculations.  

Furthermore, $\cL_{\mathrm{caps}}$ and $\cL_{\mathrm{cups}}$ are Lagrangian submanifolds, and the braid group acts by symplectomorphisms of $\CP_{m,n}$ (the classical limit of automorphisms on the quantum group). The augmentation variety is thus an intersection of Lagrangian correspondences, which is automatically a Lagrangian submanifold of $(\bC^*)^{2s}$.

We will now work our way through some illustrative examples. 

\subsection{The \texorpdfstring{$\CP_{2,1}$}{P(2,1)} phase space}
\begin{figure}[h]
    \centering
    \begin{equation*}
        \begin{tikzpicture}[baseline=40]
        \Erung{1}{1}
        \Frung{0}{1}
        \neutralmodule{0}{$k_1$,$k_2$}
        \neutralmodule{1}{,}
        \neutralmodule{2}{$k_1$,$k_2$}
        \end{tikzpicture}\quad
        = 
        \begin{tikzpicture}[baseline=40]
        \Erung{0}{1}
        \Frung{1}{1}
        \neutralmodule{0}{$k_1$,$k_2$}
        \neutralmodule{1}{,}
        \neutralmodule{2}{$k_1$,$k_2$}
        \end{tikzpicture}\quad
        + [k_1-k_2]_\fq 
        \begin{tikzpicture}[baseline=40]
        \neutralmodule{0}{$k_1$,$k_2$}
        \neutralmodule{1}{,}
        \neutralmodule{2}{$k_1$,$k_2$}
        \end{tikzpicture}
    \end{equation*}
    \caption{The square-switch relation.}
    \label{fig:square-switch}
\end{figure}
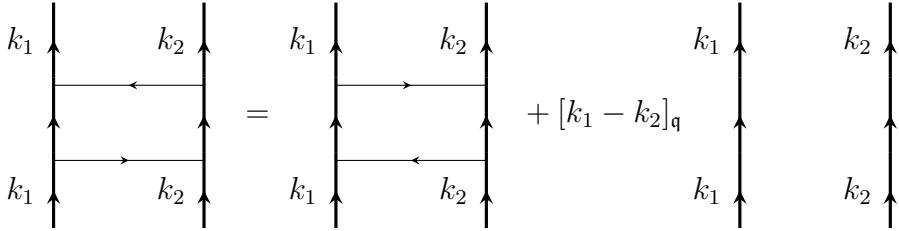

We can illustrate first the case $m=2$, $n=1$. The generators of $U_\fq(\gl_2)$ here are $K_1 = \fq^{k_1}$, 
$K_2 = \fq^{k_2}$, and the two rung operators $E$ and $F$. The action of the rung operators shifts the $k_i$:
\begin{align}
    K_1 E &= \fq E K_1  \cr
    K_2 E &= \fq^{-1} E K_2  \cr
    K_1 F &= \fq^{-1} F K_1 \cr
    K_2 F &= \fq F K_2 \, .  
\end{align}
The ``square-switch'' relation of Figure \ref{fig:square-switch} is $E F = F E + [k_1-k_2]_\fq$
and gives the final $U_\fq(\gl_2)$ relation:
\begin{equation}
    [E,F] = \frac{K_1 K_2^{-1} - K_1^{-1} K_2}{\fq - \fq^{-1}} \, .
\end{equation}

The rescaled rungs $e=(\fq - \fq^{-1}) E$ and $f=(\fq - \fq^{-1}) F$
have a good planar limit. 

The space $\CP_2$ is parameterized by (the planar limits of) $K_1$, $K_2$, $e$ and $f$, with $K_1$ and $K_2$ invertible, and non-trivial Poisson brackets:
\begin{align}
    \{K_1,e\} &= \frac12 K_1 e  \cr
    \{K_1,f\} &= -\frac12 K_1 f  \cr
    \{K_2,e\} &= -\frac12 K_2 e \cr
    \{K_2,f\} &= \frac12 K_2 f \cr
    \{e,f\} &= K_1 K_2^{-1} - K_1^{-1} K_2 \, .
\end{align}
If the strands are part of a knot or placed in a transverse $S^2$, 
the strand labels are forced to be both $0$, both $N$, or to add up to $N$. We are interested here in the last case, relevant e.g. for the recursive computation of the unknot as in Figure \ref{fig:eloun}. Then $K_1 K_2 = g$. 

The remaining relations are most easily seen from the action on cups and caps. See Appendix \ref{app:rung_algebra} for details. We denote a cup with label $k$ on the left as $|k\rangle$ and a cap with label $k$ on the left as $\langle k|$, anticipating a vector space interpretation.

In this notation, the $U_\fq(\gl_2)$ acts on cups as
\begin{align} \label{eq:gl2_cups}
    K_1 |k\rangle &= |k\rangle \fq^k  \cr
    K_2 |k\rangle &=  |k\rangle \fq^{N-k}\cr
    E |k\rangle &=  -|k+1\rangle [k+1]_\fq \cr
    F |k\rangle &=  -|k-1\rangle [N-k+1]_\fq\, .
\end{align}
These relations define the dimension $N+1$ irrep of $U_\fq(\gl_2)$. Similarly, 
\begin{align}
    \langle k| K_1 &= \fq^k \langle k|   \cr
    \langle k| K_2 &=\fq^{N-k} \langle k|  \cr
    \langle k| E &=  [N-k+1]_\fq \langle k-1| \cr
    \langle k| F &= [k+1]_\fq \langle k+1|  \, .
\end{align}
The Casimir operators of $U_\fq(\gl_2)$ act as constants: $K_1 K_2 = g$ and 
\begin{align}
    e f &= (K_1 - K_1^{-1})(\fq K_2 - \fq^{-1} K_2^{-1})  \cr
    f e &= (\fq K_1 - \fq^{-1} K_1^{-1})(K_2 - K_2^{-1}) \, .
\end{align}
as expected. This constraint defines a quotient of $U_\fq(\gl_2)$ we denote as $A_{2,1}$.

The recursive calculation of the unknot is the same as a recursive calculation of the $U_\fq(\gl_2)$-covariant inner products
\begin{equation}
    [k+1]_\fq \langle k+1|k+1\rangle = \langle k| F|k+1\rangle = -[N-k]_\fq \langle k|k\rangle \, .
\end{equation}

The braid transformation $B$ is computed in Appendix \ref{app:rung_algebra}:
\begin{align}
    K_1 B &= B K_{2}  \cr
    K_2 B &= B K_1  \cr
    K_1 E B &= B F K_1  \cr
    K_2 F B &= B E K_2 \, .
\end{align}
For example, the recursive computation of an unknot with a twist is the same as the recursive computation of non-zero matrix elements for $B$:
\begin{equation}
    \fq^{N-k} [k+1]_\fq \langle N-k-1|B|k+1\rangle = \langle N-k|K_1 E B|k+1\rangle = -\fq^{k+1}[N-k]_\fq \langle N-k|B|k\rangle \, ,
\end{equation}
i.e.
\begin{equation} \label{eq:tw_unknot_rec}
    \fq^{N-k} \frac{\langle N-k-1|B|k+1\rangle}{\langle k+1|k+1\rangle} =-\fq^{k+1}\frac{\langle N-k|B|k\rangle}{\langle k|k\rangle}  \, ,
\end{equation}
recovering the framing factor\footnote{The framing factor for $SU(N)_\kappa$ is really $\fq^{k(N-k)(1+N^{-1})}$, but we have stripped off some powers of $\fq^{N^{-1}}$ from $B$ to simplify the calculation. These will not affect the planar answers we are interested in.} $\fq^{k(N-k)}$ for $W_{\Lambda^k \bC^N}$.

In the planar limit, the Casimir constraints cut out a two-dimensional symplectic submanifold $\CP_{2,1}$ of $\CP_2$:
\begin{equation}\label{eq:casimirunknot}
  e f = (K_1 - K_1^{-1})(K_2 - K_2^{-1}), 
  \qquad
  K_1\,K_2 = g\,. 
\end{equation}
The cap enforces constraints 
\begin{equation}
    f= K_1 - K_1^{-1} \, , \qquad \qquad e = K_2 - K_2^{-1} \, , 
\end{equation}
which define a Lagrangian submanifold $\cL_{\mathrm{cap}} = \CC^t_1$ in $\CP_{2,1}$. 

We will ``keep track'' of the knot label $k$ at a point in the cup, so the cup constraints can be thought of as a Lagrangian correspondence $\cL_{\mathrm{cup}} = \CC^b_1$ between $\CP_{2,1}$ and the $\bC^* \times \bC^*$ auxiliary space parameterized by $\olambda$ and $\omu$:
\begin{equation}
    K_1 = \olambda  \qquad \qquad f = -\omu(K_2 - K_2^{-1})  \qquad \qquad e = -\omu^{-1}(K_1 - K_1^{-1}) \, .
\end{equation}
Intersecting the two constraints, we recover the augmentation variety $\cL_{\minicircle}$ in $\bC^* \times \bC^*$.

A final observation is that the $U_\fq(\gl_2)$ representations we encounter here are unitary for positive integer $N$, with $F = E^\dagger$
and $K_1^\dagger = g^{-1} K_2$, as well as $|k \rangle^\dagger = \langle k|$. This defines the Hilbert space $\cH_{2,1}$.  Recall that $\fq^\dagger =\fq^{-1}$ for real level. 

In the planar limit, these conditions define a real phase space $\CP_{2,1}^\bR$ in $\CP_{2,1}$:
\begin{equation}
    |K_1|^2 = 1  \qquad \qquad |e|^2 = g^{-1} K^2_1+ g K_1^{-2} -g -g^{-1} \, .
\end{equation}
which is a deformed version of a round $S^2$. At finite $N$, 
the representations of $U_\fq(\gl_2)$ are a natural quantization of $\CP_{2,1}^\bR$, akin to the fuzzy sphere construction.  See Appendix \ref{app:classS} for some more details.

\subsection{The \texorpdfstring{$\CP_{3,1}$}{P(3,1)} phase space}
A configuration of three strands would be relevant, say, to the recursive calculation of the expectation value of a simple network of Wilson lines, built with the help of the canonical junctions 
\begin{equation}
    W_{\Lambda^{k} \bC^N} \times W_{\Lambda^{l} \bC^N} \leftrightarrow W_{\Lambda^{k+l} \bC^N} \, ,
\end{equation}
and the duality pairing between $W_{\Lambda^{k} \bC^N}$ and $W_{\Lambda^{N-k} \bC^N}$ (``tags'' in \cite{Cautis_2014}). 

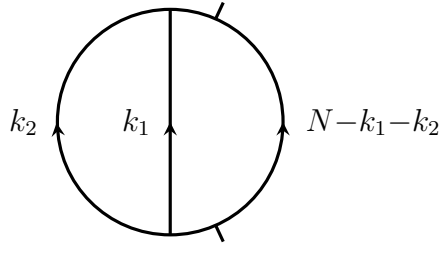
\begin{figure}[hbt!]
  \centering
  \begin{tikzpicture}[scale=1.5,>=stealth]
    \draw[very thick,
      decoration={
        markings,
        mark=at position 0.5 with {\arrow{>}}
      },
      postaction={decorate}
    ] (0,-1cm) -- (0,1cm);
    \draw[very thick,
      decoration={
        markings,
        mark=at position 0.52 with {\arrow{<}},
        mark=at position 1    with {\arrow{>}}
      },
      postaction={decorate}
    ] (0,0) circle (1cm);
    \draw[very thick] (0.4cm,0.91cm) -- (0.47cm,1.06cm);
    \draw[very thick] (0.4cm,-0.91cm) -- (0.47cm,-1.06cm);
    \node at (-0.3, 0)    {$k_1$};
    \node at (-1.3cm, 0)  {$k_2$};
    \node at ( 1.8cm,   0) {$N{-}k_1{-}k_2$};
  \end{tikzpicture}
  \caption{A network of three Wilson lines with trivalent junctions and tags.}
  \label{fig:wilnet}
\end{figure}

The simplest example of a network involves three strands with labels  $k_1$, $k_2$, $N-k_1-k_2$ joined at trivalent vertices (see Figure \ref{fig:wilnet}). It has a correlation function which is a $\fq$-trinomial
\begin{equation}
    \langle M_{\circlewithline;k_1,k_2} \rangle =  \qBinomial{N}{k_1+ k_2}\qBinomial{k_1+k_2}{k_1}  \, ,
\end{equation}
which can be derived from the recursion
\begin{equation}
    \langle M_{\circlewithline;k_1-1,k_2} \rangle = \frac{[k_1]_\fq}{[N-k_1-k_2+1]_\fq} \langle M_{\circlewithline;k_1,k_2} \rangle \, ,
\end{equation}
and analogous for $k_2$. The recursion can be derived as before by adding fundamental rungs to the network and contracting them in multiple ways. We expect the planar limit of general networks to have the same structure as for links. Here we get the augmentation variety $\cL_{\circlewithline}$:
\begin{align}
    \omu_1 &= \frac{\olambda_1 - \olambda^{-1}_1}{g \olambda_1^{-1} \olambda_2^{-1} - g^{-1} \olambda_1 \olambda_2}  \cr
    \omu_2 &= \frac{\olambda_2 - \olambda^{-1}_2}{g \olambda_1^{-1} \olambda_2^{-1} - g^{-1} \olambda_1 \olambda_2} \, .
\end{align}

\begin{figure}[h]
    \centering
    \begin{tikzpicture}[baseline=(current bounding box.center)] 
        \neutralmodule{0}{$k_1$,$k_2$,$k_3$}
        \Erung{0}{1}
        \neutralmodule{1}{,,}
        \Erung{1}{2}
        \neutralmodule{2}{,,}
        \Frung{2}{1}
        \neutralmodule{3}{,,}
        \Frung{3}{2}
        \Emrung{5}{1}{3}
        \neutralmodule{4}{,,}
        \Fmrung{6}{1}{3}
        \neutralmodule{5}{,,}
        \neutralmodule{6}{,,}
    \end{tikzpicture}
    \caption{The elementary rungs for $m=3$ and the stretched rungs. From the bottom: $E_1$, $E_2$, $F_1$, $F_2$, $E_{-;3,1}$, $F_{-;1,3}$.}\label{fig:threerung}
\end{figure}
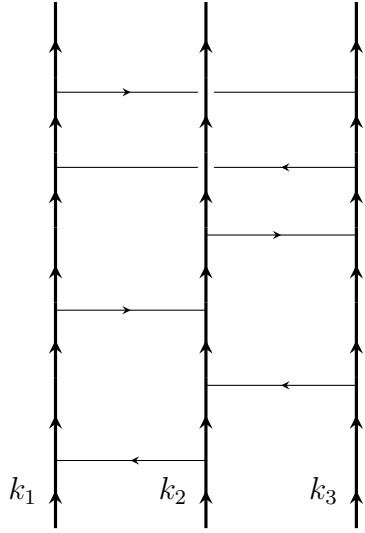

In an elongated setting, we have $K_1$, $K_2$ and $K_3$ Cartan generators as well as elementary rung generators $E_1$, $E_2$, $F_1$ and $F_2$. See Figure \ref{fig:threerung}.

The relations defining $U_\fq(\gl_3)$ include two $U_\fq(\gl_2)$ sub-algebras associated with the consecutive pairs of strands, together with $[E_1,F_2]=[E_2,F_1]=0$ and four Serre relations:
\begin{equation}
    E_1^2 E_2 - (\fq + \fq^{-1}) E_1 E_2 E_1 + E_2 E_1^2 = 0 \, ,
\end{equation}
together with $1\leftrightarrow 2$ and $E\leftrightarrow F$ permutations.

As a consequence, the algebra generated by $E_i$ rungs includes 
irreducible sequences $(E_1 E_2)^n$. It is convenient to introduce extra generators 
\begin{equation}
    E_{-;3,1}\equiv E_1 E_2 - \fq E_2 E_1  \qquad \qquad F_{-;1,3}\equiv F_2 F_1 - \fq^{-1} F_1 F_2 \, ,
\end{equation}
which have a graphical representation as rungs passing behind the 
second strand. See Figure \ref{fig:threerung}. It is easy to check that $U_\fq(\gl_3)$ then admits a ``PBW'' linear basis consisting of normal-ordered products of $K^{\pm 1}_i$, $E_i$, $F_i$, $E_{-;3,1}$ and $F_{-;3,1}$.

We expect all rescaled rung operators $e_\bullet \equiv (\fq - \fq^{-1}) E_\bullet$ and $f_\bullet \equiv (\fq - \fq^{-1}) F_\bullet$ and polynomials thereof to have a good planar limit.
The PBW basis statement then implies that the planar limit of the 
$U_\fq(\gl_3)$ rung algebra consists of polynomials in $K^{\pm 1}_i$, $e_i$, $f_i$, $e_{-;3,1}$ and $f_{-;3,1}$. We interpret them 
as holomorphic functions on a $9$-dimensional Poisson manifold $\CP_3$.

When the strands are part of a compact network or we work on a transverse $S^2$, we will get extra constraints among the generators. The analogues of the cups and caps in the $m=2$ case 
are now trivalent vertices $|k_1, k_2\rangle$ and $\langle k_1, k_2|$
between strands of labels $k_1$, $k_2$ and $N-k_1-k_2$.

With a bit of patience, one can recognize the action of $U_\fq(\gl_3)$ on trivalent vertices as an irreducible representation, the $N$-th symmetric power of the fundamental representation. Accordingly, the recursive calculation of 
\begin{equation}
    \langle k_1, k_2|k_1, k_2\rangle = \qBinomial{N}{k_1+ k_2}\qBinomial{k_1+k_2}{k_1}
\end{equation}
is simply reproducing the unique $U_\fq(\gl_3)$-covariant pairing between the irreducible representation and its dual. 

We compute some simple extra constraints:
\begin{align}
    K_1 K_2 K_3 &= g \cr
    e_1 f_1 &= (K_1-K_1^{-1})(\fq K_2- \fq^{-1} K_2^{-1})  \cr
    e_2 f_2 &= (K_2-K_2^{-1})(\fq K_3- \fq^{-1} K_3^{-1}) \, ,
\end{align}
etcetera. We denote the part of the rung algebra constrained to act on trivalent vertices as $A_{3,1}$. See Appendix \ref{app:classS} for a characterization of $A_{3,1}$ and of the 4-dimensional complex symplectic manifold $\CP_{3,1}$ which emerges in the planar limit. We will momentarily review a presentation as a space of 2d flat connections. 

The finite-dimensional irrep of $U_\fq(\gl_3)$ is unitary, leading to natural hermiticity conditions relating $E$'s and $F$'s.
In the planar limit, this defines a compact real locus $\CP^\bR_{3,1}$ (essentially, a deformed $\bCP^2$) which is naturally quantized by the finite $N$ irrep of $U_\fq(\gl_3)$. 

The braid group action computed in Appendix \ref{app:braiding_algebra} is easy to describe once we define rung operators passing in front of the second strand:
\begin{equation}
    E_{+;3,1}\equiv E_1 E_2 - \fq^{-1} E_2 E_1   \qquad \qquad F_{+;1,3}\equiv F_2 F_1 - \fq F_1 F_2 \, .
\end{equation}
Then the action of $B_1$ (braiding the first two strands) and $B_2$ (braiding the last two strands) is completely geometric, up to the extra factor of $K_i$ in the relations between $E_i$, $B_i$ and $F_i$. The planar limit of the braiding relations is defined in the same manner. 




\subsection{The \texorpdfstring{$\CP_{4,2}$}{P(4,2)} phase space}\label{sec:P42}

The case $m=4$, $n=2$ is interesting as it allows for the calculation of augmentation varieties of some simple non-trivial knots and links. We will illustrate the procedure for the Hopf link and the trefoil knot.

Again, $U_\fq(\gl_4)$ admits a PBW linear basis consisting of normal-ordered products of $K_i^{\pm 1}$, $E_{-;i,j}$ and $F_{-;i,j}$. Their planar limit can be interpreted as holomorphic functions in a 16-dimensional complex Poisson manifold $\CP_4$.

For clarity, we will use a superscript $U$ for the planar vevs of the elements of the algebra at the cups, and a superscript $A$ for their planar vevs at the caps. The shape of the braid determines the relation between the two sets. A pair of cups joining consecutive strands forces them to have labels $k_1$, $N-k_1$, $k_2$ and $N-k_2$. Accordingly,
\begin{align}
    K_1^U &= \olambda_1 \cr
    K_2^U &= g \olambda_1^{-1} \cr
    K_3^U &= \olambda_2 \cr
    K_4^U &= g \olambda_2^{-1} \, .
\end{align}
Furthermore (see Appendix \ref{app:rung_algebra} for details)
\begin{align}
    e_{-;2,1}^U &=- \omu_1^{-1} (\olambda_1 - \olambda_1^{-1})  \cr
    e_{-;3,1}^U &= \omu_1^{-1} \olambda_1 e_{2}^U  \cr
    e_{-;4,1}^U &= - \olambda_1 \olambda_2^{-1} \omu_1^{-1}\omu_2^{-1} e_{2}^U  \cr
    e_{-;4,2}^U&= - \omu_2^{-1} \olambda_2^{-1} e_{2}^U  \cr
    e_{-;4,3}^U&= - \omu_2^{-1} (\olambda_2 - \olambda_2^{-1})  \, ,
\end{align}
as well as
\begin{align}
    f_{-;1,2}^U &= - \omu_1 (g \olambda_1^{-1} - g^{-1} \olambda_1)   \cr
    f_{-;1,3}^U &= - \omu_1 g \olambda_1^{-1} f_{2}^U  \cr
    f_{-;1,4}^U &= - \omu_1 \omu_2 \olambda_1^{-1} \olambda_2 f_2^U  \cr
    f_{-;2,4}^U &= \omu_2 g^{-1} \olambda_2 f_2^U  \cr
    f_{-;3,4}^U &= - \omu_2 (g \olambda_2^{-1} - g^{-1} \olambda_2) \, .
\end{align}

For two caps with labels $k_1$ and $k_2$, we get 
\begin{align}
    K_1^A &= \olambda_1   \cr
    K_2^A &= g \olambda_1^{-1}  \cr
    K_3^A &= \olambda_2 \cr
    K_4^A &= g \olambda_2^{-1} \, ,
\end{align}
as well as 
\begin{align}
    e_{-;2,1}^A &=  (g \olambda_1^{-1} - g^{-1} \olambda_1)  \cr
    e_{-;3,1}^A &=g^{-1} \olambda_1 e_2^A  \cr
    e_{-;4,1}^A &= - \olambda_1 \olambda_2^{-1} e_2^A \cr
    e_{-;4,2}^A &= - g \olambda_2^{-1} e_2^A  \cr
    e_{-;4,3}^A &=  (g \olambda_2^{-1} - g^{-1} \olambda_2) \, ,
\end{align}
and
\begin{align}
    f_{-;1,2}^A &= (\olambda_1 - \olambda_1^{-1}) \cr
    f_{-;1,3}^A &= - \olambda_1^{-1} f_2^A  \cr
    f_{-;1,4}^A &= - \olambda_2 \olambda_1^{-1} f_2^A  \cr
    f_{-;2,4}^A &= \olambda_2 f_2^A \cr
    f_{-;3,4}^A &= (\olambda_2 - \olambda_2^{-1})  \, .
\end{align}

If we intersect the cup and cap constraints directly, we essentially recover two separate unknots. More precisely, the equations for $e_{-;2,1}$ and $e_{-;4,3}$ give the two separate augmentation varieties. The remaining equations force rungs stretched between the unknots to vanish for generic $\omu_i$. 

We can now insert some powers of $B_2$ to produce interesting knots. 
A single power of $B_2$ produces the twisted unknot of Figure \ref{fig:badun}. We immediately see that $\olambda_2 = g \olambda_1^{-1}$. We denote $\olambda_1=\olambda$. We set $\omu_2=1$, $\omu_1 = \omu$, as we only need to keep track of $k$ at one location. We can start from the cap constraint 
\begin{equation}
    e_{-;2,1}^A =  (g \olambda^{-1} - g^{-1} \olambda) \, ,
\end{equation} 
convert it across $B_2$ to $e_{-;3,1}^U$ and at the cup to
\begin{equation}
    e_{-;4,2}^U= - \omu g^{-1}  (g \olambda^{-1} - g^{-1} \olambda) \, .
\end{equation}

Next, we use
\begin{equation}
    e_{+;4,2} = e_{-;4,2} + e_{-;4,3} e_2
\end{equation}
to derive 
\begin{equation}
    e_{+;4,2}^U = -  \omu g \olambda^{-2} (g \olambda^{-1} - g^{-1} \olambda) 
\end{equation}
and convert it across $B_2$ to $e_{-;4,3}^A$ to finally obtain 
\begin{equation}
    (\olambda - \olambda^{-1}) = - \omu g \olambda^{-2} (g \olambda^{-1} - g^{-1} \olambda)  \, .
\end{equation}
The replacement $\omu \to \omu g \olambda^{-2}$ is due to the twisted framing of this unknot. 



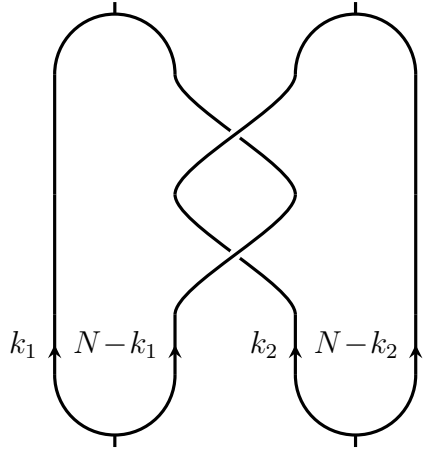
\begin{figure}[h]
    \centering
    \begin{tikzpicture}[baseline=(current bounding box.center), scale=0.8]
        \cupmodule{0}{2}
        \neutralmodule{0}{$k_1$,$N\!-\!k_1$,$k_2$,$N\!-\!k_2$}
        \braidmodule{1}{2}
        \braidmodule{3}{2}
        \capmodule{5}{2}
    \end{tikzpicture}
    \caption{The elongated Hopf link.}\label{fig:elhopf}
\end{figure}

Two powers of $B_2$ give us the Hopf link of Figure \ref{fig:elhopf}. There are two types of solutions. If the rungs stretched between the two components of the link (in either direction) vanish, then the rungs for each component can be manipulated as if the other component were absent, reproducing the combination of two unknots (``disconnected saddle''). 

We thus explore the option that some rung vevs stretched between the two components may not vanish (``connected saddle''). An $f^U_{-;1,3}$ rung can be brought across $B_2^2$ and converted into $f_{2}^A$ at the caps, but also first converted into $f_{2}^U$ at the cups and then brought across $B_2^2$. This gives the relation 
\begin{equation}
    \omu_1 f_{2}^A=g\olambda_2^{-2}f_{2}^A \, .
\end{equation}
Analogously, starting with $f_{-;2,4}^A$, we get
\begin{equation}
    \omu_2 f_{2}^A=g\olambda_1^{-2}f_{2}^A \, .
\end{equation}
Accordingly, off-diagonal rungs can be non-zero only if $\omu_1 =g\olambda_2^{-2}$ and $\omu_2=g\olambda_1^{-2}$. These relations are sufficient to satisfy all constraint equations. We compute the (charge-conserving combination of) off-diagonal vevs as well:
\begin{equation}
    e_{2}^Af_{2}^A =(g^{-1}-g) \olambda_2^{-1} \olambda_1^{-1} -g^{-1} (\olambda_1- \olambda_1^{-1})(\olambda_2 - \olambda_2^{-1})  \, .
\end{equation}



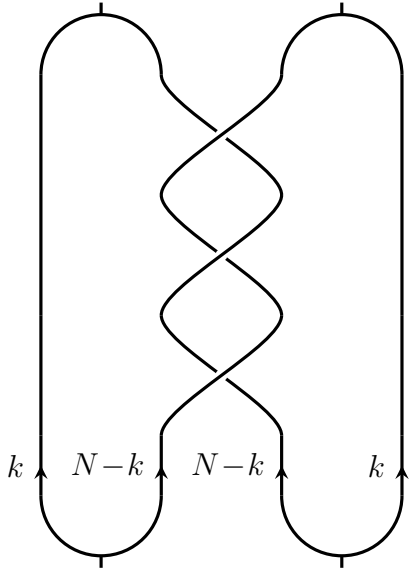
\begin{figure}[h]
    \centering
    \begin{tikzpicture}[baseline=(current bounding box.center), scale = 0.8]
        \cupmodule{0}{2}
        \neutralmodule{0}{$k$,$N\!-\!k$,$N\!-\!k$,$k$}
        \braidmodule{1}{2}
        \braidmodule{3}{2}
        \braidmodule{5}{2}
        \capmodule{7}{2}
    \end{tikzpicture}
    \caption{The elongated right-handed trefoil knot.}\label{fig:elthree}
\end{figure}

Finally, we can look at the right-handed trefoil knot of Figure \ref{fig:elthree} by inserting $B_2^3$. As in the twisted unknot example, we see that $\olambda_2 = g \olambda_1^{-1}$ and denote $\olambda_1=\olambda$ and set $\omu_2=1$, $\omu_1 = \omu$. We will derive some equations constraining $\omu$, $\olambda$, $e_{2}^A$ and $f_{2}^A$.

We start by noticing that $B_2$ converts $f_{2}$ to $e_{2}$ and vice versa, without extra factors in this case. A rung $f_{2}^U$ can then be brought to $e_{2}^A$, but can also be converted to $f_{-;1,4}^U$ at the cups, passed across the braid, and converted back to $f_{2}^A$. This gives the relation
\begin{equation}
    f_{2}^A = \omu e_{2}^A \, .
\end{equation}
We can also start with $f_{-;1,3}^U$ and pass it across $B_2^3$:
\begin{align}
    f_{-;1,3}^U \to f_{1} \to f_{+;1,3} = f_{-;1,3}-f_{1}f_{2} \to f_{1}^A - f_{+;1,3}^A e_{2}^A = f_{1}^A - f_{-;1,3}^A e_{2}^A + f_{1}^A f_{2}^A e_{2}^A\, .
\end{align}
Analogously, $f_{1}^U$ maps to $f_{-;1,3}^A(1+e_{2}^Af_{2}^A)-f_{1}^Af_{2}^A(2+e_{2}^Af_{2}^A)$. Applying the cups and caps, and converting the remaining $f_2^U$ to $e_2^A$, we get
\begin{align}
     g \omu e_{2}^A&=1-\olambda ^2 (1+e_{2}^A f_{2}^A)  \cr
    \omu  \left(g^2-\olambda ^2\right)&=g f_{2}^A \left(-1+ \olambda ^2 (2+e_{2}^A f_{2}^A)\right)\, ,
\end{align}
and eliminating the vevs we obtain the augmentation variety
\begin{equation}
    g^4 (g^2-\olambda^2)\omu^2+\left( g^4-g^4 \olambda^2 +2 g^4 \olambda^4   -2g^2 \olambda^4 -g^2 \olambda^6
    +\olambda ^8 \right) \omu+ g^2 \olambda^6(\olambda^2-1)=0\, .
\end{equation}
This matches the results in the literature after an appropriate redefinition of the variables. See the end of Appendix \ref{app:saddle} for details.

Note also
\begin{equation}
    e_2^A = \frac{g^2 - \olambda^4}{g(\olambda^4 + g^2 \omu)}\, ,
\end{equation}
so in this case every point in the augmentation variety lifts to a single point in $\cP_{4,2}$. This may not be true in more general examples.

\section{From rung vevs to character varieties}\label{sec:char}
In Appendix \ref{app:braiding_algebra} we use some extra ``spectator'' strands to give an $SU(N)$ Chern-Simons theory derivation of the relation between the planar limit $\CP_m$ of $U_\fq(\gl_m)$ and the Stokes data of an auxiliary 2d $\mathrm{\mathrm{GL}}(m)$ flat connection on the plane, with a ``rank 1'' irregular singularity at infinity and a regular singularity at the origin.  
This is a well-known property of $U_\fq(\gl_m)$ \cite{Gaiotto:2014lma,schrader2019clusterrealizationuqmathfrakslnquantum,Aamand:2019evs}, though it usually occurs in the study of (analytically continued) $\mathrm{\mathrm{GL}}(m)$ Chern-Simons theory. Here the latter interpretation will be a manifestation of holography, with the $\mathrm{\mathrm{GL}}(m)$ Chern-Simons theory being the world-volume theory of $m$ D-branes.

Concretely, we assemble the $-e_{-;i,j}$ generators into a triangular matrix $\hat e_-$ with unit diagonal. This matrix is obviously invertible. Its inverse $\hat e_+$ is a triangular matrix with off-diagonal elements $e_{+;i,j}$ that contains the vevs of rungs that pass in front of intermediate strands, instead of behind. Analogously, we can form a matrix $\hat f_-$ with unit diagonal and off-diagonal elements $f_{-;i,j}$. The inverse $\hat f_+$ has entries $-f_{+;i,j}$, with $f_{+;i,j}$ also being the vevs of rungs that pass in front of intermediate strands.

Finally, we can define a diagonal matrix $\hat K$ with entries $K_i$
and assemble a ``monodromy matrix''
\begin{equation} \label{eq:stokes}
    M = \hat K \hat f_- \hat K \hat e_+\, .
\end{equation} 
For example, for $m=2$ we have
\begin{equation}
    M = \begin{pmatrix}K_1 & 0 \cr 0 & K_2 \end{pmatrix}\begin{pmatrix}1 & f \cr 0 & 1 \end{pmatrix}\begin{pmatrix}K_1 & 0 \cr 0 & K_2 \end{pmatrix}\begin{pmatrix}1 & 0 \cr e & 1 \end{pmatrix}  = \begin{pmatrix}K_1^2 + K_1 K_2 f e & K_1 K_2 f \cr K_2^2 e  & K_2^2 \end{pmatrix}\, .
\end{equation}
The coefficients of the characteristic polynomial of this matrix 
\begin{equation}
    \det(x \mathbb{I} - M) = x^2 - K_1 K_2 (K_1 K_2^{-1} + K_2 K_1^{-1} + f e) x + K_1^2 K_2^2 \, ,
\end{equation}
coincide with the planar limit of the first and second Casimirs of $U_\fq(\gl_2)$. The constraints that define $\cP_{2,1}$, i.e. the Casimirs fixed as in \eqref{eq:casimirunknot}, give a characteristic equation
\begin{equation}
    M^2 - (1 + g^2)M + g^2 = (M - 1)(M - g^2)=0\,,
\end{equation}
i.e. the $m=2$ monodromy matrix specialized to the cap or cup constraints has eigenvalues $1$ and $g^2$.

We now make a more general claim which is supported by Appendix \ref{app:classS}, and direct calculations of cup and cap states: if the $m$ strands are part of a compact knot or network or are placed in a transverse $S^2$ geometry, the planar limit of the rung vevs is supported on the union of submanifolds defined by the equation
\begin{equation} \label{eq:M2}
    (M-1)(M-g^2) = 0\, ,
\end{equation}
i.e. $M$ has an $n$-dimensional eigenspace with eigenvalue $g^2$ and an $(m-n)$-dimensional eigenspace with eigenvalue $1$ for some $n$, which turns out to coincide with the ``baryon number'' of the system: the sum of the labels on strands is $N n$.\footnote{Note that, unlike in the $\cP_{2,1}$ case, this constraint is stronger than just fixing the Casimirs, which would allow for more general Jordan blocks.}

We identify the space of such matrices, which has dimension $2n(m-n)$, with $\CP_{m,n}$. The $\CP_{m,n}$ manifolds are thus orbits of $\mathrm{\mathrm{GL}}(m)$ equipped with the symplectic form inherited from $U_\fq(\fgl_m)$. They are an example of ``character varieties'': they parameterize the Stokes/monodromy data of a 2d flat connection with specified singularities. Here we have a general rank $1$ irregular singularity at infinity, so that the monodromy $M$ is decomposed into Stokes data as in (\ref{eq:stokes}). See \cite{Witten:2007td} for a review of Stokes phenomena and irregular character varieties. The constraint (\ref{eq:M2}) is interpreted as the presence at the origin of a regular singularity labeled by the Levi subgroup $\mathrm{\mathrm{GL}}(n) \times \mathrm{\mathrm{GL}}(m-n) \subset \mathrm{\mathrm{GL}}(m)$.

We should also observe the hermiticity conditions
\begin{align}
    K_i^\dagger &= g K_i^{-1}  \cr
    e_{+;ij}^\dagger &= - f_{+;ij} \cr
    f_{-;ij}^\dagger &= - e_{-;ij}\, ,
\end{align}
which relate $e_+$ matrix elements to $f_-^{-1}$ matrix elements and thus set $M^\dagger= g^2 M^{-1}$. In other words, $g^{-1} M$ is a unitary matrix. These reality conditions define the real symplectic manifolds $\CP^\bR_{m,n}$. We expect that the finite $N$ irreps $\cH_{m,n}$ of $U_\fq(\fgl_m)$ are a natural quantization of $\CP^\bR_{m,n}$.

\subsection{The geometric braid group action}

The braid group generators $B_i$ act on these matrices in a manner which is familiar from the theory of Stokes phenomena. Consider $m$ complex numbers $x_i$, which can be understood as transverse positions $x_i^1 + \sqrt{-1} x_i^2$ of the strands. We have phases $\vartheta_{ij}$ of $x_j - x_i$, which are the slopes of straight $ij$ rungs. We slightly perturb the original $x^2_i$ positions of the strands so that the slopes are all distinct.

Denote the vevs of the straight rungs as $e_{ij}$ and $f_{ij}$ depending on the orientation with respect to the $x^1$ direction. These will be typically different from the $e_{\pm;i,j}$ and $f_{\pm;i,j}$. If the perturbation has a convex profile, so that all straight rungs pass behind intermediate strands, then obviously $e_{ij} = e_{-;ij}$ and $f_{ij} = f_{-;ij}$. 

Consider the product of $m \times m$ matrices:
\begin{equation}
    \hat f = \prod^\leftarrow_{i<j} (1 + f_{ij} b_{ij})\, ,
\end{equation}
where $b_{ij}$ is the elementary matrix with a single element $1$ at position $ij$ and the product is taken in the order opposite to the order of the slopes. If we have a convex profile, there will not be any mixed terms in the product and $\hat f = \hat f_-$. 

The skein relations satisfied by rungs are such that this remains true even when we perturb away from the convex profile. In particular, if we take a concave profile, we will have a maximal amount of mixed terms in the product. On the other hand, 
\begin{equation}
    \hat f^{-1} = \prod^\rightarrow_{i<j} (1 - f_{ij} b_{ij}) = \hat f_+
\end{equation}
will have no mixed terms, and we see that indeed $f_{ij} = f_{+;ij}$ in this configuration, as expected. 

In the same manner, we can define 
\begin{equation}
    \hat e^{-1} = \prod^\rightarrow_{i>j} (1 - e_{ij} b_{ij})
\end{equation}
and match it with $\hat e_-$ in the convex position, $\hat e_+^{-1}$ in the concave position. 

Then 
\begin{equation}
    M = \hat K \left[\prod^\leftarrow_{i<j} (1 + f_{ij} b_{ij})\right]\hat K \left[\prod^\leftarrow_{i>j} (1 + e_{ij} b_{ij})\right]\, ,
\end{equation} 
for any perturbed positions of the strands.

As we continuously braid two strands, the corresponding slope will progressively increase and the corresponding factors in $M$ will move to the left in the product. The braid transformation ``happens'' when the slope hits the vertical and we reorganize the product: the leftmost elementary factor in $\hat e$ is conjugated across $\hat K$ and becomes the rightmost factor in $\hat f$, and vice versa. 

This is precisely the behavior of the Stokes data for a $\mathrm{\mathrm{GL}}(m)$ 2d flat connection which approaches a diagonal form at the rank $1$ singularity, with eigenvalues $x^1_i dy_1 + x^2_i dy_2$ \cite{boalch2002gbundlesisomonodromyquantumweyl}. Here we denote as $y_i$ the 2d coordinates on the auxiliary plane the connection is defined on. 

\subsection{An alternative perspective on \texorpdfstring{$\CP_{m,n}$}{P(m,n)}, cups and caps}
The construction of $M$ in Appendix \ref{app:braiding_algebra} employs an auxiliary strand, following the evolution of rungs stretched from it to the remaining strands as it is transported along a closed path at large $x$. There is no obstruction to doing the same along other paths in the $x$ plane. 

This procedure gives an alternative interpretation of the factorization of $M$: just as in the tree-level planar limit discussed earlier in the paper, the transport along different loops gives a $\mathrm{\mathrm{GL}}(m)$ local system on the $x$ plane with $m$ minimal regular singularities $x_i$, with monodromies $M_i$ such that 
\begin{equation}
    M_i = 1+ \oell_i^L \oell_i^R \, ,
\end{equation}
for a left eigenvector $\oell_i^L$ and a right eigenvector $\oell_i^R$, normalized so that 
\begin{equation}\label{eq:lines_skein}
    \oell_i^R \cdot \oell_i^L = K_i^2-1\, ,
\end{equation}
and the rung vevs are expressed as 
\begin{equation} \label{eq:lines_rungs}
    \oell^R_i \cdot \oell^L_{j} = \begin{cases}
        e_{-;i,j}\qquad\qquad i>j \\
        K_i K_j f_{-;i,j}\quad \;\;i<j
    \end{cases}
\end{equation}
computed along the appropriate open path joining $i$ and $j$. The monodromy at infinity is 
\begin{equation}
    M = M_m \cdots M_1 \, .
\end{equation}

Note that these formulae can be diagrammatically represented as in Figures \ref{fig:lines_1} and \ref{fig:lines_2}.

\begin{figure}[ht]
    \centering
    \begin{equation}
        \begin{tikzpicture}[baseline=-3]
        \node (C) at (0,1) [circle,fill, inner sep=1pt] {};
        \node (E) at (0,0) [circle,fill,inner sep=2pt] {};
        \node (a) at (0,-0.3) {$i$};
        \coordinate (D) at (0,-0.5) {};
        \coordinate (A) at (0, 0.5) {};
        \draw[mid>] (E) to[out=180,in=180,looseness=1] (C);
        \draw[mid>] (C) to[out=0,in=0,looseness=1] (E);
        \end{tikzpicture} \quad=\quad
        \begin{tikzpicture}[baseline=-3]
        \node (C) at (0,1) [circle,fill,inner sep=1pt] {};
        \node (E) at (0,0) [circle,fill,inner sep=2pt] {};
        \node (a) at (0,-0.3) {$i$};
        \coordinate (D) at (0,-0.5) {};
        \coordinate (A) at (0, 0.5) {};
        \draw[mid>] (D) to[out=180,in=180,looseness=1] (C);
        \draw[mid>] (C) to[out=0,in=0,looseness=1] (D);
        \end{tikzpicture} \quad-\quad
        \begin{tikzpicture}[baseline=-3]
        \node (C) at (0,1) [circle,fill,inner sep=1pt] {};
        \node (E) at (0,0) [circle,fill,inner sep=2pt] {};
        \node (a) at (0,-0.3) {$i$};
        \coordinate (D) at (0,-0.5) {};
        \coordinate (A) at (0, 0.5) {};
        \draw[mid>] (A) to[out=180,in=180,looseness=1] (C);
        \draw[mid>] (C) to[out=0,in=0,looseness=1] (A);
        \end{tikzpicture}
    \end{equation}
    \caption{Diagrammatic representation of \eqref{eq:lines_skein}. The small dot represents the start and end points of the path computing $M_i$.}
    \label{fig:lines_1}
\end{figure}

\begin{figure}[ht]
    \centering
    \begin{equation}
        \begin{tikzpicture}[baseline=-3]
        \node (C) at (0,1) [circle,fill,inner sep=1pt] {};
        \node (E1) at (-1,0) [circle,fill,inner sep=2pt] {};
        \node (E2) at (1,0) [circle,fill,inner sep=2pt] {};
        \node (a1) at (-1,-0.5) {$j$};
        \node (a2) at (1,-0.5) {$i$};
        \draw[mid>] (E2) to[out=180,in=0,looseness=1] (C);
        \draw[mid>] (C) to[out=180,in=0,looseness=1] (E1);
        \end{tikzpicture}\qquad \qquad 
        \begin{tikzpicture}[baseline=-3]
        \node (C) at (0,1) [circle,fill,inner sep=1pt] {};
        \node (a1) at (-1,-0.5) {$i$};
        \node (a2) at (1,-0.5) {$j$};
        \node (E1) at (-1,0) [circle,fill,inner sep=2pt] {};
        \node (E2) at (1,0) [circle,fill,inner sep=2pt] {};
        \draw[mid>] (E1) to[out=180,in=180,looseness=1] (C);
        \draw[mid>] (C) to[out=0,in=0,looseness=1] (E2);
        \end{tikzpicture}
    \end{equation}
    \caption{Diagrammatic representation of \eqref{eq:lines_rungs}.The small dot represents the location where the inner product of $v_i^R$ and $v_j^L$ is computed.}
    \label{fig:lines_2}
\end{figure}

This picture undergoes an important simplification when we restrict from $\CP_m$ to $\CP_{m,n}$: the monodromies actually happen within an $n$-dimensional subspace of $\bC^m$. Accordingly, rung vevs in $\CP_{m,n}$
map to 2d $\mathrm{GL}(n)$ local system on the plane with $m$ minimal regular singularities and monodromy $g^2$ at infinity. In the above formulae, $\oell^R_i$ and $\oell^L_i$ are replaced by $n$-dimensional vectors $\tilde s_i$ and $s_i$, defined up to an overall $\mathrm{GL}(n)$ transformation. We refer the reader to Appendix \ref{app:classS} for more details. 

The main advantage of this presentation is the extreme simplicity of the action of the braid group and of the cups and caps constraints. Braiding operations act simply by braiding the punctures:
\begin{align} \label{eq:lines_braid}
    B_i s_{i} &= M_{i}^{-1} s_{i+1} B_i \cr
    B_i s_{i+1} &= s_{i} B_i \cr 
    B_i \tilde s_{i} &=  \tilde s_{i+1} M_{i} B_i \cr
    B_i \tilde s_{i+1} &= \tilde s_{i} B_i\, ,
\end{align}
i.e. the data of the strand passing ``above'' at the braid is transformed by the monodromy associated with the strand passing ``below'' at the braid. See Figure \ref{fig:lines_braid}

\begin{figure}[ht]
    \centering
    \begin{equation}
        \begin{tikzpicture}[baseline=-3]
        \node (C) at (0,1) [circle,fill,inner sep=1pt] {};
        \node (E1) at (-0.5,0) [circle,fill,inner sep=2pt] {};
        \node (E2) at (0.5,0) [circle,fill,inner sep=2pt] {};
        \node (a1) at (-0.5,-0.3) {$i$};
        \node (a2) at (0.5,-0.3) {$i{+}1$};
        \draw[mid>] (E1) to[out=180,in=180,looseness=1] (C);
        \end{tikzpicture}\quad \xrightarrow{\; B_i\; } \quad 
        \begin{tikzpicture}[baseline=-3]
        \node (C) at (0,1) [circle,fill,inner sep=1pt] {};
        \node (a1) at (-0.5,-0.3) {$i$};
        \node (a2) at (0.5,-0.3) {$i{+}1$};
        \node (E1) at (-0.5,0) [circle,fill,inner sep=2pt] {};
        \node (E2) at (0.5,0) [circle,fill,inner sep=2pt] {};
        \coordinate (b) at (-0.5,-0.5) {};
        \draw[mid>] (E2) to[out=180,in=0,looseness=1] (b);
        \draw[mid>] (b) to[out=180,in=180,looseness=1] (C);
        \end{tikzpicture}
    \end{equation}
    \caption{Diagrammatic representation of the first equation of \eqref{eq:lines_braid}.}
    \label{fig:lines_braid}
\end{figure}

We can also express these relations in terms of the $M_i$ monodromy matrices: 
\begin{align}
    B_i M_{i} &= M_{i}^{-1} M_{i+1} M_{i} B_i \cr
    B_i M_{i+1} &= M_{i} B_i \, ,
\end{align}
fixing $M_{i+1} M_i$, just as in the presentation of the fundamental group of a knot complement, with the difference that $M$ would be the identity matrix in that case.\footnote{We apologize to the reader for landing on this convention, as opposed to the graphically more natural convention of the monodromy around the strand below jumping at the crossing, as in the introduction discussion for $g\to 1$.}

Half of the caps constraints give $K_{2i-1} K_{2i}=g$, $s_{2i} = K_{2i} s_{2i-1}$. We can take the matrix of $s_i$ vectors to be block-diagonal, with $n$ blocks of size $2\times 1$. The other half of the constraints are subsumed into $M=g^2$, so these are enough to define $\cL_{\mathrm{caps}}\subset \CP_{m,n}$. The cup constraint $\cL_{\mathrm{cups}}$ is analogously $K_{2i-1} = \olambda_i$, $K_{2i-1} K_{2i}=g$, $\tilde s_{2i-1} = - \omu_i K_{2i-1} \tilde s_{2i}$.

The cap constraints force 
\begin{equation}
    M_{2i} M_{2i-1} = 1 + s_{2i} (K_{2i} \tilde s_{2i-1}+ \tilde s_{2i})\, ,
\end{equation}
with a single non-trivial $g^2$ eigenvalue. This deforms the 
tree-level planar constraint $M_{2i}M_{2i-1}=1$. The dual vector 
$(K_{2i} \tilde s_{2i-1}+ \tilde s_{2i})$ is constrained by the remaining cup relations, or by $M=g^2$, to be orthogonal to $s_{2j}$ for $j \neq i$. For cups, 
\begin{equation}
    M_{2i} M_{2i-1} = 1 + (s_{2i-1}- \omu^{-1}  K_{2i-1} s_{2i}) \tilde s_{2i-1}\, .
\end{equation}

It is not difficult to extend these considerations to other operations on strands, such as three-way junctions. 

The simplification is both practical and conceptual. In practice, we can do calculations using the $2 n (m-n)$-dimensional space $\CP_{m,n}$ instead of keeping track of $m(m-1)$ rung vevs. Conceptually, the vevs at each horizontal ``slice'' of the elongated knot map to the data of a 2d flat connection, evolving naturally along the braid and constrained in a local way at cups and caps. We will momentarily map this data to the data of a dual D-brane.

\subsection{The trefoil augmentation variety as a family of \texorpdfstring{$\mathrm{GL}(2)$}{\mathrm{GL}(2)} flat connections}

In the $\CP_{4,2}$ case of the trefoil knot, without loss of generality, we can write at the top the matrix of $s$'s and $\tilde s$'s:
\begin{align}
    s&= \begin{pmatrix} 1 & g \olambda^{-1} & 0 & 0 \cr 0 & 0 & 1 & \olambda \end{pmatrix}  \cr
    \tilde s &= \begin{pmatrix} \olambda^2 - 1 & g \olambda^{-1} f_2 \cr
    g \olambda^{-1} - g^{-1} \olambda & g^2 \olambda^{-2} f_2 \cr
    \olambda g^{-1} e_2 & g^2 \olambda^{-2} - 1 \cr
   - \olambda^2 g^{-1} e_2 & \olambda - \olambda^{-1}\end{pmatrix}  \, ,
\end{align}
solving the cap constraints and $M_4 M_3 M_2 M_1 = g^2$. The three powers of $B_2$ produce new matrices $M'_3$, $M'_2$, $M''_3$, with 
\begin{equation}
    M_3 M_2 =  M_2 M'_3 = M'_3 M'_2 = M'_2 M''_3\, .
\end{equation}
We should then impose the cup constraints on $M_1$, $M''_3$ and $M'_2$ and $M_4$, i.e. $\tilde s_1 + \omu \olambda \tilde s''_3=0$ and $\tilde s'_2 + g \olambda^{-1} \tilde s_4=0$.

Plugging in the expressions for the $e_2$ and $f_2$ rung vevs we derived in our previous analysis, the cup constraints all vanish on the augmentation variety as expected. 

\subsection{Closed braids in \texorpdfstring{$S^2 \times S^1$}{S2 x S1}}
Although we have mostly employed our tools to study compact knots, there is another setup that can be analyzed in the same manner. Consider the manifold $M_3=S^2 \times S^1$ and a collection of $m$ strands with $\Lambda^\bullet \bC^N$ labels that extend along the $S^1$ direction while being slowly braided along $S^2$.

We can define rung operators as before. Instead of the cups and caps constraints, we now have the constraint that a rung operator above the braid must equal the same rung operator below the braid, up to factors of $\omu$ accrued when we pass by the location where we keep track of the strand labels. 

Quantum mechanically, the correlation functions are computed as a trace over $\oplus_n \cH_{m,n}$, twisted by the rescaling:
\begin{equation}
    \omu^K \equiv\quad E_i \to \omu_i^{-1} \omu_{i+1} E_i \qquad F_i \to \omu_i \omu_{i+1}^{-1} F_i \, .
\end{equation}
Recall that we will only need a non-trivial $\omu$ parameter for each of the connected components of the braid. 

The trace on $\cH_{m,n}$ is unique. Denoting as $B$ the braid operation, we expect $\Tr B \omu^K$ to be completely fixed by the relations $\Tr B \omu^K O = \Tr O B \omu^K $ for all elements $O \in A_{m,n}$. 

In the planar limit, the analogous statement is that we expect an open planar saddle to be completely determined as a point in $\CP_{m,n}$ which is fixed by the combination of braid operation and $\omu$ rescaling. As we vary $\omu$, we obtain the augmentation variety for the closed braid in $S^2 \times S^1$.

At this point, we can map $\CP_{m,n}$ to the appropriate space of $\mathrm{GL}(n)$ 2d flat connections. The augmentation variety thus describes 2d flat connections on the punctures $S^2$ which are fixed by the action of the braid group operation on the minimal regular singularities away from infinity, up to a rescaling $\omu$ of the non-trivial monodromy eigenvectors at the punctures. 

This data can be immediately promoted to the data of a 3d $\mathrm{GL}(n)$ flat connection on $S^2 \times S^1$, with a minimal regular singularity wrapping the closed braid and 
a regular singularity with monodromy $g^2$ at infinity on $S^2$. 

As an example, we can study the $m=1$ and $m=2$ cases. In the presence of a single strand, the $S^2$ Hilbert space has two states $\cH_{1,0}$ and $\cH_{1,1}$ only, with labels $k=0$ and $k=N$. Accordingly, the correlation function is just $1 + \omu^N$. We see two saddles, and recalling \eqref{eq:brane_action}, they have action $0$ and $\frac{i\pi N}{\kappa+N} \log \omu$ and thus by definition \eqref{eq:mu_def} we see $\olambda =1$ and $\olambda = g$, as expected. These correspond to a trivial connection and a $\mathrm{GL}(1)$ connection with constant holonomy $g^2$ around the origin in $S^2$. 

The Hilbert space is more interesting in the case of two strands. If we do not act with a braid operation, we have the correlation function 
\begin{equation}
    1 + \frac{\omu_1^{N+1} - \omu_2^{N+1}}{\omu_1 - \omu_2} + \omu_1^N \omu_2^N \, ,
\end{equation}
and as long as $\omu_1 \neq \omu_2$, the corresponding saddles are $(\olambda^2_1,\olambda^2_2)$ being $(1,1)$, $(1,g^2)$, $(g^2,1)$, and $(g^2,g^2)$. This is the expected planar result, as in the $n=0$ and $n=2$ cases the labels have to add up to $0$ and $2N$ so both are $0$ and $N$, respectively. For $n=1$, starting with an $ef$ configuration and bringing $e$ or $f$ around the $S^1$ we get the constraint that either $\lambda_1= \lambda_2$ or $ef=0$. The latter implies that one of the labels is 0, as expected. 

If we insert an odd power of $B_1$, the knot has only one connected component, and thus the label must be $0$, $N/2$ or $N$, which implies that $\olambda^2 = g^n$ for $n=0,1,2$. The rung vevs can be found as a function of $\omu$. 

For an even power $r$ of $B_1$, the $n=0$ and $n=2$ cases are again easy, with both labels $0$ or $N$ and saddles given by $(\olambda^2_1,\olambda^2_2)$ being $(1,1)$ or $(g^2,g^2)$, respectively. With $n=1$, we can assume the labels to be related by $k_2=N-k_1$, and then applying $B_1^2$ multiplies $e$ and $f$ by $\mu_1^4g^{-2}$. This means that either $\omu_2 = \olambda_1^{2 r} g^{-r} \lambda_1$ or $ef=0$. The latter makes one of the labels $0$ and gives saddles in which $(\olambda^2_1,\olambda^2_2)$ is $(1,g^2)$ or $(g^2,1)$. 

\section{Holography and Chern-Simons theory}\label{sec:Amodel}

A general strategy to ``derive'' holographic dualities from String Theory constructions is to embed them into open-closed duality: the principle that a theory of closed strings modified by a stack of $N$ D-branes admits a 't Hooft expansion which effectively replaces the D-branes by their back-reaction. As long as the original closed string degrees of freedom can be decoupled, one can derive a perturbative relation between the field theory limit of the D-brane world-volume theory and a closed string theory \cite{Maldacena:1997re}. In topological String Theory examples, the ``decoupling limit'' can be as simple as selecting judicious boundary conditions \cite{Costello:2018zrm}. 

A similar intuitive derivation of a holographic duality for $SU(N)_k$ Chern-Simons theory follows the identification of $U(N)_\kappa$ Chern-Simons theory as the world-volume theory on a stack of $N$ three-dimensional D-branes in the A-model topological strings \cite{Gopakumar:1998ki}. We will now discuss the A-model setup and the D-brane back-reaction. 

Roughly, the closed string sector of the A-model String Theory describes the deformations of six-dimensional symplectic manifolds. Symplectic manifolds do not have any local data, as the symplectic form can always be brought to a canonical form, so the A-model does not have local degrees of freedom.

In order to engineer Chern-Simons theory on a 3-manifold $M_3$ one considers the A-model with target $T^* M_3$ and a stack of $N$ three-dimensional D-branes supported on the base $M_3$. 
In the presence of the D-branes, the equation of motion $d\omega=0$ for the dynamical symplectic form is replaced by 
\begin{equation}
    d \omega = N \hbar \delta^{(3)}(y) dy_1 dy_2 dy_3\, ,
\end{equation}
where $y_i$ are the local coordinates on the fiber of the cotangent bundle. Here $\hbar$ is the topological string coupling. We will relate it to $\kappa^{-1}$ momentarily. 

\subsection{\texorpdfstring{$T^* \bR^3$}{T*R3} back-reaction}
If we take $M_3 = \bR^3$, with local coordinates $x^i$ and $y_i$ and undeformed symplectic form 
\begin{equation}
    \omega_0 = dx^i dy_i\, ,
\end{equation}
we can easily solve for a back-reaction that preserves all symmetries of the problem: the monopole flux 
\begin{equation}
    \omega = dx^i dy_i + \frac{N \hbar}{4 \pi} \frac{y_1 dy_2 dy_3 + y_2 dy_3 dy_1 + y_3 dy_1 dy_2}{|y|^3}\, .
\end{equation}
We can locally reabsorb the shift in $\omega$ by a redefinition of the $x^i$ coordinates. Essentially, we shift $x^i$ by a primitive of $\omega_{S^2}$. For example, we can do so away from the $y_1=y_2=0$, $y_3<0$
ray or away from the opposite ray $y_1=y_2=0$, $y_3>0$.
More explicitly, 
\begin{align}
    x_\pm^1 &= x^1 - \frac{N \hbar}{4 \pi}\frac{y_2}{y_1^2+y_2^2}\left(\frac{y_3}{|y|} \mp 1\right)  \cr
    x_\pm^2 &= x^2 + \frac{N \hbar}{4 \pi}\frac{y_1}{y_1^2+y_2^2}\left(\frac{y_3}{|y|} \mp 1\right) \cr
    x_\pm^3 &= x^3 \, .
\end{align}

The difference between the two primitives is the angular form
\begin{equation}
    \frac{N \hbar}{2 \pi}\frac{y_1 dy_2-y_2 dy_1}{y_1^2 + y_2^2}\, .
\end{equation} 
The two coordinate patches $x_\pm^i$ are thus related as
\begin{align}
    x_+^1 &= x_-^1 + \frac{N \hbar}{2 \pi}\frac{y_2}{y_1^2+y_2^2} \cr
    x_+^2 &= x_-^2 - \frac{N \hbar}{2 \pi}\frac{y_1}{y_1^2+y_2^2}  \cr
    x_+^3 &= x_-^3 \, .
\end{align}
If we exchange the role of the $x$ and $y$ coordinates, the back-reacted geometry can be seen as a twisted cotangent bundle of $\bR^3\backslash \{0\}$: the base $\bR^3\backslash \{0\}$ is parameterized by the $y_i$ coordinates and the fiber coordinates $x^i$ in different patches on the base are glued affine-linearly. 

If we use holomorphic coordinates in the $x^1, x^2$ plane and the $y_1$, $y_2$ plane, the coordinate transformation simplifies to 
\begin{align}
    x_+ &= x_- + \frac{t}{y} \cr
    x_+^3 &= x_-^3 \, . 
\end{align}
We denoted $t = \frac{N \hbar}{2 \pi i}$.

The local deformation of $T^* \bR^3$ is not easily promoted to a deformation of a general $T^* M_3$, as 
$\omega$ does not transform naturally when $y_i$ are
multiplied by the Jacobian of a coordinate transformation. It would be interesting to identify the possible back-reacted geometries for general $M_3$.

\subsection{\texorpdfstring{$T^* S^2 \times T^*\bR$}{T*S2 x T*R} back-reaction}
When studying elongated knots, we found it natural to embed them into a $S^2 \times \bR$ geometry by compactifying spatial infinity. In complex coordinates, 
we can cover $S^2$ by two patches with local coordinates $x$ and $\tilde x$ related by $x \tilde x = 1$ away from the origin. The cotangent bundle $T^* S^2$ is described by $(x,y)$ and $(\tilde x, \tilde y)$ with $\tilde y = -x^2 y$.

There is a simple way to deform this geometry so that the deformation matches the flat space back-reaction in both patches. We introduce four coordinate systems  $(x_\pm,y)$ and $(\tilde x_\pm,\tilde y)$ with relations 
\begin{align}
    x_+ y &= x_- y + t \qquad \qquad
    \tilde x_+ x_+ = 1 \qquad \qquad
    \tilde x_- x_- = 1 \cr
    \tilde y &= - x_+ x_- y = - x_-^2 y - t x_- = - x_+^2 y + t x_+\, .
\end{align}
With these definitions, we have
\begin{equation}
    \tilde x_+ \tilde y- \tilde x_- \tilde y= t\, ,
\end{equation}
so the back-reaction of the two patches of $T^* S^2$ matches correctly.     
The $(x_+,y)$ and $(\tilde x_+,\tilde y)$ coordinate systems available at $y_3>0$ define a well-known twisted cotangent bundle deformation of $T^* \bCP^1$: the deformed $A_1$ singularity $\cM_t$. We can make it manifest by defining the global coordinate $z = x_+ y = - \tilde x_+ \tilde y+t$, so that 
\begin{equation}
    y \tilde y + z(z-t) =0 \, ,
\end{equation}
with complex symplectic form 
\begin{equation}
    \omega_\bC \equiv dz \frac{dy}{2 \pi i y}\, ,
\end{equation}
whose real part enters $\omega$.\footnote{The $\cM_t$ geometry has a real Lagrangian cycle such that the period of the complex symplectic form is $t$: a circle in the $y$ plane fibered over a segment from $0$ to $t$ in the $z$ plane, shrinking smoothly at both ends.} 

The $(x_-,y)$ and $(\tilde x_-,\tilde y)$ available at $y_3<0$ also define the same complex symplectic manifold $\cM_t$, giving a well-known alternative presentation as a twisted cotangent bundle. Now, $z=x_- y + t = - \Tilde{x}_- \tilde y$. 

The two manifolds are identified trivially away from the $y_3=0$ locus, so we identify the whole geometry with $\cM_t \times T^* \bR$. 

\subsection{Wilson lines, strings and D-branes}
The standard holographic interpretation of a Wilson line in the fundamental representation $\bC^N$ involves an extended string worldsheet in the dual space-time, with a shape that approaches the support of the Wilson line at the boundary \cite{Rey:1998ik}. This perspective holds in the context of Chern-Simons theory as well. Furthermore, the non-renormalization properties of the A-model worldsheet theory allow for instanton calculations at all orders of the genus expansion, ``counting'' appropriately the contributions from extended worldsheets of given topology. 

As we discussed before, the genus zero part of the calculation is not very sensitive to the topology of the boundary Wilson line. The two terms in the unknot vev can be ascribed to two classical worldsheets which differ in homology by the $S^2$ supporting the back-reaction. We thus identify $g^2$ with $\exp \int_{S^2} \omega = \exp N \hbar$ and thus $\hbar = \frac{2 \pi i}{\kappa+N}$. A more careful comparison with HOMFLY knot invariants requires careful calculations in higher genus. 

Wilson lines associated to 1d fermions have a standard interpretation as D-branes \cite{Gomis:2006sb,Aganagic:2012jb}. For the problem at hand, they are expected to be three-dimensional D-branes in the A-model, wrapping asymptotically the co-normal bundle $N^* K\subset T^* M_3$ to the support $K$ of the Wilson loop. 

We thus aim to match the large $N$ saddles of the Wilson loop correlation functions to D-branes with the correct asymptotic shape. In practice, we need a way to map the rung vevs to the data which deforms $N^* K\subset T^* M_3$ to a D-brane in the back-reacted geometry. 

D-branes in the A-model also do not have local degrees of freedom. A typical way to present a D-brane is to specify a Lagrangian submanifold equipped with a $U(1)$ flat connection, but this data can be deformed without changing the actual D-brane. If we encode an infinitesimal deformation of the Lagrangian support into a 1-form on the Lagrangian, any exact form gives an equivalent D-brane. 

Additionally, the A-model depends holomorphically on the 
complex combination of the 1-form describing shape deformations and the connection 1-form for the $U(1)$ bundle, up to $\mathrm{GL}(1)$ gauge transformations. 
The world-volume theory of an A-model D-brane is really an analytically continued Chern-Simons theory in the sense of \cite{Witten:2010cx}.

There are two general complementary perspectives on D-branes: 
\begin{itemize}
    \item One can describe a collection of recipes to construct D-branes. The main challenge is to recognize different recipes giving the same D-brane. Furthermore, there may exist D-branes which cannot be realized by any recipe in the collection.
    \item One can describe a procedure to extract a mathematical signature from any D-brane. The main challenge is to insure that no two D-branes have the same signature. Furthermore, certain signatures may not be associated to any D-branes.
\end{itemize}
In an ideal situation, we can close the circle by building a D-brane with any given mathematical signature.
A typical way to build D-branes is to deform some small collection of generating D-branes. A typical way to extract mathematical signatures is to look at the open string stretched from a generic D-brane to some small collection of probe D-branes.   

In a cotangent bundle ambient space $T^* M_3$, many D-branes can be built by deforming a stack of D-branes in the base by a $\mathrm{GL}(k)$ flat connection on $M_3$. Intuitively, these represent Lagrangian submanifolds in $T^* M_3$ 
which intersect each fiber $k$ times. Co-dimension 2 regular singularities can represent situations where some of the $k$ sheets reach infinity along the cotangent fiber. 

For example, minimal regular singularities supported on a knot $K$ represent a single sheet extending to infinity along $N^* K$. This is precisely the data we extracted from the tree-level planar saddles in the QFT analysis, giving immediately a candidate dual D-brane in $T^* M_3$. 

\subsection{Holomorphic tricks}
We will first consider D-branes dual to a collection of $m$ parallel Wilson lines, mimicking the QFT analysis.
The $S^2 \times \bR$ setup is particularly promising, 
as the expected back-reacted geometry $\cM_t \times T^* \bR$ factors nicely and we can focus on D-branes of the form $\Sigma \times \bR$ defined from a 2d D-brane $\Sigma$ in $\cM_t$ sitting at $y_3=0$ for all $x^3$.

A further simplification arises from the fact that $\cM_t$ is a complex symplectic manifold (actually, hyperk\"ahler) and $\omega$ is the real part of the complex symplectic form $\omega_\bC$.
This allows one to focus on complex Lagrangian submanifolds with no (expected) loss of generality. 
This statement is analogous to the observation that a
2d flat connection can be represented as a holomorphic  
bundle equipped with a meromorphic connection. See also \cite{Gaiotto:2016hvd}.

Consider first a single Wilson line placed at $x=x_1$. Being on $S^2$, the label of the Wilson line will ultimately have to be $0$ or $N$ unless some other Wilson line is present. Before back-reaction, we expect a D-brane dual with equation $x=x_1$. After back-reaction, that may become something like $x_+=x_1$
or $x_- = x_1$. Both options are nice complex submanifolds of $\cM_t$:
\begin{align}
    L_+: \, x_1 y - z &= 0  \qquad \qquad x_1 (z-t) + \tilde y = 0 \cr
    L_-: \, x_1 y -z + t = &= 0  \qquad \qquad x_1 z + \tilde y = 0\, .
\end{align}
Observe that from the point of view of $L_+$, $L_-$ is obtained as a deformation by the $\mathrm{GL}(1)$ connection $\frac{\delta z}{y} = \frac{t}{y}$, which has 
holonomy $g^2$ around the origin in the $y$ plane. The same is true in the $\tilde y$ plane. 
The reader should be reminded of the conditions we encountered on $M$ for $m=1$, $n=1$. We tentatively identify $L_+$ and $L_-$ with the $k=0$ and $k=N$ Wilson lines. 

These curves are remarkably rigid. For example, a deformation 
\begin{equation}
    z = x_1 y + a
\end{equation}
naively interpolates between them, but has a secret problem: as $y \to 0$, $\tilde y$ diverges:
\begin{equation}
    \tilde y = x_1 \frac{z(z-t)}{z-a} \, .
\end{equation}
Accordingly, this curve has an extra branch reaching infinity in $\cM_t$ along the $\tilde y$ direction.

On the other hand, consider the equation
\begin{align}
    \tilde y + x_1 (z-t) + x_2 z - x_1 x_2 y =0\, .
\end{align}
Secretly, this is the union of $L_+$ and $L_-$: 
if $y \neq 0$ it is equivalent to $(z-x_1 y)(z-t-x_2 y)=0$ and if $\tilde y \neq 0$ it is equivalent to 
$(\tilde y + x_1 (z-t))(\tilde y + x_2 z)=0$. We can deform it to 
\begin{align}
    L_{2,1}[a]: \, \tilde y + (x_1 + x_2) z - x_1 x_2 y - a = 0 \, .
\end{align}
If $y \neq 0$, the equation 
\begin{equation}
   z(z-t) - (x_1 + x_2) z y + x_1 x_2 y^2  + a y= 0 
\end{equation}
reaches infinity along $z=x_1 y$ and $z = x_2 y$. 
If $\tilde y \neq 0$, we get 
\begin{equation}
    \tilde y^2 - (x_1 + x_2) z \tilde y + x_1 x_2 z(z-t) - a \tilde y= 0\, ,
\end{equation}
and still only reach infinity along the same directions 
$\tilde y = - x_1 z$ and $\tilde y = - x_2 z$.

We thus get a one-parameter family of D-branes which are potentially dual to two parallel $\Lambda^{\bullet} \bC^N$ Wilson lines. To do better, we need to learn how to put coordinates on these families of D-branes: the $a$ coordinate is not a complex coordinate on the space of D-branes. Instead, the real and imaginary parts of $a$ should be combined with $U(1)$ connection data into new holomorphic coordinates. 

This process is well-understood for families of holomorphic Lagrangian branes in $T^*C$, where $C$ is a Riemann surface. These are identified with moduli spaces of Higgs bundles via a spectral curve construction in $C$. A hyperk\"ahler rotation converts that into moduli spaces of flat connections on $C$, parameterized by monodromy data. This is the correct set of holomorphic coordinates to describe the D-brane moduli space \cite{Witten:2015dta,Kontsevich:2024esg}. 

Here, we can restrict the Lagrangian branes to $T^* \bC$ patches in $\cM_t$ and apply the procedure to the patch. We do so in Appendix \ref{app:classS},
with a simple result: the moduli space of D-branes in $\cM_t$ which deform the combination of $(m-n)$ copies of the $L_+$ line and $n$ copies of $L_-$ is precisely $\CP_{m,n}$. The restriction to a patch parameterized by $x_+$ or $y$ gives respectively the two presentations of $\CP_{m,n}$ as a $\mathrm{\mathrm{GL}}(m)$ character variety on the $y$ plane or as a $\mathrm{\mathrm{GL}}(n)$ character variety on the $x_+$ plane. 

In other words, $\CP_{m,n}$ is the phase space for D-branes in $\cM_t \times T^* \bR$ with the expected asymptotics. This gives a direct proof of the relation between rung vevs on straight Wilson lines and the holographic dual D-branes.

\subsection{Branes in \texorpdfstring{$\cM_t \times T^* \bR$}{Mt x T*R}}
The relation extends readily to braids. A 3d flat connection on $C \times \bR$ is the same as a covariantly constant family of connections on $C$ evolving along the $x^3$ direction. In the presence of singularities, ``covariantly constant family'' really means ``isomonodromic family''. In the case at hand, the braid group acts on $\CP_{m,n}$ precisely by isomonodromic deformation of the $x_i$ parameters. 

As a consequence, a sequence of rung vevs related by braid transformations is the same as the choice of a holomorphic D-brane evolving under the braiding of the $x_i$ asymptotic parameters. 

In order to obtain a complete match with our augmentation variety calculations, we need to show that the cup and cap constraints match the requirements for a cylindrical D-brane with holomorphic cross section to be able to ``end'' at sufficiently large and positive or negative $x^3$.

We expect the constraint to be ``local'' on $C\times \bR$. Indeed, cups and caps impose a local constraint on the pair of left- or right-eigenvectors associated with the merging strands. Accordingly, the main question is how one can end a D-brane with cross-section $L_{2,1}[a]$. 

We can gain some insight by decomposing a cup/cap into a ``junction''  merging the two strands, combined with an ``endpoint'' which forces the new strand to have label $\Lambda^N \bC^N$. These junctions and their planar limit are discussed in Appendices \ref{app:rung_algebra} and \ref{app:braiding_algebra}.

Remarkably, the junctions impose constraints which are $g$-independent. 
They require either the left- or right-eigenvectors of the merging strands to have a specific proportionality and identify one of them with the eigenvector of the new strand. The monodromy around the pair of merging strands becomes the monodromy around the new strand. The left vs right choice is determined by the junction being $2 \to 1$ or $1 \to 2$. 

The $g$-independence of the junction constraints suggests that the holographic interpretation of the junctions should involve properties of D-branes in $T^* \bR^3$. On the other hand, the endpoints obviously depend on $g^2$ and should thus be sensitive to the back-reaction. 
They should not be sensitive, though, to the $S^2$ vs $\bR^2$ choice of transverse geometry. 

\subsection{Twisted connections on \texorpdfstring{$\bR^3\backslash \{0\}$}{R3\textbackslash\{0\}} and endpoints.}
As the local back-reacted geometry takes the form of a twisted cotangent bundle of $\bR^3\backslash \{0\}$, we can build D-branes as {\it twisted connections} on $\bR^3\backslash \{0\}$. In practice, this is just a connection with constant diagonal curvature $t \omega_{S^2}$.

The simplest option for a single D-brane is to just take the primitive of $t \omega_{S^2}$ we discussed earlier in this Section to define $x^i_-$. This has a regular singularity at positive $y_3$, $y_1=y_2=0$, with holonomy $g^2$. It tells us that this configuration has a ``spike'' which extends towards large positive $y_3$ and large $x$, with holonomy $g^2$ around the origin of the $x$ plane. This looks like an $L_-$ brane combined with a ray in the positive $y_3$ direction.

Such a configuration could be modified to have a different shape at large $y_3$ in the $x^3$, $y_3$ plane while keeping the same $L_-$ transverse shape. In particular, it could be bent to have asymptotically constant $y_3$ and extend towards large positive or negative $x_3$. Then the asymptotic part of the D-brane would precisely look like the dual to a strand of label $\Lambda^N \bC^N$. 

This is our conjectural realization of the future and past endpoints for individual strands, denoted as $\Pi_N$ and $I_N$ in Appendices \ref{app:rung_algebra} and \ref{app:braiding_algebra}.

When the ending strands are part of a collection of $m$ strands, these endpoints force the 2d connection to take a block triangular form, implemented by setting to zero the last component of all left- or right- eigenvectors except the one for the ending strand, with one block being the rank $n-1$ connection associated with the surviving $m-1$ strands. 

This sort of triangular structure is a typical ingredient in the definition of constructible sheaves. It should fit into a general ``microlocal'' mathematical framework to build D-branes in the back-reacted geometry. 

It seems possible to extend this discussion to directly deal with cups and caps. We can consider a primitive of $t \omega_{S^2}$ which has regular singularities at both poles, with holonomies $\olambda^2$ and $g^2 \olambda^{-2}$. This represents spikes with cross-section $L_-$ and $L_+$ 
projecting towards positive and negative $y_3$. 

If we ``bend'' the spikes towards, say, positive $x_3$, the configuration can be used to end a pair of strands with labels adding up to $N$, i.e. to produce a cup. Again, we expect the cup constraints to describe how such a gadget can be glued to 2d flat connections associated to $m$ and $m-2$ 
strands to make a D-brane in the back-reacted geometry.

\subsection{Branes without end}
We can briefly discuss D-branes in a 
\begin{equation}
    \cM_t\times T^* S^1 
\end{equation}
geometry. Picking a convenient patch of $\cM_t$, a D-brane can be built from a 3d $\mathrm{\mathrm{GL}}(n)$ twisted flat connection on $S^2 \times S^1$, i.e. a 3d flat connection on $\bR^2\times S^1$ with a holonomy $g^2$ around infinity in the $\bR^2$ factor. 

As discussed in the Introduction and in Section \ref{sec:char}, this is precisely the data associated with an open planar saddle for a closed braid in $S^2 \times S^1$. We thus have a complete holographic match. 

\section{Conclusions and open directions}\label{sec:conclude}
In the main text and Appendices of this paper, we have given a detailed derivation of the following aspects of $SU(N)_\kappa$ Chern-Simons Holography:
\begin{itemize}
    \item For any knot/link $K$ with $\Lambda^\bullet \bC^N$ representations and 3d manifold $M_3$, we have given a complete match between large $N$ saddles and dual D-branes in a $g \to 1$ limit.
    \item For any number $m$ of parallel $\Lambda^\bullet \bC^N$ strands in $S^2 \times \bR$ (or $\bR^3$), we have matched the space of possible planar vevs of mesonic operators to the classical phase space of the dual D-branes. Both sides of the correspondence have a natural finite $N$ ``quantization'', which also matches. 
    \item Given the Schubert presentation of a knot/link $K$ with $\Lambda^\bullet \bC^N$ representations as a capped braid, we give a geometric definition of the space of large $N$ saddles (augmentation variety) as a continuously evolving 2d flat connection jumping in a specific manner at the cups/caps where strands merge. This determines a continuously evolving dual D-brane, but we have not proven that the cups and caps conditions match the geometric constraints on the dual D-brane.
    \item The analysis of closed braids in $M_3 = S^2 \times S^1$ does not require cups or caps. We can thus fully recover the dual D-branes. 
    \item We have applied our construction to determine the augmentation variety of a large collection of knots. Reassuringly, 
    different Schubert presentations of the same knot give consistent answers.
\end{itemize}
Our work leaves many open questions:
\begin{itemize}
    \item It would be nice to show that a proper application of the theory of constructible sheaves as descriptions of three-dimensional A-model D-branes \cite{nadler2009microlocalbranesconstructiblesheaves,nadler2008constructiblesheavesfukayacategory} to the back-reacted A-model geometry 
    matches our cups and caps gluing rules for families of two-dimensional D-branes. This would complete the proof of planar holographic duality for compact knots/links with $\Lambda^\bullet \bC^N$ labels. A detailed comparison to the DGA construction in \cite{Ekholm_2012} should also be straightforward.
    \item We have not explored the categorical structure of our construction. At the 2d level, there is a natural map \begin{equation}
        \CP_{m_1,n_1} \times \CP_{m_2,n_2} \to \CP_{m_1+m_2, n_1+n_2}
    \end{equation}
    mapping a $\mathrm{\mathrm{GL}}(n_1)$ and a $\mathrm{\mathrm{GL}}(n_2)$ 2d flat connection to their direct sum. The normal bundle to the image describes infinitesimal deformations of the direct sum of two D-branes and thus describes spaces of open strings in ghost number $1$. This construction plays a role in the description of gluing rules. It should be an ingredient in recovering the category of three-dimensional D-branes. 
    \item As detailed in the introduction, it should be possible to characterize the large $N$ saddles for the partition function on a general three-manifold $M_3$ in terms of a planar Skein module $\mathrm{Sk}^p_{M_3}[g]$. Extra Wilson lines labeled by $\Lambda^\bullet \bC^N$ may allow a characterization in terms of a dual category of D-branes. 
    \item As a particular case, it would be interesting to study the holographic duality on $\Sigma_2 \times \bR$ for any Riemann surface $\Sigma_2$, with or without extra $\Lambda^\bullet \bC^N$ strands placed at points in $\Sigma_2$. Chern-Simons theory attaches to each such surface a Skein algebra, which should have a planar limit $\mathrm{Sk}^p_{\Sigma_2}[g]$. 
    \item Appendices \ref{app:classS} and \ref{app:inter} implicitly lift the geometric calculation of the augmentation variety to the calculation of the space of circle-compactified vacua of a certain 3d ${\cal N}=2$ SQFT. The SQFT is not the same as the SQFTs that appear in the 3d-3d correspondence \cite{Dimofte:2011ju, Dimofte:2011py,Dimofte:2013iv}, but we expect to reduce to these in a $g \to 1$ limit. This observation could likely explain the knot-quiver correspondence \cite{Ekholm:2018eee}. The ellipsoid partition functions and superconformal indices of these 3d SQFTs may provide interesting knot invariants. Categorified knot invariants  may also arise from the Holomorphic Topological twist of the 3d SQFTs \cite{Aganagic:2017tvx,Costello:2020ndc}.
    \item Our characterization of the moduli spaces of A-model D-branes in a deformed $A_1$ singularity can be generalized to deformed $A_k$ singularities and possibly to $D$ and $E$ type singularities. It would be interesting to give that a holographic interpretation as well.
\end{itemize}
As observed in the introduction, 1d defects barely scratch the surface of possible fundamental modifications of 3d CS theory, which include 
non-topological 2d and 3d modifications. We hope to report on these soon. 

\subsection{A surprising formula}
Finally, we would like to comment on a surprising formula, Theorem $1.2$ of \cite{Ekholm:2021osm}. This formula is interpreted in the reference as the mathematical manifestation of the holographic duality between the colored HOMFLY polynomials of a knot $K$ and an A-model D-brane supported on a specific deformation $L_K$ of the knot conormal, which has the topology of a solid torus. 

The formula is exact at finite $N$ and does not involve any large $N$ analysis. The left hand side of the formula collects all possible instanton corrections to a world-volume calculation on $L_K$. It is presented as an element of a Skein algebra of $L_K$, i.e. a direct sum of Wilson lines on $L_K$ which should be inserted in a $\mathrm{\mathrm{GL}}(r)$ Chern-Simons correlation function on $L_K$ in order to produce a String Theory answer for $r$ D-branes wrapping $L_K$. 

The correlation function will depend on a choice of $r$ longitudinal holonomies $\rho_a$. Inserting the right hand side of the formula in $\mathrm{\mathrm{GL}}(r)$ Chern-Simons theory, we will get a generating function of colored HOMFLY polynomials, weighed by certain functions of $\rho_a$, $q$ and $g$. 

This statement is clearly much stronger than the standard expectation from holography. It is also not immediately suited for a large $N$ expansion, as the right-hand side includes HOMFLY polynomials labeled by Young Tableaux with any number of boxes. It thus includes objects which do not admit a 't Hooft expansion. 

A preliminary step in a large $N$ analysis would be to find a linear functional $F$ on the Skein module of $L_K$ which selects a specific HOMFLY polynomial or linear combination thereof with a good 't Hooft expansion, such as our $\langle W_{K}(\omu)$ or a symmetric power analogue. Then the left hand side evaluated on $F$ could be analyzed in a planar limit, and one may try to figure out how the expected augmentation variety of large $N$ saddles will emerge from the calculation. See e.g. \cite{Ekholm:2024ceb} for the Hopf link. 

This kind of result strongly reminds us of the ``giant graviton expansion'' proposed in \cite{Gaiotto:2021xce}. 
That reference also offers a formula with a holographic flavor which is nevertheless exact at finite $N$ and involves a sum over terms computed by an auxiliary $\mathrm{\mathrm{GL}}(r)$ problem and interpreted as contributions from $r$ copies of a specific dual D-brane. 

In that situation, it is also the case that the D-brane which appears to contribute to the formula is only one of many possible large $N$ saddles. A large $N$ expansion of the formula must somehow reproduce the other D-brane saddles as well as non-trivial geometries, but the details are unclear. 

In the case of the giant graviton expansion, the formula has the flavor of an equivariant localization result, with the 
relevant D-branes being the equivariant fixed points of a much more mysterious String Theory phase space. It would be interesting to explore a similar interpretation for the results of \cite{Ekholm:2024ceb}.

\section*{Acknowledgments}
We would like to thank Mina Aganagic, Alexander Braverman, Kevin Costello, Joel Kamnitzer, David Nadler, Vivek Shende, and Ben Webster for useful conversations and feedback on the draft. This research was supported in part by a grant from the Krembil Foundation. DG is supported by the NSERC Discovery Grant program and by the Perimeter Institute for Theoretical Physics. Research at Perimeter Institute is supported in part by the Government of Canada through the Department of Innovation, Science and Economic Development and by the Province of Ontario through the Ministry of Colleges and
Universities.

\appendix

\section{Some auxiliary SQFTs} \label{app:classS}
In the course of our analysis in the main text, we encounter complex symplectic manifolds $\CP_{m,n}$ which admit multiple presentations as {\it character varieties} ${\mathcal M}[G,C]$, i.e. parameterize the monodromy/Stokes data of flat connections with structure group $G$ on a surface $C$ with prescribed singular behavior at marked points. We also need to identify $\CP_{m,n}$ with a moduli space of A-model D-branes in a deformed $A_1$ singularity. 

In this Appendix we provide an alternative interpretation of $\CP_{m,n}$ in terms of auxiliary Supersymmetric Quantum Field Theories with eight supercharges, as well as
auxiliary brane constructions in IIA String Theory and M-theory. This extra structure guarantees the equivalence of different characterizations of $\CP_{m,n}$ and simplifies the detailed match between them. 

As an extra bonus, the constructions also provide matching canonical quantizations of $\CP_{m,n}$ in all of its presentations. This includes both the quantum algebras $A_{m,n}$ of observables, which admit multiple presentations as {\it skein algebras} $\mathrm {Sk}_\fq[G,C]$ that ``quantize'' the monodromy/Stokes data, and finite $N$ Hilbert spaces. 
This gives a finite $N$ holographic match for time-independent brane configurations. 

\subsection{Classical representations from three-dimensional SCFTs}
The quantum group $U_\fq(\gl_m)$ and its representations play a key role in the main text.
We will see that the quantization of $\CP_{m,n}$ is a $\fq$-deformed analogue of the quantization of the Grassmannian $\mathrm{Gr}_{m,n}$, which produces specific irreducible $U(\gl_m)$ representations. In turn, this is a generalization of the fuzzy sphere construction. 

Accordingly, we start from an SQFT associated with the Grassmannian: the three-dimensional 
${\cal N}=4$ $U(n)$ SQCD with $m$ flavors. This theory is a special case of a class of 3d quiver gauge theories denoted as $T_\rho[U(m)]$ in \cite{Gaiotto:2008ak}, which have a 
manifest $U(m)$ symmetry and are labeled by a partition $\rho$ of $m$. Here, $\rho$ is the partition $(n,m-n)$. Much of our discussion below can be extended in a straightforward way to a generic partition $\rho$. 

The theory $T_{n,m-n}[U(m)]$ has a hyperk\"ahler Higgs branch which can be presented as 
\begin{equation}
    \CP^{\mathrm{3d}}_{m,n} \equiv T^* \bC^{n m}/\!\!/\mathrm{\mathrm{GL}}(n) \, ,
\end{equation} 
as a complex symplectic manifold: the Hamiltonian reduction of $T^* \bC^{n m}$ by the obvious $\mathrm{\mathrm{GL}}(n)$ action. The reduction depends on a choice of FI parameter $t$ and can be presented in two different ways as a twisted cotangent bundle of the Grassmannian $\mathrm{Gr}_n(\bC^m)$, with twist proportional to $t$. 

If we denote the matter fields in $T^* \bC^{n m}$ as $X$, $Y$, the Hamiltonian reduction enforces the F-term relation
\begin{equation}\label{eq:moment}
    Y X = t \, 1_{n \times n}\, ,
\end{equation}
and quotients by the $\mathrm{\mathrm{GL}}(n)$ symmetry acting on $X$ from the right and $Y$ from the left. The gauge-invariant operators are generated by the $m \times m$ $\mathrm{\mathrm{GL}}(m)$ moment maps:
\begin{equation}
    Z = X Y \, .
\end{equation}
They satisfy the relation $Z^2 = t Z$, defining a special orbit in $\gl_m$: the matrix $Z$ has $n$ eigenvalues ``$t$'' and $m-n$ eigenvalues ``$0$''.

The presentation as a twisted cotangent bundle proceeds as follows. The moment map relations imply that $X$ has rank $n$. We can define charts where a subset of $n$ rows in $X$ is linearly independent, and use the $\mathrm{\mathrm{GL}}(n)$ symmetry to make it the identity. Then \eqref{eq:moment} can be solved in terms of the corresponding $n \times n$ block in $Y$. The remaining components of $X$ parameterize the base, and those of $Y$ the fiber. Under a change of coordinates, the fiber coordinates transform affine-linearly. A second description switches the roles of $X$ and $Y$. 

The Higgs branch $\CP^{\mathrm{3d}}_{m,n}$ has a natural and physically motivated quantization \cite{Yagi:2014toa}: the quantum Hamiltonian reduction of the Weyl algebra associated to $T^* \bC^{n m}$. The reduction still depends on a parameter $t$ and can be also described (in two ways) in terms of twisted holomorphic differential operators on $\mathrm{Gr}_n(\bC^m)$. The resulting algebra $A^{\mathrm{3d}}_{m,n}$ is generated by the quantum $\mathrm{\mathrm{GL}}(m)$ moment maps, which satisfy the relations of $U(\gl_m)$ but also extra relations deforming $Z^2 = t Z$. It is thus a quotient of $U(\gl_m)$.

It is interesting to consider the unitary oscillator representation of the Weyl algebra, with $Y = X^\dagger$. This requires a quantization of $t=N \hbar$. The resulting Hilbert space involves $\mathrm{SL}(n)$-invariant states built from $n N$ raising operators, giving an irreducible representation of $\mathrm{\mathrm{GL}}(m)$ labelled by a Young Tableau with $N$ columns of height $n$. This is the natural quantization of the $Y = X^\dagger$ real locus $\CP^{\mathrm{3d},\bR}_{m,n}\subset \CP^{\mathrm{3d}}_{m,n}$, which is isomorphic to 
$\mathrm{Gr}_n(\bC^m)$ under a hyperk\"ahler rotation. 

The simplest case $m=2$, $n=1$ is the familiar ``fuzzy sphere'' quantization of $S^2$, giving the spin $N/2$ representation of $U(\gl_2)$.

\subsection{Classical representations from Coulomb branches}

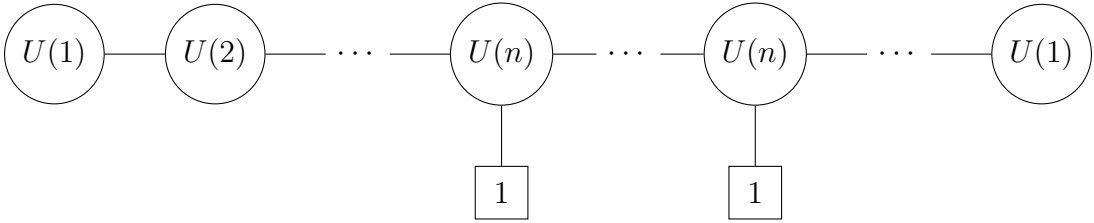
\begin{figure}[h]
    \centering
\begin{tikzpicture}[node distance=0.8cm and 0.8cm, auto]
  \tikzset{
    gauge/.style={circle, draw, minimum size=0.7cm},
    flavor/.style={rectangle, draw, minimum size=0.7cm}
  };

  \node[gauge] (L1) {$U(1)$};
  \node[gauge, right=of L1] (L2) {$U(2)$};
  \node[right=of L2] (Ldots) {$\cdots$};
  \node[gauge, right=of Ldots] (Ln) {$U(n)$};

  \node[gauge, right=2cm of Ln] (Rn) {$U(n)$};
  \node[right=of Rn] (Rdots) {$\cdots$};
  \node[gauge, right=of Rdots] (R1) {$U(1)$};

  \node[flavor, below=of Ln] (fL) {1};
  \node[flavor, below=of Rn] (fR) {1};

  \draw (L1) -- (L2);
  \draw (L2) -- (Ldots);
  \draw (Ldots) -- (Ln);

  \draw (Rn) -- (Rdots);
  \draw (Rdots) -- (R1);

  \node (middots) at ($(Ln)!0.5!(Rn)$) {$\cdots$};
  \draw (Ln) -- (middots);
  \draw (middots) -- (Rn);

  \draw (Ln) -- (fL);
  \draw (Rn) -- (fR);
\end{tikzpicture}
\caption{The quiver for $T^{(n,m-n)}[U(m)]$. There are $m-2n+1$ nodes with gauge group $U(n)$.}\label{fig:Trho}
\end{figure}

Three-dimensional mirror symmetry gives an alternative presentation of $\CP^{\mathrm{3d}}_{m,n}$ as a Coulomb branch of another quiver gauge theory $T^\rho[U(m)]$ with $m-1$ gauge nodes. From this point on, we pick $2n\leq m$ without loss of generality, so $n \leq m-n$. The nodes of the quiver have ranks which start with $1$ at the leftmost node, increase by one at each subsequent node until they reach $n-1$, followed by $m-2n+1$ nodes of rank $n$ and then a sequence with rank decreasing by one at each step, from $n-1$ to $1$. There is a single fundamental flavor at the leftmost and rightmost nodes of rank $n$, or two if these coincide. They have mass parameters $0$ and $t$. See Figure \ref{fig:Trho}. 

The Coulomb branch description hides the $\mathrm{\mathrm{GL}}(m)$ global symmetry of $\CP^{\mathrm{3d}}_{m,n}$ and $A^{\mathrm{3d}}_{m,n}$, but highlights a second hidden integrable structure. It presents them as BFN algebras \cite{Bullimore:2015lsa,Braverman:2016wma,Braverman:2016pwk}. The base of the integrable system is parameterized by the coefficients of the $m-1$ polynomials defined as the top left $a \times a$ minors of $z-Z$. 

As the rank of $Z$ is $n$, the polynomials for $a>n$ have an overall factor of $z^{a-n}$. Furthermore, as the rank of $Z-t$ is $m-n$, there is an additional factor of $(z-t)^{a-m+n}$. We thus recognize the structure of the quiver: 
\begin{enumerate}
    \item If $a\leq n$ the minors are generic monic degree $a$ polynomials and have $a$ independent coefficients.
    \item If $n\leq a\leq m-n$ the minors have a factor of $z^{a-n}$ and only contain $n$ independent coefficients. 
    \item If $m-n\leq a$ the minors have a factor of $z^{a-n}(z-t)^{a-m+n}$ and have $m-a$ independent coefficients. 
\end{enumerate}
The total number of interesting coefficients in the polynomials is $n(m-n)$, and they are known to Poisson-commute, giving $\CP^{\mathrm{3d}}_{m,n}$ an integrable system structure. 
Upon quantization, these Hamiltonians become the Gelfand-Tsetlin basis: Casimirs of $U(\gl_a)$ blocks inside $U(\gl_m)$. The above truncation structure persists. 

\subsection{Quantum group representations from Coulomb branches and multiplicative Higgs branches}

At this point we are ready to move to quantum groups. The Coulomb branch description of the 3d theory has a ``K-theoretic'' trigonometric deformation associated with the circle compactification of a 4d ${\cal N}=2$ gauge theory defined by the same quiver data. The K-theoretic Coulomb branch will be identified with $\CP_{m,n}$ from the main text, and the K-theoretic algebra $A_{m,n}$ with the corresponding quotient of $U_\fq(\gl_m)$. 

Classically, the relation to $U_\fq(\gl_m)$ will be expressed by parameterizing $\CP_{m,n}$ 
in terms of an analogue to $\omu$: an $m\times m$ matrix $M$ such that $M-1$ has rank $n$ and $M-g^2$ has rank $m-n$, together with a decomposition of $M$ into the product of an upper-triangular, a diagonal, and a lower-triangular matrix. We will also derive a ``multiplicative Higgs branch'' description, where the elements of the three sub-matrices 
are expressed as inner products of $m$ vectors $Y_i$ and $m$ co-vectors $X_i$ in $\bC^n$, constrained by a multiplicative F-term relation:
\begin{equation}
    (1_{n\times n}+Y_m X_m) \cdots (1_{n\times n}+Y_1 X_1) = g^2 1_{n \times m}\, .
\end{equation}
These descriptions identify $\CP_{m,n}$ with two character varieties.

\subsection{Brane engineering}

The path to the character varieties descriptions passes through a remarkable String Theory construction \cite{Witten:1997sc,Gaiotto:2009hg,Gaiotto:2009we,Gaiotto:2010be}. Namely, the 4d ${\cal N}=2$ linear quiver gauge theories can be engineered by a brane construction in IIA string theory, involving a variable number of ``D4 branes'', $m$ ``NS5 branes'' and $2$ ``D6 branes''. An M-theory lift of the problem maps the system to a single ``M5 brane'' wrapping a family of complicated, non-compact  holomorphic curves in a two-center Taub-NUT manifold. 

As a complex symplectic manifold, the two-center Taub-NUT manifold is the same as the deformed $A_1$ singularity $\cM_t$:  
\begin{equation}
    y \tilde y = z (z-t)\, ,
\end{equation}
with complex symplectic form $dz\,d\log y_\pm$. In the main text, we encounter $\cM_t$ first as a factor in the A-model background which is holographically dual to Chern-Simons theory on $S^2 \times \bR$. We also focus on A-model D-branes supported on certain holomorphic curves in $\cM_t$. This is not a coincidence: the K-theoretic Coulomb branch of the 4d ${\cal N}=2$ quiver gauge theories is recovered from the family of holomorphic curves by the same sequence of operations which we employ in the main text to describe the moduli space of A-branes:
\begin{enumerate}
    \item Restrict to a convenient $T^* C$ patch.
    \item Identify the holomorphic curves as spectral curves of Higgs bundles on $C$.
    \item Hyperk\"ahler rotate the moduli space of Higgs bundles on $C$ to a moduli space of flat connections on $C$, parameterized by monodromy/Stokes data.
\end{enumerate}
The choice of patch can be tracked back to the original IIA setup as different ways to organize the collection of D6 and NS5 branes, related by ``Hanany-Witten'' transitions \cite{Hanany:1996ie}.

The quiver data is converted into a polynomial equation for the spectral curve following \cite{Witten:1997sc}. The ranks of the gauge groups indicate the number of ``D4 segments'' between consecutive NS5 branes. Here they grow linearly from 1 to $n$, remain at $n$ until the $(m-n)$-th interval and then decrease linearly to $1$. The two D6 branes are placed in the intervals for the gauge nodes which carry a fundamental flavor: $n$-th and $(m-n)$-th. 

If we use Hanany-Witten moves to bring the D6 branes all the way to the left or to the right of the NS5 branes, the number of D4 brane segments changes to grow linearly from $1$ to $m$. The associated character variety description involves the Stokes/monodromy data of a $\mathrm{\mathrm{GL}}(m)$ flat connection on the $y$ (or $\tilde y$) plane with a rank $1$ irregular singularity at infinity and a regular singularity at the origin, with monodromy $M$
which satisfies $(M-1)(M-g^2)=0$ and has $n$ eigenvalues $g^2$ (or $1$). This makes contact with the classical limit of $U_\fq(\gl_m)$. 

If we bring the D6 branes to opposite sides, the number of D4 brane segments becomes constant $n$. The associated character variety description involves the monodromy data of a $\mathrm{\mathrm{GL}}(n)$ flat connection on the $z/y$ plane, with minimal regular singularities at $m$ points and monodromy $g^2$ at infinity. This is the ``multiplicative symplectic quotient'' description. 

The support of the M5 brane in $\cM_t$ is described by the polynomial equation 
\begin{equation}\label{eq:spectralyz}
\begin{split}
    \tau_0 (z-t)^n z^{m-n}+ \tau_1 (z-t)^{n-1} z^{m-n-1} P_{1,\ell}(z) y + \ldots +\tau_{n} z^{m-2n} P_{n,1}(z) y^n & \\+ \ldots+ \tau_{m-n} P_{n,m-2n+1}(z) y^{m-n} + \ldots + \tau_{m-1} P_{1,r}(z) y^{m-1}+ \tau_{m} y^{m}&=0 
\end{split}
\end{equation}
in the $(y,z)$ patch. We should supplement it with an equation for the $(\tilde y, z)$ patch: 
\begin{equation}
\begin{split}
    \tau_0 \tilde y^m+ \tau_1 P_{1,\ell}(z) \tilde y^{m-1}+ \ldots \tau_{n} P_{n,1}(z) \tilde y^{m-n}+ \ldots+ \tau_{m-n} P_{n,m-2n+1}(z) (z-t)^{m-2n} \tilde y^n& \\ + \ldots +\tau_{m-1} P_{1,r}(z) z^{n-1} (z-t)^{m-n-1}\tilde y+ \tau_{m} z^n (z-t)^{m-n} &=0 \, .
\end{split}
\end{equation} 

At large $z$ and $y$, the equation \eqref{eq:spectralyz} reduces to 
\begin{equation}
    \tau_0 z^m+ \tau_1 z^{m-1} y + \ldots + \tau_{m} y^{m}=0 \, ,
\end{equation}
with $m$ roots for $z/y$ we identify with the $x_i$ parameters.

The equation gives an $m$-sheeted cover of the $y$ plane, i.e. $m$ possible values for the dual coordinate $z/y = x_+$. The coordinate goes to $x_i$ at large $y$, giving the rank $1$ irregular singularity whose Stokes data will be braided as we braid the $x_i$. At $y \to 0$, we have a regular singularity with $n$ branches with $x_+ \sim t/v$, i.e. monodromy eigenvalue 
\begin{equation}
    g^2 = e^{2 \pi i t} \, .
\end{equation}
If we replace $x_+$ with $x_- = (z-t)/y$, we simply shift the whole connection by $t/y$, redefining the monodromy matrix $M$ around the origin with $g^{-2} M$.

Because the gauge groups in the quiver theory are $U$ instead of $SU$, the Stokes data which parametrizes $\CP_{m,n}$ consists of two triangular Stokes matrices, one upper triangular and one lower triangular, together with a diagonal formal monodromy which is part of the data. The formal monodromy parameters $\olambda_i^2$ are multiplicative moment maps for a $U(1)^m$ global symmetry of $\CP_{m,n}$ which conjugates the Stokes data by a diagonal matrix. 

Their product is $M$. For later convenience, we split the formal monodromy into two factors and write
\begin{equation}
    M = \hat K \hat f_- \hat K \hat e_+ \, ,
\end{equation}
constrained to have $n$ $g^2$ eigenvalues and $m-n$ $1$ eigenvalues.

In the $x_+=z/y$ plane, with fiber coordinate $y$, we write
\begin{equation}\label{eq:spectraltwo}
\begin{split}
    \tau_0 (y x_+ -t)^n x_+^{m-n}+ \tau_1 (y x_+-t)^{n-1} x_+^{m-n-1} P_{1,\ell}(y x_+) + \ldots \tau_{n}x_+^{m-2n} P_{n,1}(y x_+) & \\+ \ldots+ \tau_{m-n} P_{n,m-2n+1}(y x_+) + \ldots + \tau_{m-1} P_{1,r}(y x_+) y^{n-1}+ \tau_{m} y^{n}&=0 \, ,
\end{split}
\end{equation}
and find an $n$-sheeted cover, leading to a $\mathrm{\mathrm{GL}}(n)$ flat connection with minimal regular singularities at the locations $x_i$ and a monodromy $g^2$ at infinity. 


We can package the monodromy eigenvalue information into a choice of non-trivial right eigenvectors $\tilde s_i$. The monodromy around the punctures can then be written as
\begin{equation}
    M_i = 1+s_i \tilde s_i\, .
\end{equation}
This implies $\tilde s_i \cdot s_i = \olambda_i^2-1$, where $\mu_i^2$ is now the monodromy eigenvalue. We see that the data of $s_i$ and $\tilde s_i$ completely specifies the monodromy data. The total monodromy is constrained:
\begin{equation}
    M_m M_{m-1} \cdots M_1 = g^2 \, .
\end{equation}

Such a connection can be equivalently interpreted as a {\it twisted} connection on $\bCP^1$ 
with $m$ minimal regular singularities, which is precisely the natural description for an A-model D-brane in a twisted cotangent bundle of $\bCP^1$, e.g. $\cM_t$.

\subsection{Translating between descriptions}
Consider a 2d $\mathrm{\mathrm{GL}}(n)$ flat connection with $m$ minimal regular singularities placed roughly along the real axis. The monodromy eigenvalue is not fixed and a choice of non-trivial (left) eigenvector at each singularity is part of the definition of the moduli space. Denote as $s_i$ the left eigenvectors transported to infinity along the positive $y$ axis and $\tilde s_i$ the right eigenvectors, normalized so that the monodromy $M_i$ along a path which descends to the left of the $i$-th singularity goes around it counterclockwise once and goes back to infinity takes the form specified above:
\begin{equation}
    M_i = 1+s_i \tilde s_i\, ,
\end{equation}
with $\tilde s_i \cdot s_i = K_i^2-1$. This implies 
\begin{equation}
    M^{-1}_i = 1- K_i^{-2} s_i \tilde s_i\, .
\end{equation}
We impose 
\begin{equation}
    M_m M_{m-1} \cdots M_1 = g^2\, ,
\end{equation}
and quotient by overall $\mathrm{\mathrm{GL}}(n)$ conjugation, leaving $2 n m - 2 n^2 = 2 n(m-n)$ degrees of freedom. The monodromy condition for $g^2 \neq 1$ implies that the matrix $s$ with entries $s_i$ has rank $n$.

Compare two sets of vectors:
\begin{align}
    s'_i &=  M_1^{-1} \cdots M_{i-1}^{-1}s_i  \cr
    s''_i &= M_m \cdots M_{i+1} M_i s_i\, .
\end{align}
Clearly, $g^2 s'_i = s''_i$. We can expand 
\begin{align}
    s'_1 &= s_1  \cr
    s'_2 &= s_2 - s_1 (\tilde s_1\cdot s_2) K_1^{-2} \cr
    s'_3 &= s_3 - s_1 (\tilde s_1 \cdot s_3) K_1^{-2}  - s_2 (\tilde s_2\cdot s_3) K_2^{-2} +s_1 (\tilde s_1\cdot s_2) K_1^{-2} (\tilde s_2\cdot s_3)K_2^{-2} \cr
    \cdots &= \cdots\, .
\end{align}
Collecting the vectors into matrices, we can write that as 
\begin{equation}
    s' = s \left(1-K_1^{-2} (\tilde s_1\cdot s_2)b_{12} \right) \left(1-K_1^{-2} (\tilde s_1\cdot s_3)b_{13} \right)\left(1-K_2^{-2} (\tilde s_2\cdot s_3)b_{23} \right) \cdots \, .
\end{equation}
Similarly, 
\begin{align}
    s''_m =&\; s_m \olambda^2_m \cr
    s''_{m-1} =&\; s_{m-1} \olambda^2_{m-1} + s_m (\tilde s_m\cdot s_{m-1})  \olambda^2_{m-1} \cr
    s''_{m-2} =&\; s_{m-2} \olambda^2_{m-2}+ s_{m-1} (\tilde s_{m-1} \cdot s_{m-2}) \olambda^2_{m-2} - s_m (\tilde s_m\cdot s_{m-2}) \olambda^2_{m-2} \cr
    &+s_m (\tilde s_m\cdot s_{m-1}) (\tilde s_{m-1}\cdot s_{m-2}) \olambda^2_{m-2} \cr
    \cdots =&\; \cdots\, ,
\end{align}
i.e. 
\begin{equation}
    s'' = s \left(1+(\tilde s_m\cdot s_{m-1})b_{m,m-1} \right) \left(1+(\tilde s_m\cdot s_{m-2})b_{m,m-2} \right)\left(1+(\tilde s_{m-1}\cdot s_{m-2})b_{m-1,m-2} \right) \cdots \hat K^2\, .
\end{equation}
Accordingly, 
\begin{align}
    g^2 s &= s \left(1+(\tilde s_m\cdot s_{m-1})b_{m,m-1} \right) \left(1+(\tilde s_m\cdot s_{m-2})b_{m,m-2} \right)\left(1+(\tilde s_{m-1}\cdot s_{m-2})b_{m-1,m-2} \right) \cdots \cr &\hat K \cdots \left(1+K_2^{-1}K_3^{-1} (\tilde s_2\cdot s_3)b_{23} \right)\left(1+K_1^{-1}K_3^{-1} (\tilde s_1\cdot s_3)b_{13} \right)\left(1+K_1^{-1} K_2^{-1}(\tilde s_1\cdot s_2)b_{12} \right) \hat K\, .
\end{align}
The $m \times m$ matrices on the right-hand side are identified with the Stokes data of a connection with a rank $1$ irregular singularity. In the notation used elsewhere in the paper, we identify the matrices as $\hat e_+ = \hat e_-^{-1}$ and $\hat f_-$. Accordingly, $f_{-;i,j} = K_i^{-1}K_j^{-1} (\tilde s_i\cdot s_j)$ and 
$e_{-;i,j} = (\tilde s_i\cdot s_{j})$.

As $s \hat e_+ \hat K \hat f_- \hat K = g^2 s$, the matrix $s$ is identified with the matrix of right eigenvectors of $\hat e_+ \hat K \hat f_- \hat K$ with eigenvalues $g^2$. We can repeat the exercise for $\tilde s$. Namely, define 
\begin{align}
    \tilde s'_i &= \tilde s_i M_{i+1}^{-1} \cdots M_m^{-1}  \cr
    \tilde s''_i &= \tilde s_i M_i \cdots M_1\, ,
\end{align}
so that $g^2 \tilde s'_i = \tilde s''_i$. We can expand 
\begin{align}
    \tilde s'_m &= \tilde s_m \cr
    \tilde s'_{m-1} &= \tilde s_{m-1} - K_m^{-2} (\tilde s_{m-1}\cdot s_m) \tilde s_m  \cr
    \cdots &= \cdots\, .
\end{align}
Collecting the vectors into matrices, we can write that as 
\begin{equation}
    \tilde s' = \cdots  \left(1-K_m^{-2} (\tilde s_{m-1}\cdot s_m)b_{m-1,m} \right) \tilde s \, .
\end{equation}
Similarly, 
\begin{align}
   \tilde s''_1 &= \olambda^2_1 \tilde s_1\cr
    \tilde s''_2 &= \olambda^2_2 \tilde s_2  +  (\tilde s_2\cdot s_{1})  \olambda^2_2 \tilde s_1 \cr
    \cdots &= \cdots\, ,
\end{align}
i.e. 
\begin{equation}
    \tilde s'' =  \hat K^2 \cdots \left(1+(\tilde s_2\cdot s_1)b_{2,1} \right)  \tilde s\, .
\end{equation}
Accordingly, 
\begin{align}
    \tilde s g^2  &= \hat K \left(1+K_{m-1}^{-1}K_m^{-1} (\tilde s_{m-1}\cdot s_m)b_{m-1,m} \right)  \cdots \hat K \cdots \left(1+(\tilde s_2\cdot s_1)b_{2,1} \right)  \tilde s\, .
\end{align}
As $\hat K \hat f_- \hat K \hat e_+ \tilde s= g^2 \tilde s$, the matrix $\tilde s$ is identified with the matrix of left eigenvectors of $\hat K \hat f_- \hat K \hat e_+$ with eigenvalues $g^2$.

In particular, this implies that $M = \hat K \hat f_- \hat K \hat e_+$ has both $n$ independent left eigenvectors and $n$ independent right eigenvectors with eigenvalues $g^2$. 
In order to complete the match to the Stokes description of $\CP_{m,n}$, we need to show that $M$ acts as the identity from the right on the orthogonal complement to $\tilde s$. 

In order to do so, be $c_i$ a collection of numbers such that $\sum_i c_i \tilde s_i =0$, i.e. $c \tilde s =0$. We want to compare $c \hat K \hat f_- \hat K$ and $c \hat e_-$. 
The former has entries 
\begin{align}
    &c_1 K_1^2  \cr
    &c_2 K_2^2 + c_1 (\tilde s_1\cdot s_2)  \cr
    &c_3 K_3^2 + c_2 (\tilde s_2\cdot s_3)+c_1 (\tilde s_1\cdot s_3)  \cr
    &\cdots \, .
\end{align}
But this is the same as 
\begin{align}
    &c_1 - c_2 (\tilde s_2\cdot s_1)- c_3 (\tilde s_3\cdot s_1)- \ldots \cr
    &c_2 - c_3 (\tilde s_3\cdot s_2) - \ldots \cr
    &c_3 - \ldots \cr
    &\cdots \, ,
\end{align}
i.e. the entries of $c \hat e_-$. This shows that $c M = c$, as desired.

This identification can be inverted. Starting from the Stokes data $\hat K$, $\hat e_+$, $\hat f_-$ we can define $s$ and $\tilde s$ as matrices of eigenvectors, defined up to two separate $\mathrm{\mathrm{GL}}(m)$ actions. We can further rigidify that by constraining the entries of $\tilde s s$ to have the above expressions in terms of $\hat K$, $\hat f_-$ and $\hat e_-$. This is an over-determined set of linear equations on $n^2$ variables, but it should be possible to mirror the manipulations above in order to verify that it can be solved. It is a problem completely analogous to reconstructing $X$ and $Y$ from the data of $\omu$ in $\CP_{m,n}^{\mathrm{3d}}$.

The relation we proposed here between the two presentations of $\CP_{m,n}$ can be 
derived in a more rigorous fashion as a generalized Nahm transform \cite{Bullimore:2015lsa}.
One takes a family of auxiliary $\mathrm{\mathrm{GL}}(1)$ flat connections, either $y_0 dx_+$ on the $x_+$ plane or $x_0 dy$ on the $y$ plane, and looks at the space of extensions between such $\mathrm{\mathrm{GL}}(1)$ flat connections and the original $\mathrm{\mathrm{GL}}(m)$ or $\mathrm{\mathrm{GL}}(n)$ flat connections. The space of extensions 
fibered over the $y_0$ or $x_0$ planes gives the transformed connection, and vice versa. 

The transform can be expressed in the language of monodromy data. For example, 
we can express the Nahm transform of the $\mathrm{\mathrm{GL}}(m)$ connection in terms of the space $\CP_{m+1,n}/\!\!/\mathrm{\mathrm{GL}}(1)$ of $\mathrm{\mathrm{GL}}(m+1)$ connections with the same singularity structure and parameters $x_i$, $x_0$ at infinity, with fixed $K_0=1$ and quotiented by the corresponding $\mathrm{\mathrm{GL}}(1)$ symmetry. There is an obvious embedding $\CP_{m,n} \subset \CP_{m+1,n}/\!\!/\mathrm{\mathrm{GL}}(1)$ where the $\mathrm{\mathrm{GL}}(m+1)$ connection is the combination of a $\mathrm{\mathrm{GL}}(m)$ connection and a trivial $\mathrm{\mathrm{GL}}(1)$ connection. The normal bundle to this embedding has dimension $2n$ and splits into the two types of extensions, e.g. the extra rows/columns in $\hat e_+$ and $\hat f_-$. If we vary $x_0$ and braid it around the $x_i$, the $\mathrm{\mathrm{GL}}(m+1)$ Stokes data will be braided as well. This acts linearly on the extensions, giving the desired $\mathrm{\mathrm{GL}}(m)$ flat connection on the $x_0$ plane.

\subsection{More about the D-brane interpretation}
Recall the twisted cotangent bundle interpretation of $\cM_t$: we take local coordinates 
$x_+ = \frac{z}{y}$ and $\tilde x_+ = x_+^{-1} = \frac{z-t}{\tilde y}$ on the base $\bCP^1$, 
$y$ and $-\tilde y$ on the fiber. Then $\tilde y = -x_+^2 y + t x_+$, an affine deformation of the usual cotangent bundle definition. 

A twisted connection on $\bCP^1$ is modeled on the twisted cotangent bundle: it is considered trivial at infinity if it behaves as $\frac{t}{x_+}$ in the $x_+$ patch. Passing to the monodromy data, a twisted $\mathrm{\mathrm{GL}}(n)$ connection on $\bCP^1$ is simply a $\mathrm{\mathrm{GL}}(n)$ connection 
which has monodromy $g^2$ around infinity. 

A-model D-branes in $\cM_t$ are expected to be described precisely by the (derived category of) such twisted connections. We conclude that $\CP_{m,n}$ is a moduli space of A-model D-branes in $\cM_t$. We will now sharpen the geometric characterization of these D-branes. 

Consider the spectral curves for $m=1$, $n=0$ and $m=1$, $n=1$:
\begin{align}
    &L_+[x]: \quad z - x y =0  \qquad \qquad \tilde y - x(z-t) =0 \cr
    &L_-[x]: \quad z -t - x y =0  \qquad \qquad \tilde y - x z = 0\, .
\end{align}
These are respectively a generic fiber of the above twisted cotangent bundle presentation of $\cM_t$, and a generic fiber of the second twisted cotangent bundle presentation of $\cM_t$.
They both reach to the same direction $x$ at large $z$ and $y$. These are the support of rigid D-branes $D_+$ and $D_-$.

Next, consider the union of $m-n$ copies of $D_+$ for various $x$'s and $n$ copies of $D_-$ at other $x$'s, all distinct. This can be deformed to a more general D-brane with the same asymptotic shape. $\CP_{m,n}$ is the moduli space of such D-branes. 

If we project the D-brane to the base of the twisted cotangent bundle presentation, then the $D_+$ branes will be ``vertical'', while the $D_-[x_i]$ branes will cover $\bCP^1$ while also spiking along the fiber at $x_i$. This is why the corresponding twisted connection on $\bCP^1$ has rank $n$ and $m$ minimal regular singularities. On the other hand, the $\mathrm{\mathrm{GL}}(m)$ description arises from a projection onto the $y$ plane.

\subsection{Quantization}
Finally, we can comment on the quantization of these classical relations. The character variety descriptions of $\CP_{m,n}$ have a natural quantization as skein algebras. The quantization maps holonomies to Wilson loops in Chern-Simons theory. The different presentations are expected to give equivalent quantizations, as the resulting algebra $A_{m,n}$ is an intrinsic property of the 4d ${\cal N}=2$ SQFT: the fusion algebra of BPS line defects \cite{Gaiotto:2010be}. 

In the $\mathrm{\mathrm{GL}}(n)$ description, the rung vevs are mapped to open fundamental Wilson lines in a $\mathrm{\mathrm{GL}}(n)$ Chern-Simons theory, i.e. essentially to rung operators, albeit stretched between a different type of co-dimension $2$ defects. The skein relations themselves do not really know about $n$, and so naturally reproduce the rung algebra from the original $SU(N)_\kappa$ CS theory. This can be seen as a finite $N$ (but not yet restricted to be integer) holographic match between the operator algebra on the $m$ $\Lambda^\bullet \bC^{N}$ Wilson lines and the quantization of the phase space of the dual D-branes. 

\subsection{Real loci and Hilbert spaces}
In order to push the analysis to the actual Hilbert space $\cH_{m,n}$ of the system, we need to discuss reality/hermiticity conditions. 

In the original $SU(N)_\kappa$ CS theory, the adjoint of a rung operator is a rung in the opposite direction: $E_{-;ij}^\dagger = F_{-;j,i}$. We also have $\fq^\dagger = \fq^{-1}$ and $g^\dagger = g^{-1}$. As $\fq-\fq^{-1}$ is pure imaginary, we get $e_{-;ij}^\dagger =- f_{-;j,i}$.

Classically, we have $\hat f_-^\dagger = \hat e_- = e_+^{-1}$ and $\hat K^\dagger = K^{-1}$. As a result, $M^\dagger = M^{-1}$ and $M$ is unitary. 
This characterizes the real locus $\CP^\bR_{m,n}$ in the $\mathrm{\mathrm{GL}}(m)$ character variety description. Notice that $\CP^\bR_{m,n}$ is diffeomorphic to the Grassmannian $\mathrm{Gr}_{m,n}$ and is in particular compact. 

The quantization of a compact real phase space is expected to produce a finite-dimensional Hilbert space $\cH_{m,n}$, which we identify with the irreducible unitary representation of $U_\fq(\gl_m)$ labeled by a rectangular 
Tableau of sides $n$ and $N$, i.e. a summand in the Hilbert space of the $SU(N)_\kappa$  CS theory on the two-sphere in the presence of $m$ $\Lambda^\bullet \bC^{N}$ Wilson lines.

In the $\mathrm{\mathrm{GL}}(n)$ description, the reality condition seems compatible with the flat connection being unitary, i.e. $M_i^\dagger = M_i^{-1}$, i.e. 
$s_i^\dagger = \sqrt{-1} K_i^{-1} \tilde s_i$. Recall that $K_i^\dagger = K_i^{-1}$. In other words, we identify 
$\CP^\bR_{m,n}$ with a space of twisted $U(n)$ flat connections with $m$ minimal regular singularities. 

This is the phase space of a $U(n)_{\kappa+N-n}$ Chern-Simons theory in the presence of $m$ vortex defects and with a total Abelian charge $N$ at infinity. There is an obvious resemblance to the original $SU(N)_\kappa$ CS theory setup: if we promoted the gauge group to $U(N)_\kappa$, we would need a total Abelian charge $n$ at infinity. The holographic duality at the level of Hilbert spaces seems to be the composition of two Howe dualities: we map the $SU(N)_\kappa$ CS theory Hilbert space in the presence of $m$ defects with ``total charge'' $n$ to a unitary irrep of $U_\fq(\gl_m)$ and then to the $U(n)_{\kappa+N-n}$ CS theory Hilbert space in the presence of $m$ defects with total charge $N$. This perspective could be further justified by identifying the vortex defects with $U(n)$ Wilson lines transforming in the symmetric powers of the fundamental representation. 

\subsection{Example: \texorpdfstring{$\CP_{2,1}$}{CP(2,1)}}
The 3d version of the manifold $\CP^{\mathrm{3d}}_{2,1} = T^* \bC^2/\!\!/\mathrm{\mathrm{GL}}(1)$ is itself a deformed $A_1$ singularity
\begin{equation}
    \omu_{12}\omu_{21} = \omu_{11}(\omu_{11}-t) \, .
\end{equation} 
The real locus $\CP^{\mathrm{3d},\bR}_{2,1}$ is defined by 
\begin{equation}
    |\omu_{12}|^2 = \omu_{11}(\omu_{11}-N \hbar) \, ,
\end{equation}
with real $\omu_{11}$, and it is a sphere $S^2$. The Hilbert space $\cH^{\mathrm{3d}}_{2,1}=\bC^{2N+1}$ is the spin $N/2$ irrep of $U(\gl_2)$,
the natural ``fuzzy sphere'' quantization of $S^2$. 

The 3d theory SQED${}_2$ is self-mirror, so the manifold $\CP^{\mathrm{3d}}_{2,1}$ is also presented as the Coulomb branch of
SQED${}_2$. 

The K-theoretic Coulomb branch $\CP_{2,1}$ of SQED${}_2$ is a multiplicative version of the deformed $A_1$ singularity:
\begin{equation}
    e f = (K_1 - K_1^{-1})(g/K_1-K_1/g) \, .
\end{equation} 
The multiplicative Higgs branch description is simple: we start from 
\begin{equation}
    (1+s_1 \tilde s_1)(1+s_2 \tilde s_2) = g^2\, ,
\end{equation}
and $e = s_2\tilde s_1$, $f = K_1^{-1} K_2^{-1}s_1\tilde s_2$, and get 
\begin{equation}
    e f = K_1^{-1} K_2^{-1} (K_1^2-1)(K_2^2-1) \, ,
\end{equation} 
as desired. 

The real locus $\CP^{\bR}_{2,1}$ is 
\begin{equation}
    |e|^2 + (K_1 - K_1^{-1})(g/K_1-K_1/g)=0 \, .
\end{equation} 
It consists of a circle fibered over an arc in the $K_1$ plane, shrinking at the endpoints. Topologically, it is still a sphere. 

Geometrically, the M5 brane support is 
\begin{align}
    z(z-t) + (x_1 + x_2) (z-a) y + x_1 x_2 y^2 &= 0 \cr
    \tilde y^2 + (x_1 + x_2) (z-a) \tilde y + x_1 x_2 z(z-t) &= 0\, ,
\end{align}
but we can also write it as the affine-linear constraint
\begin{equation}
    \tilde y + (x_1 + x_2) z + x_1 x_2 y =(x_1 + x_2) a\, .
\end{equation}

\section{Geometric correspondences and BPS interfaces}\label{app:inter}
We begin with some representation theory. Much as $\cH_{m,n}$, the Hilbert space $\cH^{\mathrm{3d}}_{m,n}$ 
can be described either as the $U(N)$-invariant part of a Fock space of $m N$ fermionic oscillators with total fermion number $n N$, or as the part of a Fock space of $m n$ bosonic oscillators which transform as the $N$-th power of the determinant representation of $U(n)$. These statements are derived from general Howe duality statements: 
\begin{align}
    \Lambda^\bullet \bC^{m N} &= \sum_T R_T[U(N)] \otimes R_{T^t}[U(m)]  \cr
    S^\bullet \bC^{m n} &= \sum_T R_T[U(n)] \otimes R_{T}[U(m)]\, ,
\end{align}
where both sums run over Young Tableaux $T$. In the first case 
we pick the rectangular $T$ with $n$ columns of height $N$, in the second the rectangular $T$ with $N$ columns of height $n$.

In the first description, we could split the $m$ fermionic oscillators into groups of size $m_1$ and $m_2$. This gives an obvious map 
\begin{equation}
    \cH^{\mathrm{3d}}_{m_1,n_1} \times \cH^{\mathrm{3d}}_{m_2,n_2} \to \cH^{\mathrm{3d}}_{m,n}\, , 
\end{equation}
for every $n_1 + n_2=n$. We would like to describe this map in terms of bosonic oscillators. In order to do so, we need a stronger result. Write:
\begin{equation}
    \Lambda^\bullet \bC^{m N} = \sum_{T_1,T_2} R_{T_1}[U(N)] \otimes R_{T_2}[U(N)] \otimes R_{T_1^t}[U(m_1)]\otimes R_{T_2^t}[U(m_2)] \, .
\end{equation}
Projection over $SU(N)$ invariants requires $T_1$ and $T_2$ to be vertically complementary in the $n N$ rectangle. So we can decompose 
$\cH^{\mathrm{3d}}_{m,n}$ into a sum of products of $U(m_1)$ and $U(m_2)$
irreps labelled by $T_1^t$ and $T_2^t$ complementary horizontally in the $N n$ rectangle. In particular, we can find $\cH^{\mathrm{3d}}_{m_1,n_1} \times \cH^{\mathrm{3d}}_{m_2,n_2}$ exactly once in $\cH^{\mathrm{3d}}_{m,n}$.

Accordingly, any non-zero map will do up to a scale. The bosonic oscillators decompose into four blocks each: lowering operators $X_{11}$, $X_{12}$, $X_{21}$, $X_{22}$ and raising operators $Y_{11}$, etc. Consider states in the bosonic Fock space which are annihilated by 
$X_{12}$ and $X_{21}$ and built from $n_1 N$ of the $Y_{11}$ and $n_2 N$ of the $Y_{22}$ raising operators, and are $U(n_1) \times U(n_2)$ invariant. This is a copy of $\cH^{\mathrm{3d}}_{m_1,n_1} \times \cH^{\mathrm{3d}}_{m_2,n_2}$. We can then average over $U(n)$ to get a map to $\cH^{\mathrm{3d}}_{m,n}$.

The goal of this Section is to identify a classical limit of this map as a Lagrangian correspondence between $\CP^{\mathrm{3d}}_{m_1,n_1} \times \CP^{\mathrm{3d}}_{m_2,n_2}$ and $\CP^{\mathrm{3d}}_{m,n}$ and then lift it to a Lagrangian correspondence between $\CP_{m_1,n_1} \times \CP_{m_2,n_2}$ and $\CP_{m,n}$. This is our candidate for the classical limit of a map 
\begin{equation}
    \cH_{m_1,n_1} \times \cH_{m_2,n_2} \to \cH_{m,n} \, ,
\end{equation}
which plays the role of a ``generalized cup'' in the main text: it describes states for $m$ strands created by two disconnected networks ending the $m_1$ and $m_2$ groups of strands independently. We will verify this conjecture for actual cups and caps by explicit calculations. 

\subsection{Correspondences and half-BPS interfaces}
The above bosonic construction of the map 
\begin{equation}
    \cH^{\mathrm{3d}}_{m_1,n_1} \times \cH^{\mathrm{3d}}_{m_2,n_2} \to \cH^{\mathrm{3d}}_{m,n} 
\end{equation}
has a simple classical limit: we impose 
\begin{align}
    X_{11} &= X^{(1)}  \cr
    X_{12} &=0 \cr
    X_{21} &= 0 \cr
    X_{22} &= X^{(2)} \cr
    Y_{11} &= Y^{(1)} \cr
    Y_{22} &= Y^{(2)}\, , \cr
\end{align}
where $X$ and $Y$ are the variables from $\CP^{\mathrm{3d}}_{m,n}$, $X^{(1)}$, $Y^{(1)}$ from $\CP^{\mathrm{3d}}_{m_1,n_1}$, and 
$X^{(2)}$ and $Y^{(2)}$ from $\CP^{\mathrm{3d}}_{m_2,n_2}$. The F-term equations are compatible with this but further impose 
$Y_{12} X^{(2)} =0$ and $Y_{21} X^{(1)} =0$.

We have 
\begin{align}
    2n_1 m_1 + 2n_2 m_2 + n_1 m_2 + n_2 m_1 - 2 n_1^2 - 2 n_2^2 - 2 n_1 n_2 =&\; n_1(m_1-n_1) + n_2(m_2-n_2)\cr &+(n_1 + n_2)(m_1 + m_2 - n_1 - n_2)
\end{align}
degrees of freedom, as appropriate for a Lagrangian correspondence. We denote this as $\CC^{\mathrm{3d}}_{m_1,n_1;m_2,n_2}$.

In terms of gauge-invariant quantities, we have
\begin{align}
    \omu_{11} &= \omu^{(1)} \cr
    \omu_{12} &= X^{(1)} Y_{12}  \cr
    \omu_{21} &= X^{(2)} Y_{21}  \cr
    \omu_{22} &= \omu^{(2)} \, .
\end{align}

The above equations also work to define a supersymmetric, half-BPS interface between $U(n)$ SQCD with $m$ flavors and the product of  $U(n_1)$ SQCD with $m_1$ flavors and $U(n_2)$ SQCD with $m_2$ flavors.
We only need to specify in addition that the $U(n)$ gauge group is broken to $U(m_1) \times U(m_2)$ at the interface \cite{Bullimore:2016nji}. 

The interface formulation also gives a quantization $\CC^{\mathrm{3d}}_{m_1,n_1;m_2,n_2}$ as a bimodule $B^{\mathrm{3d}}_{m_1,n_1;m_2,n_2}$
between $A^{\mathrm{3d}}_{m,n}$ and $A^{\mathrm{3d}}_{m_1,n_1} \otimes A^{\mathrm{3d}}_{m_2,n_2}$. See \cite{Bullimore:2016nji} for the general recipe.

It would be interesting to identify a mirror description of the supersymmetric interfaces, which could then be lifted to an interface between 4d ${\cal N}=2$ SQFTs. We do not know how to do so. 

Instead, we can tentatively apply our strategy to the multiplicative Higgs branch description of $\CP_{m,n}$: we constrain $\tilde s$ in the same manner as $X$, leading to 
\begin{align}
    \tilde s_{11} &= \tilde s^{(1)} \cr
    \tilde s_{12} &=0 \cr
    \tilde s_{21} &= 0  \cr
    \tilde s_{22} &= X^{(2)} \cr
    s_{11} &= s^{(1)}  \cr
    s_{22} &= s^{(2)}\, . 
\end{align}
More concretely, $\tilde s_1 \cdots \tilde s_{m_1}$ annihilate the $\bC^{n_2}$ summand in $\bC^{n_1} \oplus \bC^{n_2}$ and $\tilde s_{m_1+1} \cdots \tilde s_{m_1+m_2}$ annihilate the first. This is enough to guarantee that the block-diagonal part of the total monodromy $M_\infty$ 
acting on a vector $v$ in $\bC^{n_2}$ will map it to $g^2$ times itself plus an extra term in $\bC^{n_1}$, a bilinear which should be set to $0$ in analogy with the $Y_{12} X^{(2)} =0$ conditions above. 

Similarly, $M^{-1}_\infty$ acting on a vector in $\bC^{n_1}$ will map it to 
 $g^{-2}$ times itself plus an extra term in $\bC^{n_2}$, a bilinear which should be set to $0$ in analogy with the $Y_{21} X^{(1)} =0$ conditions above.

The conditions imply that $K_1 = K_1^{(1)}$, etc. and $K_{m_1+1} = K_1^{(2)}$, etc. Also, $e$'s and $f$'s within each group of strands match. On the other hand, 
\begin{align}
    f_{-;i,m_1+j} &= K_1^{-1} K_{m_1+1}^{-1} \tilde s_j \cdot s_{m_1+j}  \cr
    e_{-;m_1+i,j} &= \tilde s_{m_1+i} \cdot s_j\, 
\end{align}
are matrices of rank $n_1$ and $n_2$ respectively. 

\subsection{Preparing a cup}
Now we specialize to $m_2=2$, $n_2=1$. Then the last component of 
$\tilde s_1 \cdots \tilde s_{m-2}$ is zero and only the last component of $\tilde s_{m-1}$ and $\tilde s_m$ are non-zero. Accordingly, inner products of bilinears can be rearranged:
\begin{align}
    e_{-;m,m-1} f_{-;m-1,m} &= (K_{m-1}- K_{m-1}^{-1})(K_{m}- K_{m}^{-1})\cr
    (K_m^2-1) e_{-;m-1,j} &= K_{m-1} K_m f_{-;m-1,m} e_{-;m,j} \cr
    (K_{m-1}^2-1) e_{-;m,j} &= e_{-;m,m-1} e_{-;m-1,j} \cr
    e_{-;m,j}e_{-;m-1,k} &= e_{-;m-1,j}e_{-;m,k}\, .
\end{align}

We also have $K_m K_{m-1} = g$ and 
\begin{equation}
    g^2 = (1 + K_{m-1}^2 s_m \tilde s_m + s_{m-1} \tilde s_{m-1})M_{m-2} \cdots M_1\, .
\end{equation}
Acting on a unit vector $b_m$ in the last component, that gives simply
\begin{equation}
    (g^2-1) b_m = K_{m-1}^2 s_m \tilde s_m \cdot b_m + s_{m-1} \tilde s_{m-1} \cdot b_m\, ,
\end{equation}
constraining a linear combination of $s_m$ and $s_{m-1}$. That gives  
\begin{equation}
    K_{m-1}K_m f_{-;i,m} \tilde s_m \cdot b_m + f_{-;i,m-1} \tilde s_{m-1} \cdot b_m=0\, .
\end{equation}

On the other hand, 
\begin{equation}
    g^{-2} = M_1^{-1} \cdots M_{m-2}^{-1} (1 - K_{m-1}^{-2} K_{m}^{-2} s_{m-1} \tilde s_{m-1}- K_{m}^{-2} s_{m} \tilde s_{m})
\end{equation}
acting on a vector orthogonal to $\tilde s_m$ gives a constraint of the form $\sum_i \tilde s_m \cdot s_i \tilde s_i =0$.

In order to specialize to a cup, we further define
\begin{align}
    e_{-;m,m-1} &= \omu (K_{m-1} - K_{m-1}^{-1}) \cr
    f_{-;m-1,m} &= \omu^{-1} (K_{m} - K_{m}^{-1}) \, ,
\end{align}
so that 
\begin{align}
    e_{-;m,j} &= \omu K_{m-1}^{-1}e_{-;m-1,j} \cr
    K_m f_{-;i,m} \omu &= -f_{-;i,m-1} \, .
\end{align}

\section{The \texorpdfstring{$U_\fq(\fgl_m)$}{Uq(glm)} rung algebra and the braid group}
\label{app:rung_algebra}

The reference \cite{Cautis_2014} describes diagrammatic rules for the manipulation of tensor products of $\Lambda^\bullet \bC^N$ representations of $U_\fq(\sl_N)$. In particular, we can represent the tensor product 
\begin{equation}
    \Lambda^{k_1} \bC^N \otimes \ldots \otimes \Lambda^{k_m} \bC^N
\end{equation}
as a collection of vertical lines with labels $k_1, \ldots, k_m$. 

\subsection{The rung algebra as a representation of \texorpdfstring{$U_\fq(\fgl_m)$}{Uq(glm)}}
``Rung'' operators are defined by combining canonical ``junction'' maps $\Lambda^{r+s}\bC^N \leftrightarrow \Lambda^{r}\bC^N \otimes \Lambda^{s}\bC^N$ to maps:
\begin{align}
    E^{(r)}_i &: \Lambda^{k_i}\bC^N \otimes \Lambda^{k_{i+1}}\bC^N \to \Lambda^{k_i+r}\bC^N \otimes \Lambda^{k_{i+1}-r}\bC^N \cr
    F^{(r)}_i &: \Lambda^{k_i}\bC^N \otimes \Lambda^{k_{i+1}}\bC^N \to \Lambda^{k_i-r}\bC^N \otimes \Lambda^{k_{i+1}+r}\bC^N\, .
\end{align}
We will mostly use $E_i\equiv E_i^{(1)}$ and $F_i\equiv F_i^{(1)}$. 

Collections of rungs can be manipulated according to the rules reviewed in Appendix \ref{app:rules}. For example, 
\begin{align}
    E^{(r)}_i E^{(s)}_i &= \qBinomial{r+s}{r} E^{(r+s)}_i \cr
    F^{(r)}_i F^{(s)}_i &= \qBinomial{r+s}{r} F^{(r+s)}_i\, ,
\end{align}
so e.g. $E_i^s = [s]_\fq! E_i^{(s)}$ and $F_i^s = [s]_\fq! F_i^{(s)}$.

These algebra of rung operators together with $K_i = \fq^{k_1}$ gives a representation of $U_\fq(\gl_m)$. By definition, 
\begin{align}
    K_i E_i &= \fq E_i K_i \cr
    K_{i+1} E_i &= \fq^{-1} E_i K_{i+1} \cr
    K_i F_i &= \fq^{-1} F_i K_i \cr
    K_{i+1} F_i &= \fq F_i K_{i+1} \, .
\end{align}
Furthermore, the square-switch relation gives 
\begin{equation}
    E_i F_i = F_i E_i + [k_i - k_{i+1}]_\fq \, ,
\end{equation}
e.g. 
\begin{equation}
    [E_i, F_i] = \frac{K_i K_{i+1}^{-1} - K_i^{-1} K_{i+1}}{\fq - \fq^{-1}}\, .
\end{equation}
All other pairs of symbols except $(E_i,E_{i+1})$ and $(F_i, F_{i+1})$ commute, and we have Serre relations
\begin{align}
    E_i^2 E_{i+1} - [2]_\fq E_i E_{i+1} E_i + E_{i+1} E_i^2 &=0 \cr
     E_i E_{i+1}^2 - [2]_\fq E_{i+1} E_{i} E_{i+1} + E_{i+1}^2 E_i &=0 \cr
      F_i^2 F_{i+1} - [2]_\fq F_i F_{i+1} F_i + F_{i+1} F_i^2 &=0 \cr
        F_i F_{i+1}^2 - [2]_\fq F_{i+1} F_{i} F_{i+1} + F_{i+1}^2 F_i &=0 \, ,
\end{align}
which follow, say, from the partial fusion of the two $E_i$ rungs in $E_i E_{i+1} E_i$ followed by the square-switch rule producing $E_i^{(2)} E_{i+1}$
and $E_{i+1} E_i^{(2)}$.

\subsection{Rungs and braids}
The reference also provides a diagram $\beta$ which represents the braiding of two consecutive strands with labels $k$ and $l$. We will use a simplified expression $B$ with 
\begin{equation}
    \beta = (-1)^{k l} \fq^{-\frac{k l}{N}} B\, ,
\end{equation}
and denote as $B_i$ the diagram inserted between strands $i$ and $i+1$.\footnote{The simplification is motivated as follows:
\begin{itemize}
    \item In the physical setup, the fundamental $\bC^N$ representation is taken to be Grassmann odd. Hence $\Lambda^k \bC^N$ has Grassmann parity $(-1)^k$ and we should add a Koszul sign $(-1)^{k l}$  to the braiding.
    We can restrict to even $N$ to ensure $\Lambda^N \bC^N$ is still a trivial representation.
    \item The factor $\fq^{-\frac{k l}{N}}$ will not affect the commutation 
    rules of $B$ and rung generators in the planar level. It could also be eliminated by promoting the $\mathrm{SL}(N)$ gauge group to $\mathrm{\mathrm{GL}}(N)$. 
    \item The two above statements could be combined by taking the gauge group to be a Spin version of $\mathrm{\mathrm{GL}}(N)$.
\end{itemize}}

Commuting $B_i$ across rungs, we find relations 
\begin{align}
    K_i B_i &= B_i K_{i+1} \cr
    K_{i+1} B_i &= B_i K_{i} \cr
   K_i E_i B_i &= B_i F_i K_i  \cr
   K_{i+1} F_i B_i &= B_i E_i K_{i+1}  \cr
   (E_i E_{i+1} - \fq E_{i+1} E_i) B_i &= B_i E_{i+1}
   \cr
   (E_{i-1} E_i - \fq^{-1} E_i E_{i-1} ) B_i &= B_i E_{i-1} \cr 
  ( F_{i+1} F_i- \fq^{-1} F_i F_{i+1}) B_i &= B_i F_{i+1}\cr  
   (F_i F_{i-1} - \fq F_{i-1} F_i) B_i &= B_i F_{i-1} \cr
   E_{i+1} B_i &= B_i (E_i E_{i+1}- \fq^{-1} E_{i+1} E_i)
   \cr
   E_{i-1} B_i &= B_i (E_{i-1}E_i - \fq E_i E_{i-1})
   \cr
   F_{i+1} B_i &= B_i (F_{i+1}F_i - \fq F_i F_{i+1})
   \cr
   F_{i-1} B_i &= B_i (F_i F_{i-1} - \fq^{-1} F_{i-1} F_i )\, .
\end{align}

\subsection{Long rungs and a PBW basis}
Inspired by these formulae and with the help of $B$ specialized to $k=1$ or $l=1$, we recognize as 
\begin{equation}
    E_{-;i+2,i} \equiv E_i E_{i+1} - \fq E_{i+1} E_i
\end{equation} 
a left-pointing rung extending from the $(i+2)$-th strand to the $1$-th strand while passing behind the $(i+1)$-th strand,
\begin{equation}
    E_{+;i+2,i} \equiv E_i E_{i+1} - \fq^{-1} E_{i+1} E_i
\end{equation} 
a similar rung passing in front and 
\begin{align}
    F_{-;i,i+2} &\equiv F_{i+1} F_i- \fq^{-1} F_i F_{i+1} \cr
    F_{+;i,i+2} &\equiv F_{i+1} F_i- \fq F_i F_{i+1} 
\end{align} 
right-pointing rungs from the $i$-th strand to the $(i+2)$-th strand passing behind and in front of the $(i+1)$-th strand. 

More generally, we define recursively rung operators between any pairs of strands:
\begin{align}
    E_{\pm;i+1,i} &= E_i \cr
    E_{-;i+2,i} &= E_i E_{i+1} - \fq E_{i+1} E_i \cr
    E_{+;i+2,i} &= E_i E_{i+1} - \fq^{-1} E_{i+1} E_i \cr
    E_{-;j,i} &= E_i E_{-;j,i+1} - \fq E_{-;j,i+1} E_i \qquad \qquad j>i+1\cr
    E_{+;j,i} &= E_i E_{+;j,i+1} - \fq^{-1} E_{+;j,i+1} E_i \qquad \qquad j>i+1\cr
    F_{\pm;i,i+1} &= F_i \cr
    F_{-;i,i+2}&= F_{i+1} F_i- \fq^{-1} F_i F_{i+1} \cr
    F_{+;i,i+2}&= F_{i+1} F_i- \fq F_i F_{i+1} \cr
    F_{-;i,j} &= F_{-;i+1,j} F_{i} - \fq^{-1} F_{i}F_{-;i+1,j} \qquad \qquad j>i+1\cr
    F_{+;i,j} &= F_{+;i+1,j} F_{i} - \fq F_{i}F_{+;i+1,j}  \qquad \qquad j>i+1 \, ,
\end{align}
passing behind or in front of intermediate strands. 

The $B_i$ permutations act in the obvious way on the endpoints of these rung operators, except for the extra factors when acting on $E_i$ or $F_i$, i.e. when the direction a rung points to is flipped. This defines an action of the braid group of $m$ strands onto $U_\fq(\gl_m)$ as algebra automorphisms. It can be thought of as a $\fq$-deformed version of the Weyl group action on $U(\gl_m)$.

Lexicographically ordered monomials of $K^\pm_i$, $E_{-;i,j}$ and $F_{-;i,j}$ generators form a linear (PBW) basis for $U_\fq(\gl_m)$. Defining
\begin{align}
    e_{\pm;i,j} &= (\fq-\fq^{-1}) E_{\pm;i,j} \cr
    f_{\pm;i,j} &= (\fq-\fq^{-1}) f_{\pm;i,j} \, ,
\end{align}
etcetera, a crucial observation is that the rescaled operators will all have a good planar limit. This is expected from the large $N$ combinatorics of the planar limit, but can also be verified by computing commutators. E.g. 
\begin{align}
    [e_i, e_{-;i+2,i}] &= (\fq^{-1}-1) e_{-;i+2,i}e_i \cr
    [e_{-;i+2,i},e_{i+1}] &= (\fq^{-1}-1) e_{i+1} e_{-;i+2,i} \cr
    [e_{-;i+2,i},f_i] &= -(\fq - \fq^{-1})K_i e_{i+1} K_{i+1}^{-1} \cr
    [e_{-;i+2,i},f_{i+1}] &= -(\fq - \fq^{-1})K_i e_iK_{i+1}^{-1} \, .
\end{align}
Accordingly, monomials of $K^\pm_i$, $e_{-;i,j}$ and $f_{-;i,j}$ generators form a linear basis for the planar limit of $U_\fq(\gl_m)$. They can be understood as holomorphic functions (polynomials) on an auxiliary Poisson complex manifold $\CP_m$. The manifold itself is a product of $\bC$ and $\bC^*$ factors, but the Poisson bracket is complicated. 

The $SU(N)_\kappa$ CS theory is unitary. The adjoint of a Wilson line is a Wilson line for the conjugate representation. We expect Hermiticity conditions $\fq^\dagger = \fq^{-1}$, $K_i^\dagger= K_i^{-1}$, and $E^\dagger_{-;i,j} = F_{-;j,i}$. In the planar limit, $e^\dagger_{-;i,j} = f_{-;j,i}$. These reality conditions define a real submanifold $\CP^\bR_m$
of $\CP_m$.

The braid operations $B_i$ have an obvious planar limit. They act as Poisson automorphisms of $\CP_m$. The Hermiticity conditions are compatible with $B_i^\dagger = B_i^{-1}$.

\subsection{Cups and caps}
In the current setup, a ``cup'' or a ``cap'' joins smoothly consecutive strands of labels $k$ and $N-k$ with the help of a ``tag''. We will use conventions where cups have a down-pointing tag and caps an up-pointing tag. These conventions, together with the signs we included in the definition of $B_i$, ensure that a knot or link defined by a collection of cups, caps and braids has a vev which equals the standard $SU(N)_\kappa$ Chern-Simons answer times a factor of $(-1)^k$ for each closed loop of label $k$, at least for even $N$. 

A cup $\cup_i$ between the $i$-th and $(i+1)$-th strands constrains 
\begin{align}
    K_{i+1} K_i \cup_i &= g \cup_i \cr
    E_i \cup_i &= -\frac{K_i-K_i^{-1}}{\fq-\fq^{-1}} \cup_i \cr
    F_i \cup_i &= -\frac{g K^{-1}_i-g^{-1} K_i}{\fq-\fq^{-1}} \cup_i \, .
\end{align}
We also have 
\begin{align}
    E_{-;j,i}\cup_i &=K_i E_{-;j,i+1} \cup_i\cr
    E_{+;j,i}\cup_i &=K_i^{-1} E_{+;j,i+1} \cup_i\cr
    F_{-;j,i+1} \cup_i &=K_{i+1}^{-1} F_{-;j,i} \cup_i \cr
    F_{+;j,i+1} \cup_i &=K_{i+1} F_{+;j,i} \cup_i \cr
    E_{-;i+1,j}\cup_i &=-K_i^{-1} E_{-;i,j} \cup_i\cr
    E_{+;i+1,j}\cup_i &=-K_i E_{+;i,j} \cup_i\cr
    F_{-;i,j} \cup_i &=-K_{i+1} F_{-;i+1,j} \cup_i \cr
    F_{+;i,j} \cup_i &=-K_{i+1}^{-1} F_{-;i+1,j} \cup_i\, .
\end{align}

Analogously, a cap between the $i$-th and $(i+1)$-th strands constrains
\begin{align}
   \cap_i K_{i+1} K_i  &= \cap_i g \cr
   \cap_i E_i &= \cap_i\frac{g K_i^{-1}-g^{-1} K_i}{\fq-\fq^{-1}} \cr
    \cap_i F_i  &= \cap_i\frac{K_i- K_i^{-1}}{\fq-\fq^{-1}}\, . 
\end{align}
We also have  
\begin{align}
   \cap_i E_{-;j,i} &=\cap_i E_{-;j,i+1}K_{i+1}^{-1} \cr
   \cap_i E_{+;j,i}&=\cap_i E_{+;j,i+1}K_{i+1} \cr
   \cap_i F_{-;j,i+1}  &=\cap_i F_{-;j,i} K_{i}   \cr
   \cap_i F_{+;j,i+1}  &=\cap_i F_{-;j,i} K_{i}^{-1}  \cup_i \cr
   \cap_i E_{-;i+1,j} &=- \cap_i  E_{-;i,j}K_{i+1}\cr
   \cap_i E_{+;i+1,j} &=-\cap_i  E_{+;i,j}K_{i+1}^{1}\cr
   \cap_i F_{-;i,j} &=- \cap_i F_{-;i+1,j} K_{i}^{-1} \cr
   \cap_i F_{+;i,j} &=- \cap_i  F_{-;i+1,j} K_{i}\, .
\end{align}

The relations found above for cups and caps have a natural planar limit. 
They define Poisson correspondences $\cL_{\cup_i}$ and $\cL_{\cap_i}$ between $\CP_m$ and $\CP_{m-2}$, i.e. the difference of Poisson brackets acts on functions on $\CP_m \times \CP_{m-2}$ which vanish on the correspondences. 

We will typically use one of the cups as the place where we keep track of the label $k$ of a loop in the knot/link. This adds the extra constraint 
$K_i = \olambda$ and inserts factors of $\omu^\pm$ in the remaining relations: operators which raise the $k$ label by one must be multiplied by $\omu$ and operators which lower it by $1$ must be multiplied by $\omu^{-1}$.
This promotes $\cL_{\cap_i}$ to a correspondence between $\CP_m$ and 
$\CP_{m-2} \times (\bC^*)^2$.

\subsection{Junctions}
We now have enough information for the analysis of knot and link invariants. For completeness, we will also discuss the effect of junctions, which can be employed to study the augmentation varieties of Wilson line networks. 

We can list some natural relations, denoting as $Y_i$ a junction splitting the $i$-th strand:
\begin{align}
    Y_i K_i &= K_i K_{i+1} Y_i \cr
    Y_i K_j &= K_j Y_i \qquad j<i \cr
    Y_i K_j &= K_{j+1} Y_i \qquad j>i \cr
    Y_i E_j &= E_j Y_i \qquad j<i \cr
    Y_i E_{j} &= E_{j+1} Y_i \qquad j>i \cr
    Y_i F_{j} &= F_j Y_i \qquad j<i-1 \cr
    Y_i F_{j} &= F_{j+1} Y_i \qquad j>i-1 \cr
    E_{i} Y_i &= \frac{K_i - K_i^{-1}}{\fq- \fq^{-1}} Y_i\cr
    F_i Y_i &= \frac{K_{i+1} - K_{i+1}^{-1}}{\fq- \fq^{-1}} Y_i\, .
\end{align}
There are other Serre-like relations. It should be possible to express them in terms of longer rungs. E.g. we expect longer rungs which do not end on the legs of the junction to go through the junction in the obvious way. Some other relations:
\begin{align}
    (E_{-;i+1,j}-K_i^{-1} E_{-;i,j}) Y_i &= 0 \qquad j<i\cr
    (E_{+;i+1,j}-K_i E_{+;i,j}) Y_i &= 0 \qquad j<i\cr
    (F_{-;i,j}-K_{i+1} F_{-;i+1,j}) Y_i &= 0 \qquad j>i+1\cr
    (F_{+;i,j}-K_{i+1}^{-1} F_{+;i+1,j}) Y_i &=  0 \qquad j>i+1 \cr
    (E_{-;j;i}+\fq K_i E_{-;j,i+1})Y_i &= Y_i E_{-;j,i} \qquad j>i+1\cr
    (E_{+;j;i}+\fq^{-1} K_i^{-1} E_{+;j,i+1})Y_i &= Y_i E_{+;j,i} \qquad j>i+1\cr
    (F_{-;j;i+1}+\fq^{-1} K_{i+1}^{-1} F_{-;j,i})Y_i &= Y_i F_{-;j,i}\qquad j<i \cr
    (F_{+;j;i+1}+\fq K_{i+1} F_{+;j,i})Y_i &= Y_i F_{+;j,i}\qquad j<i \, .
\end{align}
In particular, we have enough relations to bring every element of $U_\fq(gl_{m-1})$ acting below the junction to some element of $U_\fq(gl_{m})$ above the junction.

We can denote the opposite junction merging consecutive strands as $\Lambda_i$. We can derive very similar relations: 
\begin{align}
    K_i \Lambda_i &= \Lambda_i K_i K_{i+1} \cr
    K_j\Lambda_i &= \Lambda_i K_j  \qquad j<i \cr
    K_j \Lambda_i&= \Lambda_i K_{j+1} \qquad j>i \cr
    E_j \Lambda_i &= \Lambda_i E_j \qquad j<i-1 \cr
    E_{j}\Lambda_i &= \Lambda_i E_{j+1}  \qquad j>i-1 \cr
    F_{j} \Lambda_i &=\Lambda_i F_j  \qquad j<i \cr
    F_{j} \Lambda_i&= \Lambda_i F_{j+1}  \qquad j>i \cr
   \Lambda_i E_{i} &= \Lambda_i \frac{K_{i+1} - K_{i+1}^{-1}}{\fq- \fq^{-1}} \cr
    \Lambda_i F_i &= \Lambda_i \frac{K_{i} - K_{i}^{-1}}{\fq- \fq^{-1}} \cr
    \Lambda_i (E_{-;i+1,j}+ E_{-;i,j}K_{i+1}) &= E_{-;i,j} \Lambda_i \qquad j<i\cr
     \Lambda_i (E_{+;i+1,j}+ E_{+;i,j}K_{i+1}) &= E_{+;i,j} \Lambda_i \qquad j<i\cr
    \Lambda_i(F_{-;i,j}- F_{-;i+1,j}K_{i}^{-1}) &= F_{-;i,j} \Lambda_i \qquad j>i+1\cr
    \Lambda_i(F_{+;i,j}- F_{+;i+1,j}K_{i}^{-1}) &= F_{+;i,j} \Lambda_i\qquad j>i+1 \cr
    \Lambda_i(E_{-;j;i}-E_{-;j,i+1} K_{i+1}^{-1} ) &= 0 \qquad j>i+1\cr
    \Lambda_i(E_{+;j;i}-E_{+;j,i+1} K_{i+1} ) &= 0 \qquad j>i+1\cr
    \Lambda_i(F_{-;j;i+1}-F_{-;j,i}K_{i}) &= 0\qquad j<i \cr
    \Lambda_i(F_{+;j;i+1}-F_{+;j,i}K_{i}) &= 0\qquad j<i\, .
\end{align}
All of these relations have an immediate planar limit. They define Poisson correspondences $\cL_{Y_i}$ and $\cL_{\Lambda_i}$ between $\CP_m$ and $\CP_{m-1}$.

\subsection{Endpoints}
Another simple way to reduce the number of strands is simply to set $k_i=0$ for a strand and forget it. We denote this operation as $\Pi_{0;i}$ or $I_{0;i}$ depending on the orientation. It sets $K_i=1$ and sets to $0$ any rung starting at the $i$-th position. Rungs which do not end on $i$ simply go through. 

We can also set strand labels to $N$. We denote this operation as $\Pi_{N;i}$ or $I_{N;i}$ depending on the orientation. Because of the overall factor we removed from the braiding, a strand with label $N$ is not completely transparent. Instead, a rung which does not end on $i$ will have to be multiplied by $\fq^{\pm 1}$ when brought across such an endpoint. Rungs ending at that position will be set to $0$, and $K_i$ to $\fq^N$.

Cups and caps are combinations of junctions and endpoints. 

\section{The planar quantum group as a character variety}
\label{app:braiding_algebra}
In this Appendix, we continue the planar analysis from the previous Appendix \ref{app:rung_algebra}. We refer to it for notations. Recall 
that we expect $K_i$ to have a well-defined classical limit, as well as $e_{-;i,j} \equiv (\fq-\fq^{-1}) E_{-;i,j}$
and $f_{-;i,j} \equiv (\fq-\fq^{-1}) F_{-;i,j}$. We define $e_i=e_{-;i+1,i}=e_{+;i+1,i}$ and $f_i=f_{-;i,i+1}=f_{+;i,i+1}$ as special cases. 

The analogous $e_{+;i,j}\equiv (\fq-\fq^{-1}) E_{+;i,j}$ and $f_{+;i,j}\equiv (\fq-\fq^{-1}) F_{+;i,j}$ limits are not independent. For example, we compute 
\begin{align}
    e_{+;i+2,i}-e_{-;i+2,i} &= e_i e_{i+1} \cr
    e_{+;i+3,i}-e_{-;i+3,i} &= e_{i+2} e_{-;i+2,i} + e_{+;i+3,i+1} e_i \cr
    e_{+;i+4,i}-e_{-;i+4,i} &= e_{+;i+4,i+1}e_i+ e_{+;i+4,i+2} e_{-;i+2,i}+e_{i+3} e_{-;i+3,i}\, ,
\end{align}
which have a nice planar limit. More generally, we have 
\begin{equation}
    e_{+;j,i} - e_{-;j,i} = \sum_{k=i+1}^{j-1} e_{+;j,k}e_{-;k,i} \, ,
\end{equation}
which can be proven by bringing the rung across intermediate strands one at a time. 

Similarly, 
\begin{align}
    f_{+;i,i+2}-f_{-;i,i+2} &= - f_i f_{i+1} \cr
    f_{+;i,i+3}-f_{-;i,i+3} &= - f_i f_{+;i+1,i+3} - f_{-;i,i+2} f_{i+2}\, .
\end{align}
More generally, 
\begin{equation}
    f_{+;j,i} - f_{-;j,i} = -\sum_{k=i+1}^{j-1} f_{-;i,k} f_{+;k,j}\, .
\end{equation}

If we define triangular matrices $\hat e_\pm$ with unit diagonal and $\pm e_{\pm;i,j}$ off-diagonal elements, the equations take the form 
\begin{equation}
    \hat e_+ \hat e_- = \hat e_- \hat e_+ = 1 \, .
\end{equation}
Similarly, if we define $\hat f_\pm$ with unit diagonal and $\mp f_{\pm;i,j}$ off-diagonal elements, 
\begin{equation}
    \hat f_+ \hat f_- = \hat f_- \hat f_+ = 1 \, . 
\end{equation}

\subsection{Framing strands}
We can improve our geometric understanding by adding two ``spectator'' strands at positions $0$ and $m+1$. This gives useful collections of auxiliary quantities: $e_{\pm;i,0}$, $e_{\pm;m+1,i}$, 
$f_{\pm;0,i}$, $f_{\pm;i,m+1}$. 

Manipulations of the original set of strands will give us linear operations on these $8$ collections of $m$ quantities, which in turn will imply specific transformations of the $\hat e_\pm$ and $\hat f_\pm$ matrices. 

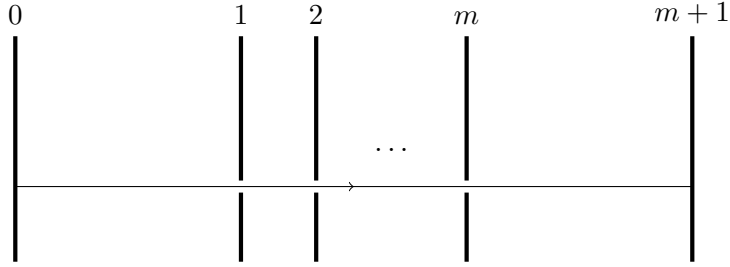
\begin{figure}[h]
    \centering
    \begin{tikzpicture}[scale=1, ultra thick, every node/.style={font=\small, align=center}]
        \def\ythin{1}    
        \def\gap{0.08}   

        \draw (-2,0) -- (-2,3) node[above] {$0$};

        \foreach \x/\lbl in {1/1,2/2,4/m}{
            \draw (\x,0) -- (\x,\ythin-\gap);
            \draw (\x,\ythin+\gap) -- (\x,3) node[above] {$\lbl$};
        }

        \node at (3,1.5) {$\hdots$};

        \draw (7,0) -- (7,3) node[above] {$m+1$};

        \draw[thin,
      postaction={
        decorate,
        decoration={
          markings,
          mark=at position 0.5 with {\arrow{>}}
        }
      }
    ]
    (-2,\ythin) -- (7,\ythin);
    \end{tikzpicture}
    \caption{Auxiliary strands $0$ and $m+1$ with a $f_{+;0,m+1}$ rung connecting them. We see that any braiding of the intermediate rungs does not affect this rung.}
    \label{fig:auxrung}
\end{figure}

As
\begin{equation}
    e_{+;m+1,0} - e_{-;m+1,0} = \sum_{k=1}^{m} e_{+;m+1,k}e_{-;k,0} = \sum_{k=1}^{m} e_{-;m+1,k}e_{+;k,0}\, ,
\end{equation}
and this quantity is unaffected by various manipulations of the intermediate strands (see Figure \ref{fig:auxrung}), we can think about $e_{-;i,0}$ and $e_{+;m+1,i}$ as having a natural pairing, and $e_{+;i,0}$ and $e_{-;m+1,i}$ as well. Similarly for $f_{-;0, m+1} - f_{+;0,m+1}$ giving a natural pairing on the $f$'s. 

We will now define a monodromy matrix which encodes the transformation of, say, $e_{-;i,0}$ as the two spectator strands are braided simultaneously around the original strands. 

\begin{figure}[hbt!]
    \centering
    \begin{tikzpicture}[baseline=(current bounding box.center), scale = 0.6]
        \neutralmodule{0}{0,1,2,3}
        \braidmodule{1}{1}
        \braidmodule{3}{2}
        \braidmodule{5}{3}
        \braidmodule{7}{2}
        \braidmodule{9}{1}
        \neutralmodule{11}{0,1,2,3}
    \end{tikzpicture}
    \caption{The half braid of strands at position $0$ and $m+1$ illustrated for $m=2$.}
    \label{fig:Halfbraid}
    \end{figure}
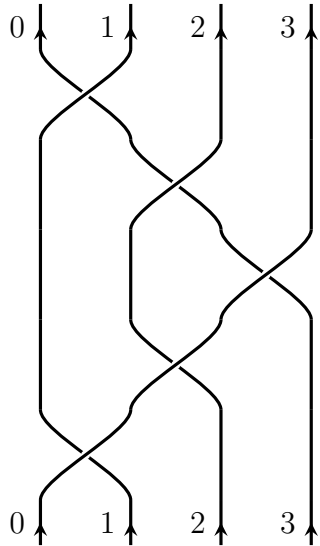

\begin{itemize}
    \item The first step is to map them to a basis of $e_{+;i,0}$ elements. This is implemented by the $\hat e_+$ matrix.
    \item Next, we braid the extra strands from position $0$ to position $m+1$ and vice versa, by applying the sequence $B_0 \cdots B_{m-1} B_m B_{m-1} \cdots B_0$.See Figure \ref{fig:Halfbraid}. This converts 
    \begin{align}
        e_{+;1,0}\to K_1 f_{+;0,1} K_0^{-1}  \to K_{m+1} K_1^{-1} f_{+;1,m+1} \cr
        e_{+;2,0}\to e_{+;2,1} \to K_2 f_{+;1,2} K_1^{-1} \to K_{m+1} K_2^{-1} f_{+;2,m+1} \cr
        e_{+;i,0} \to K_{m+1} K_i^{-1} f_{+;i,m+1}\, .
    \end{align}
    So if we use a new basis $K_{m+1} f_{+;i,m+1}$, it is expressed in terms of the old one by multiplication by a diagonal matrix with entries $K_i$. 
    \item Now we map this to a basis $K_{m+1} f_{-;i,m+1}$. The change of basis is $\hat f_-$.
    \item We finally braid again the extra strands by $B_0 \cdots B_{m-1} B_m B_{m-1} \cdots B_0$. E.g. $f_{-;m,m+1} \to K_0 e_{-;m,0} K_{m}^{-1}$. If we use a new basis $K_0^2 e_{-;i,0}$, this is implemented by a diagonal matrix with entries $K_i$ again. 
\end{itemize}
Overall, we build a {\it monodromy matrix}
\begin{equation}
  M = \hat K \hat f_- \hat K \hat e_+\, ,
\end{equation}
decomposed into two half-monodromies $\hat K \hat e_+$ and $\hat K \hat f_-$. 

This sheds some light on the geometric meaning of individual $B_i$ transformations. They act in a very specific manner on the $e_{-;j,0}$ elements: 
\begin{align}
    e_{-;i+1,0} &\to e_{-;i,0} \cr
    e_{-;i,0} &\to e_{-;i+1,0} + e_i e_{-;i,0}\, ,
\end{align}
and all other unchanged. The entries of $M$ must change by conjugation with this simple linear transformation. In the main text, we give a Stokes data interpretation of this transformation.
\subsection{The bilinear parameterization}
We now introduce a redundant parameterization of $\CP_m$ which will be helpful in describing cups, caps, junctions and endpoints. The parameterization: 
\begin{align}
    \oell^R_i \cdot \oell^L_i &= K_i^2 -1 \cr
    \oell^R_i \cdot \oell^L_j &= e_{-;i,j} \qquad \qquad \;\,i>j \cr
    \oell^R_i \cdot \oell^L_j &= K_i K_j f_{-;i,j} \quad \quad i<j \,
\end{align}
involves $m$ vectors $\oell_i^L$ in some $\bC^r$ and 
$m$ dual vectors $\oell_i^R$. It is invariant under the $\mathrm{\mathrm{GL}}(r)$ action on $\bC^r$. Without loss of generality we can take $r=m$, though we will see that cups, junctions, etc. impose constraints which allow for smaller values of $r$ to be employed. 

We also define matrices
\begin{align}
    M_i &= 1 + \oell^L_i \oell^R_i \cr
    M^{-1}_i &= 1 - K_i^{-2} \oell^L_i \oell^R_i \, .
\end{align}
They appear in a natural lift of the action of $B_i$: 
\begin{align}
    B_i \oell^L_i &= M_i^{-1} \oell^L_{i+1} B_i \cr
    B_i \oell^L_{i+1} &=\oell^L_{i} B_i \cr
    B_i \oell^R_i &= \oell^R_{i+1} M_i B_i \cr
    B_i \oell^R_{i+1} &=\oell^R_{i} B_i\, .
\end{align}

We can now look at the cap constraints in this language. We find in particular that 
\begin{equation}
     \cap_i \oell^R_{j} \cdot \oell^L_{i} = \cap_i K_{i+1}^{-1} \oell^R_{j} \cdot \oell^L_{i+1}\, ,
\end{equation}
for {\it all} $j$, which we can solve without loss of generality by 
\begin{equation}
     \cap_i \oell^L_{i} = \cap_i K_{i+1}^{-1} \oell^L_{i+1}\, .
\end{equation}
Most of the remaining relations assert 
\begin{equation}
   \cap_i \oell^R_i \cdot \oell^L_j =- \cap_i \oell^R_{i+1} \cdot \oell^L_j K_{i+1}^{-1} \qquad \qquad j \neq i,i+1\, ,
\end{equation}
i.e. the cap forces $\oell^R_i + K_{i+1}^{-1} \oell^R_{i+1}$ to be orthogonal to all $\oell^L_j$ except for $\oell^L_i$ and $\oell^L_{i+1}$, which we have just taken to be proportional to each other. We also have the usual $\cap_i K_{i+1} K_i  = \cap_i g$.

Note that 
\begin{equation}
    \cap_i (\oell^R_i + K_{i+1}^{-1} \oell^R_{i+1})\cdot \oell^L_i = \cap_i K_i^2(1-g^{-2})\, ,
\end{equation}
so as long as $g^2 \neq 1$, neither $\oell^L_i$ nor $\oell^R_i + K_{i+1}^{-1} \oell^R_{i+1}$ can be non-zero. Without loss of generality, we could set them to be proportional to something like $(0, \cdots, 0,1)$ and set to $0$ the last components of all $\oell^L_j$ except for $\oell^L_i$ and $\oell^L_{i+1}$. 

Accordingly, the cap constrains the matrix of $\oell^L_i$'s to have a block-diagonal form, with $2 \times 1$ and $(m-2) \times (r-1)$-dimensional blocks. In order for remaining generators to pass through the cap, we simply need to identify 
the remaining $m-2$ vectors above the cap with the $(r-1)$-dimensional blocks in the $\oell_i^L$ and $\oell_i^R$ below the cap. 

We can also compute
\begin{equation}
   \cap_i M_{i+1} M_i = \cap_i(1 + K_{i+1}^2 \oell^L_{i} (\oell^R_i+K_{i+1}^{-1}\oell^R_{i+1}))\, ,
\end{equation}
and thus 
\begin{align}
   \cap_i M_{i+1} M_i (M_{i-1} \cdots M_1)(M_m^{-1} \cdots M_{i+2})= \cap_i\big(&(M_{i-1} \cdots M_1)(M_m^{-1} \cdots M_{i+2}) \cr
   & + K_{i+1}^2 \oell^L_{i} (\oell^R_i+K_{i+1}^{-1}\oell^R_{i+1})\big)\, .
\end{align}
In particular, $(\oell^R_i+K_{i+1}^{-1}\oell^R_{i+1})$ is a right eigenvector with eigenvalue $g^2$. 

This matrix also acts as $(M_{i-1} \cdots M_1)(M_m^{-1} \cdots M_{i+2})$ on the $\oell^L_j$ with $j$ different from $i$ and $i+1$. The matrix $(M_{i-1} \cdots M_1)(M_m^{-1} \cdots M_{i+2})$ has a block-triangular form, with a block identified with the analogous matrix above the cap. 

The cup analysis is analogous, with the roles of left- and right-eigenvectors permuted.

If we take $m = 2n$ and consider a full collection of caps
joining consecutive pairs of strands, the above analysis can be done simultaneously for all. The $\oell^L_{2i}$ and $\oell^L_{2i-1}$ will be pairwise parallel and the $\oell^L$
matrix will consist of a collection of $2 \times 1$ blocks. 

The ``monodromy at infinity'' matrix $M_{2n} \cdots M_1$
takes the form 
\begin{equation}
    (1 + K_{2n}^2 \oell^L_{2n-1} (\oell^R_{2n-1}+K_{2n}^{-1}\oell^R_{2n}))\cdots (1 + K_{2}^2 \oell^L_{1} (\oell^R_1+K_{2}^{-1}\oell^R_{2})\, ,
\end{equation}
and the various factors act independently. It thus restricts to $g^2$ times the identity matrix acting on the span of the $\oell^L_{2i-1}$ and, dually, on the span of the $\oell^R_{2i-1}+K_{2i}^{-1}\oell^R_{2i}$. 

Without loss of generality, we can set $r=n$ and use the $\oell^R_i$ to encode generic rung vevs compatible with the full cap state. The same is true for the full cup state. 
This makes contact with the discussion in Appendices \ref{app:classS} and \ref{app:inter}.

\subsection{Planar junctions and ends}
Besides $Y_i K_i = K_i K_{i+1} Y_i$, etc., part of the relations can be satisfied by imposing
\begin{equation}
    \oell^R_{i+1} Y_i = K_i^{-1} \oell^R_i Y_i \, ,
\end{equation}
as for the case of cups. The combination $\oell^L_i+ K_i \oell^L_{i+1}$ also plays an important role. Observe
\begin{equation}
    \oell^R_i \cdot (\oell^L_i+ K_i \oell^L_{i+1}) Y_i = (K_i^2 K_{i+1}^2-1) Y_i\, .
\end{equation}
The remaining relations are consistent with 
\begin{align}
    \oell^{L,R}_j Y_i &= Y_i \oell^{L,R}_j \qquad j<i \cr
    \oell^{L,R}_j Y_i &= Y_i \oell^{L,R}_{j-1} \qquad j>i+1 \cr
    \oell^R_i Y_i &= Y_i \oell^R_i \cr
    (\oell^L_i+ K_i \oell^L_{i+1}) Y_i &= Y_i \oell^L_i\, .
\end{align}
In other words, we align $\oell^R_{i}$ and $\oell^R_{i+1}$ above the $Y_i$ junction with a specific ratio $K_i$, so that the combination 
\begin{equation}
   M_{i+1} M_i = 1 + (\oell^L_i + K_i \oell^L_{i+1})\oell^R_i
\end{equation}
is identified with the new $M_i$, while the rest of the data goes through unchanged. 

The $\Lambda_i$ junction has analogous effects, setting $\oell^L_{i} = K_{i+1}^{-1} \oell^L_{i+1}$ and identifying $\oell^R_{i+1} + K_{i+1} \oell^R_i$ as the new $\oell^R_i$, $\oell^L_{i+1}$ as the new $\oell^L_i$.

Endpoints are also easily reproduced. The $I_{0;i}$ endpoint, for example, is annihilated by all rungs which emanate from the $i$-th strand. It can be reproduced by setting $\oell^R_i=0$ and leaving all other data unaffected. 
This is one way to make $M_i=0$. The opposite endpoint $\Pi_{0;i}$ 
instead sets $\oell^L_i=0$.

The endpoint $I_{N;i}$, instead, sets to zero all rungs which end on the $i$-th strand, but requires $\oell_i^R \cdot \oell_i^L = g^2-1 \neq 0$.
Accordingly, it can be implemented by making all $\oell^R_j$ for $j \neq i$ 
be orthogonal to $\oell^L_i$. Without loss of generality, we can take $\oell^{L,R}_i$ to be proportional to $(0, \cdots, 0,1)$ and set to zero the last component of all $\oell^R_j$ for $j \neq i$. Across the endpoint, we take the $(r-1) \times (m-1)$ parts of the data as the new $\oell^{L,R}$'s.

The opposite happens for the $\Pi_{N,i}$ endpoint.

\section{Diagrammatic rules}\label{app:rules}

The diagrammatic rules for the fusing of antisymmetric powers of the fundamental representation of $SU(N)$ are extracted from \cite{Cautis_2014}. Below is a list of the rules relevant to our purposes.

\begin{gather}
  \tikz[baseline]{\draw[->] (0,0.5) node[above] {$k$} arc (45:-315:0.5cm);}
  = \qBinomial{n}{k}
  \label{eq:loop}
  \\[1em]
  \begin{tikzpicture}[baseline=20]
    \foreach \n in {0,...,3} {
      \coordinate (z\n) at (0.4*\n, 0.8*\n);
    }
    \draw[mid>] (z0) -- node[right] {$k+l$} (z1);
    \draw[mid>] (z2) -- node[right] {$k+l$} (z3);
    \draw[mid>] (z1) to[out=150,in=-190] node[left] {$k$} (z2);
    \draw[mid>] (z1) to[out=-30,in=0]   node[right] {$l$} (z2);
  \end{tikzpicture}
  = \qBinomial{k+l}{l}
  \tikz[baseline=20]{\draw[mid>] (0,0) -- node[right] {$k+l$} (1,2);}
  \qquad
  \begin{tikzpicture}[baseline=20]
    \foreach \n in {0,...,3} {
      \coordinate (z\n) at (0.4*\n, 0.8*\n);
    }
    \draw[mid>] (z0) -- node[right] {$k$} (z1);
    \draw[mid>] (z2) -- node[right] {$k$} (z3);
    \draw[mid<] (z1) to[out=150,in=-190] node[left] {$l$} (z2);
    \draw[mid>] (z1) to[out=-30,in=0]   node[right] {$k+l$} (z2);
  \end{tikzpicture}
  = \qBinomial{n-k}{l}
  \tikz[baseline=20]{\draw[mid>] (0,0) -- node[right] {$k$} (1,2);}
  \label{eq:bigon2}
  \displaybreak[1]
  \\[1em]
  \begin{tikzpicture}[baseline]
    \foreach \x/\y in {0/0,1/0,2/0,0/1,1/1,0/2} {
      \coordinate (z\x\y) at (\x+\y/2,\y/1.5);
    }
    \coordinate (z03) at (1,2);
    \draw[mid>] (z00) node[below] {$k$}      -- (z01);
    \draw[mid>] (z01)              -- node[left]  {$k+l$} (z02);
    \draw[mid>] (z10) node[below] {$l$}      -- (z01);
    \draw[mid>] (z20) node[below] {$m$}      -- (z02);
    \draw[mid>] (z02) -- node[left] {$k+l+m$} (z03);
  \end{tikzpicture}
  =
  \begin{tikzpicture}[baseline]
    \foreach \x/\y in {0/0,1/0,2/0,0/1,1/1,0/2} {
      \coordinate (z\x\y) at (\x+\y/2,\y/1.5);
    }
    \coordinate (z03) at (1,2);
    \draw[mid>] (z00) node[below] {$k$} -- (z02);
    \draw[mid>] (z10) node[below] {$l$} -- (z11);
    \draw[mid>] (z20) node[below] {$m$} -- (z11);
    \draw[mid>] (z11)              -- node[right] {$l+m$} (z02);
    \draw[mid>] (z02) -- node[left] {$k+l+m$} (z03);
  \end{tikzpicture}
  \label{eq:IH}
  \displaybreak[1]
  \\
  \label{eq:id1b}
  \tikz[baseline=40]{
    \laddercoordinates{1}{2}
    \ladderEn{0}{0}{$k-s$}{$l+s$}{$s$}
    \ladderEn{0}{1}{$k-s-r$}{$l+s+r$}{$r$}
    \node[left]  at (l00) {$k$};
    \node[right] at (l10) {$l$};
  }
  =
  \qBinomial{r+s}{r}
  \tikz[baseline=20]{
    \laddercoordinates{1}{1}
    \ladderEn{0}{0}{$k-s-r$}{$l+s+r$}{$r+s$}
    \node[left]  at (l00) {$k$};
    \node[right] at (l10) {$l$};
  }
  \displaybreak[1]
  \\
  \label{eq:commutation}
  \begin{tikzpicture}[baseline=40]
    \laddercoordinates{1}{2}
    \node[left]  at (l00) {$k$};
    \node[right] at (l10) {$l$};
    \ladderEn{0}{0}{$k{-}s$}{$l{+}s$}{$s$}
    \ladderFn{0}{1}{$k{-}s{+}r$}{$l{+}s{-}r$}{$r$}
  \end{tikzpicture}
  = \sum_t \qBinomial{k-l+r-s}{t}
  \begin{tikzpicture}[baseline=40]
    \laddercoordinates{1}{2}
    \node[left]  at (l00) {$k$};
    \node[right] at (l10) {$l$};
    \ladderFn{0}{0}{$k{+}r{-}t$}{$l{-}r{+}t$}{$r{-}t$}
    \ladderEn{0}{1}{$k{-}s{+}r$}{$l{+}s{-}r$}{$s{-}t$}
  \end{tikzpicture}
  \displaybreak[1]
  \\
  \begin{tikzpicture}[baseline=10]
    \node (C) at (-1,2) {$l$};
    \node (D) at (1,-1) {$l$};
    \node (A) at (1,2)  {$k$};
    \node (B) at (-1,-1){$k$};
    \node (i) at (intersection of A--B and D--C) {};
    \draw[mid>] (B) -- (A);
    \draw[mid>] (D) -- (i);
    \draw[mid>] (i) -- (C);
  \end{tikzpicture}
  = (-1)^{k+kl} q^{k-\frac{kl}{n}} \sum_{\substack{a,b\geq 0 \\ b-a = k-l}} (-q)^{-b}
  \begin{tikzpicture}[baseline=40]
    \laddercoordinates{1}{2}
    \node[left]  at (l00) {$k$};
    \node[right] at (l10) {$l$};
    \ladderEn{0}{0}{$k{-}b$}{$l{+}b$}{$b$}
    \ladderFn{0}{1}{$l$}{$k$}{$a$}
  \end{tikzpicture}
  \label{eq:braid}
\end{gather}

In particular, (\ref{eq:commutation}) has an important special case: if $k-s+r=l$ and thus $l+s-r=k$, as on the right-hand side of (\ref{eq:braid}), the sum on the right-hand side has a single term $t=0$. In particular, in (\ref{eq:braid}) we can permute the horizontal rungs. 

\begin{figure}[h]
    \centering
    \begin{tikzpicture}[baseline=40]
        \laddercoordinates{1}{2}
        \node[left] at (l00) {$k$};
        \node[right] at (l10) {$l$};
        \ladderEn{0}{0}{}{}{$b$}
        \ladderFn{0}{1}{$l$}{$k$}{$a$}
    \end{tikzpicture}
    \quad 
    {$=$} 
    \quad 
    \begin{tikzpicture}[baseline=40]
        \laddercoordinates{1}{2}
        \node[left] at (l00) {$k$};
        \node[right] at (l10) {$l$};
        \ladderFn{0}{0}{}{}{$a$}
        \ladderEn{0}{1}{$l$}{$k$}{$b$}
    \end{tikzpicture}
    \caption{A special case of the square-switch rule.}
    \label{fig:rung-permutation}
\end{figure}
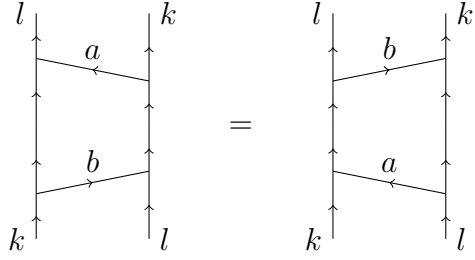

Tags are introduced as convenient notation to represent the following vertices:
\begin{align}
\fuse{k}{n-k}{n} & = \tikz[baseline=0.5cm]{\draw[mid>] (0,0) node[below] {$k$} arc (180:90:0.6) node[coordinate] (c) {}; \draw[mid<] (c) arc (90:0:0.6) node[below] {$n-k$}; \draw (c) -- +(0,0.2);} \\
\fork{k}{n-k}{n} & = \tikz[baseline=-0.5cm]{\draw[mid<] (0,0) node[above] {$k$} arc (-180:-90:0.6) node[coordinate] (c) {}; \draw[mid>] (c) arc (-90:0:0.6) node[above] {$n-k$}; \draw (c) -- +(0,0.2);}
\end{align}
They satisfy the following relations:
\begin{align}
\tikz[baseline=0.4cm]{
\foreach \n in {0,1,2} {
	\coordinate (a\n) at (0.4*\n, 0.8*\n);
}
\draw[mid>] (a0) -- node[right] {$k$} (a1);
\draw[mid<] (a1) -- node[right] {$n-k$} (a2);
\draw (a1) -- +(-0.2,0.1);
}
& = (-1)^{k(n-k)}
\tikz[baseline=0.4cm]{
\foreach \n in {0,1,2} {
	\coordinate (a\n) at (0.4*\n, 0.8*\n);
}
\draw[mid>] (a0) -- node[right] {$k$} (a1);
\draw[mid<] (a1) -- node[right] {$n-k$} (a2);
\draw (a1) -- +(0.2,-0.1);
} \\
\begin{tikzpicture}[baseline=20]
\coordinate (z1) at (0,0);
\coordinate (z2) at (1,0);
\coordinate (c) at (0.5,0.5);
\coordinate (ce) at (0.5,1);
\coordinate (e) at (0.5,1.45);
\draw[mid>] (z1) node[below] {$k$} -- (c);
\draw[mid>] (z2) node[below] {$l$} -- (c);
\draw[mid>] (c) -- node[right] {$k{+}l$} (ce);
\draw[mid<] (ce) -- (e) node[above] {$n{-}k{-}l$};
\draw (ce) -- +(0.2,0);
\end{tikzpicture}
&=
\begin{tikzpicture}[baseline=20]
\coordinate (z1) at (0,0);
\coordinate (z2) at (1.5,0);
\coordinate (cze) at (1.25,0.333);
\coordinate (c) at (0.75,1);
\coordinate (e) at (0.75,1.45);
\draw[mid>] (z1) node[below] {$k$} -- (c);
\draw[mid>] (z2) node[below] {$l$} -- (cze);
\draw[mid<] (cze) -- node[right] {$n{-}l$} (c);
\draw (cze) -- + (0.15,0.15);
\draw[mid<] (c) -- (e) node[above] {$n{-}k{-}l$};
\end{tikzpicture} \\
\begin{tikzpicture}[baseline=20]
\coordinate (z1) at (0,0);
\coordinate (z2) at (1,0);
\coordinate (c) at (0.5,0.5);
\coordinate (ce) at (0.5,1);
\coordinate (e) at (0.5,1.45);
\draw[mid>] (z1) node[below] {$k{+}l$} -- (c);
\draw[mid<] (z2) node[below] {$k$} -- (c);
\draw[mid>] (c) -- node[right] {$l$} (ce);
\draw[mid<] (ce) -- (e) node[above] {$n{-}l$};
\draw (ce) -- +(-0.2,0);
\end{tikzpicture}
&=
\begin{tikzpicture}[baseline=20,x=-1cm]
\coordinate (z1) at (0,0);
\coordinate (z2) at (1.5,0);
\coordinate (cze) at (1.25,0.333);
\coordinate (c) at (0.75,1);
\coordinate (e) at (0.75,1.45);
\draw[mid<] (z1) node[below] {$k$} -- (c);
\draw[mid>] (z2) node[below] {$k{+}l$} -- (cze);
\draw[mid<] (cze) -- node[left] {$n{-}k{-}l$} (c);
\draw (cze) -- + (0.15,0.15);
\draw[mid<] (c) -- (e) node[above] {$n{-}l$};
\end{tikzpicture} \\
\tikz[baseline=0.6cm]{
\foreach \n in {0,1,2,3} {
	\coordinate (a\n) at (0.4*\n, 0.8*\n);
}
\draw[mid>] (a0) -- node[right] {$k$} (a1);
\draw[mid<] (a1) -- node[right] {$n-k$} (a2);
\draw[mid>] (a2) -- node[right] {$k$} (a3);
\draw (a1) -- +(-0.2,0.1);
\draw (a2) -- +(0.2,-0.1);
}
&= \tikz[baseline=0.6cm]{\draw[mid>] (0,0) -- node[right] {$k$} (1.2,2.4);}
\end{align}

\begin{figure}[h]
    \centering
\begin{equation*}
\begin{tikzpicture}[baseline=40]
\laddercoordinates{1}{2}
\node[left] at (l00) {$k$};
\node[right] at (l10) {$1$};
\ladderEn{0}{0}{$a$}{$k{-}a{+}1$}{$k{-}a$}
\ladderFn{0}{1}{$a{+}1$}{$k{-}a$}{$1$}
\end{tikzpicture}
= 
\begin{tikzpicture}[baseline=40]
\laddercoordinates{1}{2}
\node[left] at (l00) {$k$};
\node[right] at (l10) {$1$};
\ladderFn{0}{0}{$k{+}1$}{$0$}{$1$}
\ladderEn{0}{1}{$a{+}1$}{$k{-}a$}{$k{-}a$}
\end{tikzpicture}
+ [a]_\fq \times
\begin{tikzpicture}[baseline=40]
\laddercoordinates{1}{2}
\node[left] at (l00) {$k$};
\node[right] at (l10) {$1$};
\ladderFn{0}{0}{$k$}{$1$}{$0$}
\ladderEn{0}{1}{$a{+}1$}{$k{-}a$}{$k{-}a{-}1$}
\end{tikzpicture}
\end{equation*}
    \caption{Another special case of square-switch.}
    \label{fig:ladder-transformation-3}
\end{figure}
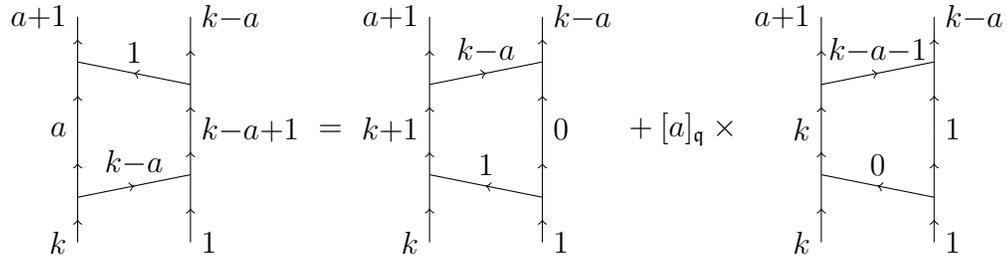

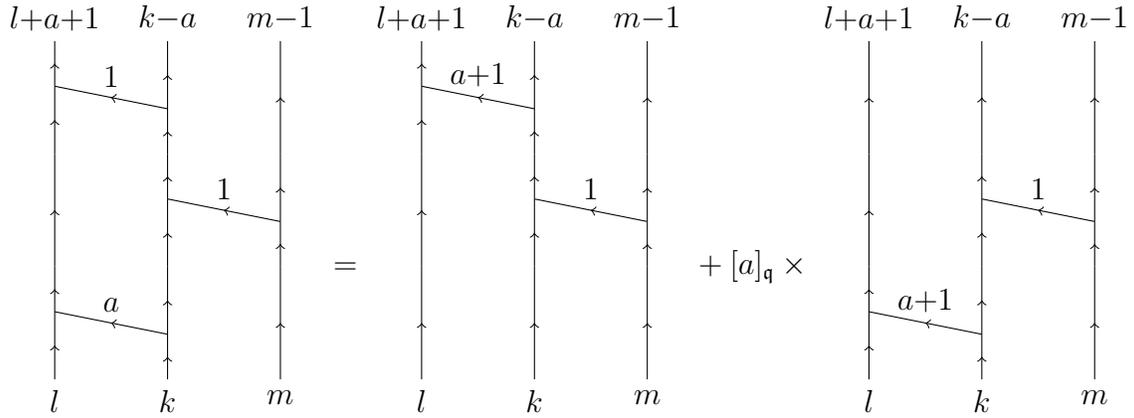
\begin{figure}[h]
    \centering
\begin{equation*}
\begin{tikzpicture}[baseline=40]
    \laddercoordinates{3}{3} 

    \node[below] at (l00) {$l$};
    \node[below] at (l10) {$k$};
    \node[below] at (l20) {$m$};

    \node[above] at (l03) {$l{+}a{+}1$};
    \node[above] at (l13) {$k{-}a$};
    \node[above] at (l23) {$m{-}1$};

    \ladderFn{0}{0}{}{}{$a$} 
    \ladderFn{0}{2}{}{}{$1$} 
    \ladderFn{1}{1}{}{}{$1$} 

    \ladderIn{0}{1}{1}
    \ladderIn{2}{0}{1}
    \ladderIn{2}{2}{1}
\end{tikzpicture}
= 
\begin{tikzpicture}[baseline=40]
    \laddercoordinates{3}{3} 

    \node[below] at (l00) {$l$};
    \node[below] at (l10) {$k$};
    \node[below] at (l20) {$m$};

    \node[above] at (l03) {$l{+}a{+}1$};
    \node[above] at (l13) {$k{-}a$};
    \node[above] at (l23) {$m{-}1$};

    
    \ladderFn{0}{2}{}{}{$a{+}1$} 
    \ladderFn{1}{1}{}{}{$1$} 

    \ladderIn{0}{0}{1}
    \ladderIn{1}{0}{1}
    \ladderIn{0}{1}{1}
    \ladderIn{2}{0}{1}
    \ladderIn{2}{2}{1}
\end{tikzpicture}
+ [a]_\fq \times
\begin{tikzpicture}[baseline=40]
    \laddercoordinates{3}{3} 

    \node[below] at (l00) {$l$};
    \node[below] at (l10) {$k$};
    \node[below] at (l20) {$m$};

    \node[above] at (l03) {$l{+}a{+}1$};
    \node[above] at (l13) {$k{-}a$};
    \node[above] at (l23) {$m{-}1$};

    \ladderFn{0}{0}{}{}{$a{+}1$}  
    \ladderFn{1}{1}{}{}{$1$} 

    \ladderIn{0}{2}{1}
    \ladderIn{1}{2}{1}
    \ladderIn{0}{1}{1}
    \ladderIn{2}{0}{1}
    \ladderIn{2}{2}{1}
\end{tikzpicture}
\end{equation*}
    \caption{A useful relation.}
    \label{fig:ladder-transformation-4}
\end{figure}

Figures \ref{fig:ladder-transformation-3} to \ref{fig:ladder-transformation-5} include some important diagrammatic calculations.

\begin{figure}[h]
    \centering
    \begin{equation*}
    \begin{aligned}
        &\sum_b (-\fq)^{k-b-1} \times
        \begin{tikzpicture}[baseline=40] 
            \laddercoordinates{1}{3} 
            \node[left] at (l00) {$k$};
            \node[right] at (l10) {$l$};

            \ladderEn{0}{1}{}{}{$b$} 
            \ladderFn{0}{2}{$l+1$}{$k-1$}{$a+2$} 
            \ladderEn{0}{0}{}{}{$1$} 
        \end{tikzpicture}
        = \quad \sum_b (-\fq)^{k-b-1} [b+1]_\fq \times
        \begin{tikzpicture}[baseline=40]
            \laddercoordinates{1}{2}
            \node[left] at (l00) {$k$};
            \node[right] at (l10) {$l$};
            \ladderEn{0}{0}{}{}{$b{+}1$} 
            \ladderFn{0}{1}{$l+1$}{$k-1$}{$a+2$}
        \end{tikzpicture}= \\
        &= \quad \sum_b (-\fq)^{k-b-1} [b+1]_\fq \times \left(
        \begin{tikzpicture}[baseline=40]
            \laddercoordinates{1}{2}
            \node[left] at (l00) {$k$};
            \node[right] at (l10) {$l$};
            \ladderFn{0}{0}{}{}{$a+2$}
            \ladderEn{0}{1}{$l+1$}{$k-1$}{$b+1$}
        \end{tikzpicture}
        \quad + \quad
         \begin{tikzpicture}[baseline=40]
            \laddercoordinates{1}{2}
            \node[left] at (l00) {$k$};
            \node[right] at (l10) {$l$};
            \ladderFn{0}{0}{}{}{$a+1$}
            \ladderEn{0}{1}{$l+1$}{$k-1$}{$b$}
        \end{tikzpicture}
        \right) = \\ &=
        \quad \sum_b (-\fq)^{k-b-1} \left([b+1]_\fq - \fq [b]_\fq \right) \times 
        \begin{tikzpicture}[baseline=40]
            \laddercoordinates{1}{2}
            \node[left] at (l00) {$k$};
            \node[right] at (l10) {$l$};
            \ladderFn{0}{0}{}{}{$a+2$}
            \ladderEn{0}{1}{$l+1$}{$k-1$}{$b+1$}
        \end{tikzpicture}
        \quad = \\&=\quad \sum_b (-1)^{k+b+1} \fq^{k-2b-1}
         \begin{tikzpicture}[baseline=40]
            \laddercoordinates{1}{2}
            \node[left] at (l00) {$k$};
            \node[right] at (l10) {$l$};
            \ladderFn{0}{0}{}{}{$a+1$}
            \ladderEn{0}{1}{$l+1$}{$k-1$}{$b$}
        \end{tikzpicture}
    \end{aligned}
    \end{equation*}
    \caption{Acting with $E$ at the bottom of the braid.}
    \label{fig:ladder-transformation-1}
\end{figure}
\clearpage
\begin{figure}[h]
    \centering
\begin{equation*}
    \begin{aligned}
        &\sum_b (-\fq)^{k-b} \times
        \begin{tikzpicture}[baseline=40] 
            \laddercoordinates{1}{3} 
            \node[left] at (l00) {$k$};
            \node[right] at (l10) {$l$};

            \ladderEn{0}{0}{}{}{$b$} 
            \ladderFn{0}{1}{}{}{$a$} 
            \ladderFn{0}{2}{$l{+}1$}{$k{-}1$}{$1$} 
        \end{tikzpicture}
        = \quad \sum_b (-\fq)^{k-b} [a+1]_\fq \times
        \begin{tikzpicture}[baseline=40]
            \laddercoordinates{1}{2}
            \node[left] at (l00) {$k$};
            \node[right] at (l10) {$l$};
            \ladderEn{0}{0}{}{}{$b$} 
            \ladderFn{0}{1}{$l{+}1$}{$k{-}1$}{$a{+}1$}
        \end{tikzpicture}= \\
        &= \quad \sum_b (-\fq)^{k-b} [a+1]_\fq \times \left(
        \begin{tikzpicture}[baseline=40]
            \laddercoordinates{1}{2}
            \node[left] at (l00) {$k$};
            \node[right] at (l10) {$l$};
            \ladderFn{0}{0}{}{}{$a+1$}
            \ladderEn{0}{1}{$l+1$}{$k-1$}{$b$}
        \end{tikzpicture}
        \quad + \quad
         \begin{tikzpicture}[baseline=40]
            \laddercoordinates{1}{2}
            \node[left] at (l00) {$k$};
            \node[right] at (l10) {$l$};
            \ladderFn{0}{0}{}{}{$a$}
            \ladderEn{0}{1}{$l+1$}{$k-1$}{$b-1$}
        \end{tikzpicture}
        \right) = \\ &=
        \quad \sum_b (-\fq)^{k-b} \left([a+1]_\fq - \fq^{-1} [a+2]_\fq \right) \times 
        \begin{tikzpicture}[baseline=40]
            \laddercoordinates{1}{2}
            \node[left] at (l00) {$k$};
            \node[right] at (l10) {$l$};
            \ladderFn{0}{0}{}{}{$a+1$}
            \ladderEn{0}{1}{$l+1$}{$k-1$}{$b$}
        \end{tikzpicture}
        \quad = \\ &=\quad \sum_b (-1)^{k+b+1} \fq^{2k-l-2b-2}
         \begin{tikzpicture}[baseline=40]
            \laddercoordinates{1}{2}
            \node[left] at (l00) {$k$};
            \node[right] at (l10) {$l$};
            \ladderFn{0}{0}{}{}{$a+1$}
            \ladderEn{0}{1}{$l+1$}{$k-1$}{$b$}
        \end{tikzpicture}
    \end{aligned}
    \end{equation*}
    \caption{Acting with $F$ at the bottom of the braid.}
    \label{fig:ladder-transformation-2}
\end{figure}

\begin{figure}[h]
    \centering
\begin{equation*}
\begin{aligned}
&\sum_b(-\fq)^{k-b}
\begin{tikzpicture}[baseline=40]
    \laddercoordinates{4}{3} 

    \node[below] at (l00) {$k$};
    \node[below] at (l10) {$l$};
    \node[below] at (l20) {$m$};

    \ladderEn{0}{0}{}{}{$b$} 
    \ladderFn{0}{1}{$l$}{$k$}{$a$} 
    \ladderFn{0}{3}{}{}{$1$} 
    \ladderFn{1}{2}{}{}{$1$} 

    \ladderIn{0}{2}{1}
    \ladderIn{2}{1}{1}
    \ladderIn{2}{0}{1}
    \ladderIn{2}{3}{1}
\end{tikzpicture}
= 
\sum_b(-\fq)^{k-b}
\begin{tikzpicture}[baseline=40]
    \laddercoordinates{4}{3} 

    \node[below] at (l00) {$k$};
    \node[below] at (l10) {$l$};
    \node[below] at (l20) {$m$};

    \ladderEn{0}{0}{}{}{$b$} 
    \ladderFn{0}{2}{}{}{$a+1$} 
    \ladderFn{1}{1}{}{}{$1$} 

    \ladderIn{0}{0}{1}
    \ladderIn{1}{0}{1}
    \ladderIn{0}{1}{1}
    \ladderIn{2}{0}{1}
    \ladderIn{2}{2}{1}
\end{tikzpicture}
+ \\&+\sum_b(-\fq)^{k-b}[a]_\fq \times
\begin{tikzpicture}[baseline=40]
    \laddercoordinates{4}{3} 

    \node[below] at (l00) {$k$};
    \node[below] at (l10) {$l$};
    \node[below] at (l20) {$m$};

    \ladderEn{0}{0}{}{}{$b$} 
    \ladderFn{0}{1}{}{}{$a+1$}  
    \ladderFn{1}{2}{}{}{$1$} 

    \ladderIn{0}{2}{1}
    \ladderIn{1}{1}{1}
    \ladderIn{2}{0}{1}
    \ladderIn{2}{1}{1}
\end{tikzpicture} = 
\sum_b(-\fq)^{k-b}
\begin{tikzpicture}[baseline=40]
    \laddercoordinates{4}{3} 

    \node[below] at (l00) {$k$};
    \node[below] at (l10) {$l$};
    \node[below] at (l20) {$m$};

    \ladderEn{0}{0}{}{}{$b$} 
    \ladderFn{0}{2}{}{}{$a+1$} 
    \ladderFn{1}{1}{}{}{$1$} 

    \ladderIn{0}{0}{1}
    \ladderIn{1}{0}{1}
    \ladderIn{0}{1}{1}
    \ladderIn{2}{0}{1}
    \ladderIn{2}{2}{1}
\end{tikzpicture}
+ \\&+\sum_b(-\fq)^{k-b} q [a+1]_\fq \times
\begin{tikzpicture}[baseline=40]
    \laddercoordinates{4}{3} 

    \node[below] at (l00) {$k$};
    \node[below] at (l10) {$l$};
    \node[below] at (l20) {$m$};

    \ladderEn{0}{0}{}{}{$b$} 
    \ladderFn{0}{1}{}{}{$a+1$}  
    \ladderFn{1}{2}{}{}{$1$} 

    \ladderIn{0}{2}{1}
    \ladderIn{1}{1}{1}
    \ladderIn{2}{0}{1}
    \ladderIn{2}{1}{1}
\end{tikzpicture}
-\fq^l \sum_b(-1)^{l-a}  \times
\begin{tikzpicture}[baseline=40]
    \laddercoordinates{4}{3} 

    \node[below] at (l00) {$k$};
    \node[below] at (l10) {$l$};
    \node[below] at (l20) {$m$};

    \ladderEn{0}{0}{}{}{$b$} 
    \ladderFn{0}{1}{}{}{$a+1$}  
    \ladderFn{1}{2}{}{}{$1$} 

    \ladderIn{0}{2}{1}
    \ladderIn{1}{1}{1}
    \ladderIn{2}{0}{1}
    \ladderIn{2}{1}{1}
\end{tikzpicture}
\end{aligned}
\end{equation*}
    \caption{The computation of the relation between braiding $B_2$ and $F_{23}$. The last term vanishes after application of the square-switch.}
    \label{fig:ladder-transformation-5}
\end{figure}
\clearpage

\subsection{Planar limit of rungs, braids, caps, and cups}

In the planar limit, the fundamental rungs attached to heavy strands obey a specific set of relations. Here we give a three-dimensional "geometric" picture of those relations, which we use to manipulate the rungs and compute the augmentation variety. It is understood that the $\olambda_i$ and $\omu_i$ in the formulae correspond to components of heavy strands in a knot/link.

To avoid labeling every heavy strand with its representation, we can consider knots with a \textit{basepoint}. Keeping track of the labeling is equivalent to the following relations involving the basepoint:
\begin{align}
\begin{tikzpicture}[baseline=-3]
\node (C) at (0,1) {};
\node (E) at (0,0) [circle,fill,inner sep=2pt] {};
\node (D) at (0,-1) {};
\node (A) at (-1,-0.2) {};
\coordinate (i) at (0,-0.2) {};
\draw[mid>] (A) -- (i);
\draw[mid>,very thick] (D) -- (E);
\draw[mid>,very thick] (E) -- (C);
\end{tikzpicture}\quad=&\quad \omu\begin{tikzpicture}[baseline=-3]
\node (C) at (0,1) {};
\node (E) at (0,0) [circle,fill,inner sep=2pt] {};
\node (D) at (0,-1) {};
\node (A) at (-1,0.2) {};
\coordinate (j) at (0,0.2) {};
\draw[mid>] (A) -- (j);
\draw[mid>,very thick] (D) -- (E);
\draw[mid>,very thick] (E) -- (C);
\end{tikzpicture}
\\
\begin{tikzpicture}[baseline=-3]
\node (C) at (0,1) {};
\node (E) at (0,0) [circle,fill,inner sep=2pt] {};
\node (D) at (0,-1) {};
\node (A) at (-1,-0.2) {};
\coordinate (i) at (0,-0.2) {};
\draw[mid>] (i) -- (A);
\draw[mid>,very thick] (D) -- (E);
\draw[mid>,very thick] (E) -- (C);
\end{tikzpicture}\quad=&\quad \omu^{-1}\begin{tikzpicture}[baseline=-3]
\node (C) at (0,1) {};
\node (E) at (0,0) [circle,fill,inner sep=2pt] {};
\node (D) at (0,-1) {};
\node (A) at (-1,0.2) {};
\coordinate (j) at (0,0.2) {};
\draw[mid>] (j) -- (A);
\draw[mid>,very thick] (D) -- (E);
\draw[mid>,very thick] (E) -- (C);
\end{tikzpicture}
\end{align}

The relation converting over and under crossings into ladders simplifies into a skein relation. Similar to the previous section, we rescale the rungs by a factor of $\fq - \fq^{-1}$ and define the rescaled rung operators, which have a good planar limit. Therefore, we perform the substitution,
\begin{align}
  \begin{tikzpicture}[baseline=-0.5ex]
    \draw[mid>] (-1,0) -- (1,0);
    \node[left]  at (-1,0) {$F_{i,j} = $};
  \end{tikzpicture}
  \quad\longrightarrow\quad
  \frac{1}{\fq - \fq^{-1}}\;\;
  \begin{tikzpicture}[baseline=-0.5ex]
    \draw[mid>] (-1,0) -- (1,0);
    \node[right] at (1,0) {$= \dfrac{f_{i,j}}{\fq - \fq^{-1}}$};
  \end{tikzpicture}
\end{align}
This way, the skein relations of \ref{eq:braid} becomes,
\begin{align}
\begin{tikzpicture}[baseline=-3]
\node (C) at (0,1) {};
\node (D) at (0,-1) {};
\node (A) at (1,0) {};
\node (B) at (-1,0) {};
\node (i) at (intersection of A--B and D--C) {};
\draw (B) -- (i);
\draw (i) -- (A);
\draw[very thick] (D) -- (C);
\end{tikzpicture}-
\begin{tikzpicture}[baseline=-3]
\node (C) at (0,1) {};
\node (D) at (0,-1) {};
\node (A) at (1,0) {};
\node (B) at (-1,0) {};
\node (i) at (intersection of A--B and D--C) {};
\draw[] (B) -- (A);
\draw[very thick] (D) -- (i);
\draw[very thick] (i) -- (C);
\end{tikzpicture}\quad =&\quad  \epsilon
\begin{tikzpicture}[baseline=-3]
\node (C) at (0,1) {};
\node (D) at (0,-1) {};
\coordinate (A) at (0,0) {};
\node (B) at (-1,0) {};
\draw[] (B) -- (A);
\draw[very thick] (D) -- (A);
\draw[very thick] (A) -- (C);
\end{tikzpicture}\cdot \begin{tikzpicture}[baseline=-3]
\node (C) at (0,1) {};
\node (D) at (0,-1) {};
\coordinate (A) at (0,0) {};
\node (B) at (1,0) {};
\draw[] (A) -- (B);
\draw[very thick] (D) -- (A);
\draw[very thick] (A) -- (C);
\end{tikzpicture}
\label{eq:planarskein}
\end{align}
The factor of $\epsilon$ changes depending on the direction of the rungs involved. We choose the convention for the heavy strand going up and the fundamental rung going to the right to have $\epsilon = 1$, and the heavy strand going up and the fundamental rung going left to have $\epsilon = -1$. If the heavy strand is going down, the signs accordingly flip. This corresponds to the algebraic skein relation satisfied by the fundamental rung in the planar limit given in Appendix \ref{app:braiding_algebra}.
\begin{align}
    e_{-;i+2,i}-e_{+;i+2,i} &= -e_i e_{i+1}\\
    f_{-;i,i+2}-f_{+;i,i+2} &= f_i f_{i+1}\, .
\end{align}
The bubble removal relations of Eqn. \ref{eq:loop} is also written in terms of the planar variables,
\begin{align}
\begin{tikzpicture}[baseline=-3]
\node (C) at (0,1) {};
\node (D) at (0,-1) {};
\node (A) at (0,-0.5) {};
\draw[mid>] (A) arc(270:90:0.5);
\draw[mid>,very thick] (D) -- (C);
\end{tikzpicture}\quad =&\quad \olambda-\olambda^{-1}
\\
\begin{tikzpicture}[baseline=-3]
\node (C) at (0,1) {};
\node (D) at (0,-1) {};
\node (A) at (0,0.5) {};
\draw[mid>] (A) arc(90:270:0.5);
\draw[mid>,very thick] (D) -- (C);
\end{tikzpicture}\quad =&\quad g\olambda^{-1}-g^{-1}\olambda
\end{align}

\begin{figure}[h]
    \centering
    \begin{subfigure}[b]{0.45\textwidth}
        \centering
        \begin{tikzpicture}[baseline=(current bounding box.center),scale=1]
            \neutralmodule{0}{,,,}
            \Erung{-0.2}{1}{2}
            \braidmodule{1}{2}
            \neutralmodule{3}{,,,}
        \end{tikzpicture}
        \caption{}
    \end{subfigure}
    \hfill
    \begin{subfigure}[b]{0.45\textwidth}
        \centering
        \begin{tikzpicture}[baseline=(current bounding box.center),scale=1]
            \neutralmodule{0}{,,,}
            \braidmodule{1}{2}
            \Erung{2.7}{1}
            \Erung{2.8}{2}
            \Emrung{3.4}{1}{3}
            \neutralmodule{3}{,,,}
        \end{tikzpicture}
        \caption{}
    \end{subfigure}
    \caption{An example of moving the rung up through a braid. In (a), we start with a \( e_{-;k,i} \) and move it through braid \( B_k \), which then decomposes into the rungs \( e_{-;k+1, i} + \epsilon_{k+1} \; e_{-;k+1, k} \; e_{-;k, i} \) as given by Eq.~\eqref{eq:planarbraidrules}. We depicted on the right-hand side of the picture all the rungs that are in the final answer.}
    \label{fig:braidrung1}
\end{figure}
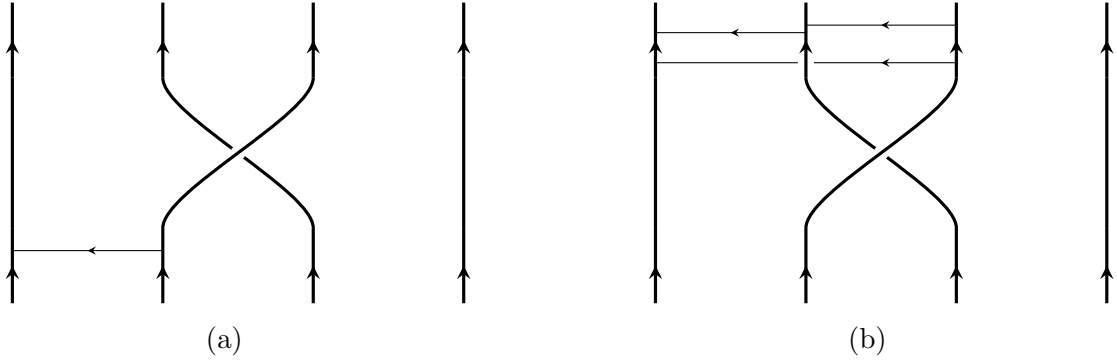

When twisting fundamental rungs around heavy strands, we obtain factors due to the change in the framing of the involved lines. The relevant formulae are,
\begin{align}
\begin{tikzpicture}[baseline=-3]
\node (A) at (-1,-0.5) {};
\node (B) at (0,0) {};
\coordinate (B1) at (-0.3,0.3) {};
\coordinate (B2) at (0.3,0.3) {};
\coordinate (B3) at (-0.3,-0.3) {};
\coordinate (B4) at (0.3,-0.3) {};
\node (C) at (1,-0.5) {};
\draw (A) to[out=0,in=225,looseness=1]  (B3);
\draw (B3) --  (B);
\draw (B) --  (B2);
\draw (B1) -- (B4);
\draw (B4) to[out=-45,in=180,looseness=1]  (C);
\draw (B2) to[out=45,in=135,looseness=3] (B1);
\end{tikzpicture}\quad =& \quad -g\begin{tikzpicture}[baseline=-3]
\node (A) at (1,0) {};
\node (B) at (-1,0) {};
\draw (B) --  (A);
\end{tikzpicture}
\end{align}

\begin{align}
\begin{tikzpicture}[baseline=-3]
\node (C) at (0,1) {};
\node (D) at (0,-1) {};
\coordinate (i) at (0,-0.2) {};
\node (j) at (0,0.2) {};
\coordinate (j2) at (0,0.2) {};
\node (A) at (1,0.2) {};
\draw (i) to[out=180,in=180,looseness=4] (j2);
\draw[mid>] (j2) -- (A);
\draw[mid>,very thick] (D) -- (i);
\draw[very thick] (i) -- (j);
\draw[mid>,very thick] (j) -- (C);
\end{tikzpicture}
\quad = \quad \olambda^{-1} \quad \begin{tikzpicture}[baseline=-3]
\node (C) at (0,1) {};
\node (D) at (0,-1) {};
\coordinate (A) at (0,0) {};
\node (B) at (1,0) {};
\draw[mid>] (A) -- (B);
\draw[mid>,very thick] (D) -- (A);
\draw[mid>,very thick] (A) -- (C);
\end{tikzpicture}
\quad \quad \quad
\begin{tikzpicture}[baseline=-3]
\node (C) at (0,1) {};
\node (D) at (0,-1) {};
\coordinate (i) at (0,-0.2) {};
\coordinate (i2) at (0,0) {};
\node (j) at (0,0.2) {};
\coordinate (j2) at (0,0.2) {};
\node (A) at (1,0.2) {};
\draw (i) to[out=180,in=180,looseness=4] (j);
\draw[mid>] (j) -- (A);
\draw[mid>,very thick] (D) -- (i);
\draw[very thick] (i) -- (j2);
\draw[mid>,very thick] (j2) -- (C);
\end{tikzpicture}
\quad = \quad \olambda \quad \begin{tikzpicture}[baseline=-3]
\node (C) at (0,1) {};
\node (D) at (0,-1) {};
\coordinate (A) at (0,0) {};
\node (B) at (1,0) {};
\draw[mid>] (A) -- (B);
\draw[mid>,very thick] (D) -- (A);
\draw[mid>,very thick] (A) -- (C);
\end{tikzpicture}
\end{align}

\begin{align}
\begin{tikzpicture}[baseline=-3]
\node (C) at (0,1) {};
\node (D) at (0,-1) {};
\coordinate (i) at (0,-0.2) {};
\node (j) at (0,0.2) {};
\coordinate (j2) at (0,0.2) {};
\node (A) at (1,0.2) {};
\draw (j2) to[out=180,in=180,looseness=4] (i);
\draw[mid>] (A) -- (j2);
\draw[mid>,very thick] (D) -- (i);
\draw[very thick] (i) -- (j);
\draw[mid>,very thick] (j) -- (C);
\end{tikzpicture}
\quad = \quad - g^{-1} \olambda \quad
\begin{tikzpicture}[baseline=-3]
\node (C) at (0,1) {};
\node (D) at (0,-1) {};
\coordinate (A) at (0,0) {};
\node (B) at (1,0) {};
\draw[mid<] (A) -- (B);
\draw[mid>,very thick] (D) -- (A);
\draw[mid>,very thick] (A) -- (C);
\end{tikzpicture}
\quad \quad \quad
\begin{tikzpicture}[baseline=-3]
\node (C) at (0,1) {};
\node (D) at (0,-1) {};
\coordinate (i) at (0,-0.2) {};
\coordinate (i2) at (0,0) {};
\node (j) at (0,0.2) {};
\coordinate (j2) at (0,0.2) {};
\node (A) at (1,0.2) {};
\draw (j) to[out=180,in=180,looseness=4] (i);
\draw[mid>] (A) -- (j);
\draw[mid>,very thick] (D) -- (i);
\draw[very thick] (i) -- (j2);
\draw[mid>,very thick] (j2) -- (C);
\end{tikzpicture}
\quad = \quad - g \olambda^{-1} \quad
\begin{tikzpicture}[baseline=-3]
\node (C) at (0,1) {};
\node (D) at (0,-1) {};
\coordinate (A) at (0,0) {};
\node (B) at (1,0) {};
\draw[mid<] (A) -- (B);
\draw[mid>,very thick] (D) -- (A);
\draw[mid>,very thick] (A) -- (C);
\end{tikzpicture}
\end{align}
\begin{align}
\begin{tikzpicture}[baseline=-3]
\node (C) at (0,1) {};
\node (D) at (0,-1) {};
\coordinate (i) at (0,-0.2) {};
\node (j) at (0,0.2) {};
\coordinate (j2) at (0,0.2) {};
\node (A) at (1,0.2) {};
\draw (i) to[out=180,in=180,looseness=4] (j2);
\draw[mid>] (j2) -- (A);
\draw[mid<,very thick] (D) -- (i);
\draw[very thick] (i) -- (j);
\draw[mid>,very thick] (C) -- (j);
\end{tikzpicture}
\quad = \quad - g^{-1} \olambda \quad 
\begin{tikzpicture}[baseline=-3]
\node (C) at (0,1) {};
\node (D) at (0,-1) {};
\coordinate (A) at (0,0) {};
\node (B) at (1,0) {};
\draw[mid>] (A) -- (B);
\draw[mid<,very thick] (D) -- (A);
\draw[mid>,very thick] (C) -- (A);
\end{tikzpicture}
\quad \quad \quad
\begin{tikzpicture}[baseline=-3]
\node (C) at (0,1) {};
\node (D) at (0,-1) {};
\coordinate (i) at (0,-0.2) {};
\coordinate (i2) at (0,0) {};
\node (j) at (0,0.2) {};
\coordinate (j2) at (0,0.2) {};
\node (A) at (1,0.2) {};
\draw (i) to[out=180,in=180,looseness=4] (j);
\draw[mid>] (j) -- (A);
\draw[mid<,very thick] (D) -- (i);
\draw[very thick] (i) -- (j2);
\draw[mid>,very thick] (C) -- (j2);
\end{tikzpicture}
\quad = \quad - g \olambda^{-1} \quad 
\begin{tikzpicture}[baseline=-3]
\node (C) at (0,1) {};
\node (D) at (0,-1) {};
\coordinate (A) at (0,0) {};
\node (B) at (1,0) {};
\draw[mid>] (A) -- (B);
\draw[mid<,very thick] (D) -- (A);
\draw[mid>,very thick] (C) -- (A);
\end{tikzpicture}
\end{align}

\begin{align}
\begin{tikzpicture}[baseline=-3]
\node (C) at (0,1) {};
\node (D) at (0,-1) {};
\coordinate (i) at (0,-0.2) {};
\node (j) at (0,0.2) {};
\coordinate (j2) at (0,0.2) {};
\node (A) at (1,0.2) {};
\draw (j2) to[out=180,in=180,looseness=4] (i);
\draw[mid>] (A) -- (j2);
\draw[mid<,very thick] (D) -- (i);
\draw[very thick] (i) -- (j);
\draw[mid>,very thick] (C) -- (j);
\end{tikzpicture}
\quad = \quad \olambda^{-1} \quad
\begin{tikzpicture}[baseline=-3]
\node (C) at (0,1) {};
\node (D) at (0,-1) {};
\coordinate (A) at (0,0) {};
\node (B) at (1,0) {};
\draw[mid<] (A) -- (B);
\draw[mid<,very thick] (D) -- (A);
\draw[mid>,very thick] (C) -- (A);
\end{tikzpicture}
\quad \quad \quad
\begin{tikzpicture}[baseline=-3]
\node (C) at (0,1) {};
\node (D) at (0,-1) {};
\coordinate (i) at (0,-0.2) {};
\coordinate (i2) at (0,0) {};
\node (j) at (0,0.2) {};
\coordinate (j2) at (0,0.2) {};
\node (A) at (1,0.2) {};
\draw (j) to[out=180,in=180,looseness=4] (i);
\draw[mid>] (A) -- (j);
\draw[mid<,very thick] (D) -- (i);
\draw[very thick] (i) -- (j2);
\draw[mid>,very thick] (C) -- (j2);
\end{tikzpicture}
\quad = \quad \olambda \quad
\begin{tikzpicture}[baseline=-3]
\node (C) at (0,1) {};
\node (D) at (0,-1) {};
\coordinate (A) at (0,0) {};
\node (B) at (1,0) {};
\draw[mid<] (A) -- (B);
\draw[mid<,very thick] (D) -- (A);
\draw[mid>,very thick] (C) -- (A);
\end{tikzpicture}
\end{align}
These relations allow us to determine how braids, cups, and caps act on the fundamental rungs. Consider the setup, a collection of heavy strands, labeled from $1$ to $m$, parallel to each other and all pointing upwards. Consider the collection of fundamental rungs $e_{-;ij}$ and $f_{-;ij}$, with $1 \leq i\neq j \leq m$ between $1$ and $m$. $e_{-;ij}$ and $f_{-;ij}$ represent left and right pointing rungs starting at the strand $i$ and ending at the strand $j$ respectively, and going below all intermediate heavy strands. This is analogous to the "PBW" basis of $U_{\fq}\gl_{m}$, which we take as a basis to do our computation. We denote by $B_k$ the element of the braid group that braids the heavy strand $k$ on top of $k+1$. The geometric process of sliding a fundamental rung across the braid from bottom to top can be expressed algebraically as an action of a braid‐group element on the rung variables, expressing them as linear combinations and products of other fundamental rungs, with the help of the skein relation \ref{eq:planarskein} (see Figure \ref{fig:braidrung1}). This action of $B_k$ on the heavy strands is given by (we have used the notation $B e = f B \iff e \mapsto f$)
\begin{equation}
    \begin{split}
        K_k & \mapsto K_{k+1}\\
        K_{k+1} & \mapsto K_{k}\, ,
    \end{split}
\end{equation}
analogous to the algebraic relation in Appendix \ref{app:braiding_algebra}.

\begin{figure}[h]
    \centering
    \begin{tikzpicture}[baseline=(current bounding box.center), scale=0.7]
        \neutralmodule{0}{$i$,$k$,$k+1$,$j$}
        \braidmodule{1}{2}
        \neutralmodule{3}{$i$,$k+1$,$k$,$j$}
    \end{tikzpicture}
    \caption{Braiding action on heavy strands}\label{fig:braidheavy}
\end{figure}
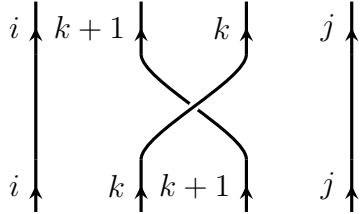

The action of $B_k$ on the rungs in the planar limit is given by 
\begin{equation}
\begin{split}
    i,j \neq k,k+1 &:\quad e_{-;i,j} \mapsto e_{-;i,j} \\ & \quad \quad f_{-;i,j} \mapsto f_{-;i,j}\\
    i < k, j = k &:\quad f_{-;i,k} \mapsto f_{-;i, k+1} - \epsilon_{k+1} \;  f_{-;i, k} \; f_{-;k, k+1} \\
    i = k, j < k &:\quad e_{-;k, j} \mapsto e_{-;k+1, j} + \epsilon_{k+1} \; e_{-;k+1, k} \; e_{-;k, j} \\
    i=k+1, j \neq k &:\quad e_{-;k+1, j} \mapsto e_{-;k, j} \\ & \quad \quad f_{-;k+1,j} \;  \mapsto f_{-;k,j}\\
    i \neq k, j = k+1 &:\quad e_{-;i, k+1} \mapsto e_{-;i, k} \\& \quad \quad f_{-;i,k+1} \;  \mapsto f_{-;i,k}\\
    i = k, j = k+1 &:\quad f_{-;k, k+1} \mapsto K_{k+1} \; K_{k}^{-1}\; e_{-;k+1, k} \\
    i= k+1, j = k &:\quad e_{-;k+1, k} \mapsto K_{k+1}^{-1} \; K_{k} \; f_{-;k, k+1} \\
    i = k, j > k+1 &:\quad f_{-;k, j} \mapsto f_{-;k+1, j} + \epsilon_{k+1} \; K_{k+1}  K_{k}^{-1} \; e_{-;k+1, k} f_{-;k, j} \\
    i > k+1, j = k &:\quad e_{-;i, k} \mapsto e_{-;i, k+1} - \epsilon_{k+1} \; K_{k+1}^{-1} \; K_{k} \; f_{-;k, k+1} \; e_{-;i, k}\, ,
\end{split}
\label{eq:planarbraidrules}
\end{equation}
where it is understood that $K_k$ corresponds to the eigenvalue of the evaluation operator $K_k$ associated with the strand $k$ before the braiding is carried out. Therefore $K_k$ can be thought of as c-number when performing subsequent braiding.

It can be checked that the action described above is, in fact, a representation of the braid group, compatible with its defining relations. The factor $\epsilon$ keeps track of whether the heavy strands at $k$ are going up or down, that is $\epsilon_{\text{up}} = 1$ and $\epsilon_{\text{down}} = -1$. But, during most of our analysis, we will assume the strands are going only up. It can also be checked that these relations are, in fact, the planar limit of the braiding action defined in Appendix \ref{app:rung_algebra}.

\begin{figure}[h]
    \centering
    \begin{subfigure}[t]{0.45\textwidth}
        \centering
        \begin{tikzpicture}[baseline=(current bounding box.center)]
            \cupmodule{0}{2}
            \Emrung{0.8}{1}{4}
            \neutralmodule{0}{,,,}
            \vphantom{\Erung{-0.9}{2}{3}} 
        \end{tikzpicture}
        \caption{}
    \end{subfigure}
    \hfill
    \begin{subfigure}[t]{0.45\textwidth}
        \centering
        \begin{tikzpicture}[baseline=(current bounding box.center)]
            \cupmodule{0}{2}
            \neutralmodule{0}{,,,}
            \Erung{-0.9}{2}{3}
            \vphantom{\Emrung{0.8}{1}{4}} 
        \end{tikzpicture}
        \caption{}
    \end{subfigure}
    \caption{An example of cup constraints: In (a), We have the rung $e_{-; 4, 1}$ which, when applied to the cup constraints, can be expressed as $- K_1 \; K_{3}^{-1} \; e_{-;3,2}$ as denoted in (b).}
    \label{fig:cups}
\end{figure}
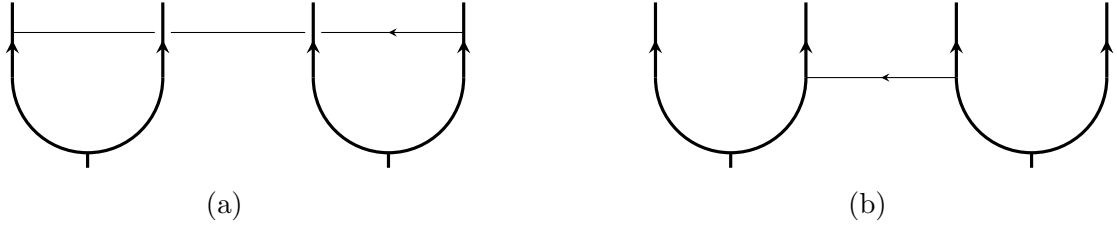

The cups and caps can be thought of geometrically as objects that reduce long rungs to the closest possible rung configuration (see Figure \ref{fig:cups}). This reduces the number of independent rung vevs to keep track of in a knot/link. We assume the cups connect strands $i$ and $i+1$, with $i$ odd. 

The planar limit of the cup and cap relation in Appendix \ref{app:braiding_algebra} then becomes (again here we use the notation $a$ $\; \mapsto \;$ $b$ to indicate that $a \; \cup_i = b \; \cup_i$ and we typically use them interchangeably)
\begin{equation}
\begin{split}
    i \; \text{odd}, \; j = i + 1 &:\quad f_{-;i,j} \mapsto -\left(g K_i^{-1} - g^{-1} K_i \right) \\
    j \; \text{odd}, \; i = j + 1 &:\quad e_{-;i,j}\mapsto -\left(K_i - K_i^{-1} \right) \\
    i < j, \quad \text{even } i, \text{ odd } j &:\quad f_{-;i,j} \mapsto  f_{-;i,j} \\
    i > j, \quad \text{even } j, \text{ odd } i &:\quad e_{-;i,j} \mapsto  e_{-;i,j} \\
    i > j, \quad \text{odd } i, \text{ odd } j &:\quad e_{-;i,j} \mapsto K_j \;  e_{-;i,j+1} \\
    i < j, \quad \text{odd } i, \text{ odd } j &:\quad f_{-;i,j} \mapsto - K_{i+1} \; f_{-;i+1,j} \\
    i > j, \quad \text{even } i, \text{ even } j &:\quad  e_{-;i,j} \mapsto - K_{i-1}^{-1} \; e_{-;i-1,j} \\
    i < j, \quad \text{even } i, \text{ even } j &:\quad f_{-;i,j} \mapsto K_j^{-1} \; f_{-;i,j-1} \\
    i > j, \quad \text{even } i, \text{ odd } j &:\quad e_{-;i,j} \mapsto - K_j \; K_{i-1}^{-1} \; e_{-;i-1,j+1} \\
    i < j, \quad \text{odd } i, \text{ even } j &:\quad f_{-;i,j} \mapsto - K_{i+1} \; K_j^{-1} \; f_{-;i+1,j-1}\, .
\end{split}
\end{equation}

Similarly, the cap constraints become
\begin{equation}
\begin{split}
    j = i + 1, \quad i \text{ odd} &:\quad f_{-;i,j} \mapsto ( K_i - K_i^{-1} )\\
    i = j + 1, \quad j \text{ odd} &:\quad e_{-;i,j} \mapsto (g K_j^{-1} - g^{-1} K_j) \\
    i < j, \quad i \text{ even}, \quad j \text{ odd} &:\quad f_{-;i,j} \mapsto f_{-;i,j} \\
    i > j, \quad i \text{ odd}, \quad j \text{ even} &:\quad e_{-;i,j} \mapsto e_{-;i,j} \\
    i > j, \quad i \text{ odd}, \quad j \text{ odd} &:\quad e_{-;i,j} \mapsto K_{j+1}^{-1} \; e_{-;i,j+1} \\
    i < j, \quad i \text{ odd}, \quad j \text{ odd} &:\quad f_{-;i,j} \mapsto -K_i^{-1} \; f_{-;i+1,j} \\
    i > j, \quad i \text{ even}, \quad j \text{ even} &:\quad  e_{-;i,j} \mapsto -K_i \;  e_{-;i-1,j} \\
    i < j, \quad i \text{ even}, \quad j \text{ even} &:\quad  f_{-;i,j} \mapsto K_{j-1} \; f_{-;i,j-1} \\
    i > j, \quad i \text{ even}, \quad j \text{ odd} &:\quad e_{-;i,j} \mapsto -K_{j+1}^{-1} \; K_i \; e_{-;i-1,j+1} \\
    i < j, \quad i \text{ odd}, \quad j \text{ even} &:\quad f_{-;i,j} \mapsto -K_i^{-1} \; K_{j-1} \; f_{-;i+1,j-1}\, .
\end{split}
\end{equation}

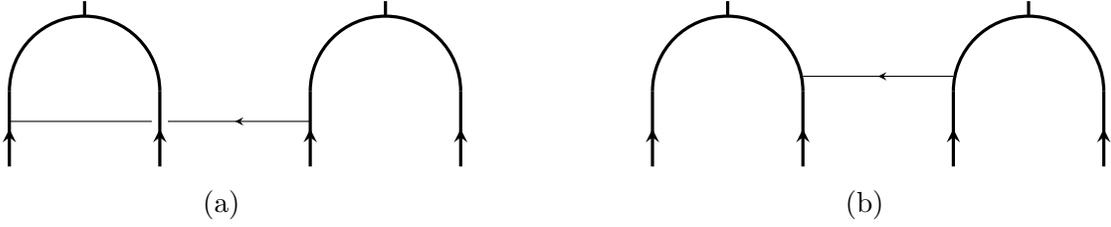
\begin{figure}[h]
    \centering
    \begin{subfigure}[t]{0.45\textwidth}
        \centering
        \begin{tikzpicture}[baseline=(current bounding box.center)]
            \Emrung{0.8}{1}{3}
            \neutralmodule{0}{,,,}
            \capmodule{1}{2}
        \end{tikzpicture}
        \caption{}
    \end{subfigure}
    \hfill
    \begin{subfigure}[t]{0.45\textwidth}
        \centering
        \begin{tikzpicture}[baseline=(current bounding box.center)]
            \neutralmodule{0}{,,,}
            \Erung{0.3}{2}{3}
            \capmodule{1}{2}
        \end{tikzpicture}
        \caption{}
    \end{subfigure}
    \caption{An example of cap constraints: In (a), we have the rung $e_{-; 3, 1}$ which, when applied to the cup constraints, can be expressed as $K_{2}^{-1} \; e_{-;3,2}$ as in (b).}
\end{figure}

\section{Examples of augmentation varieties} \label{app:knots}

\paragraph{Torus knot $T(2,2l+1)$.} Also labeled by $(2l+1)_1$, this knot has a Schubert presentation in terms of 4 strands and the insertion of the braiding $B_2^{2l+1}$. Note that this knot is chiral and this is the right-handed version. Also, the vertical framing we are considering differs in $2l+1$ units with respect to the canonical framing, in which the self-linking number is 0.

We take the same conventions as for the trefoil $T(2,3)$ in section \ref{sec:P42}, and we use a superscript $(i)$ for the value of the vevs after $i$ braidings, starting from the bottom. For example, $f_{-;1,2}^{(2l+1)}$ is the vev of $f_{-;1,2}$ at the top of the braid, and $e_2^{(1)}$ is the vev of $e_2$ located above the first braiding at the bottom.

Similarly to the case of the trefoil, braiding twice keeps the vevs of $e_2$ and $f_2$ invariant, so we will denote $f_2=f_2^{(2i+1)}$ and $e_2=e_2^{(2i+1)}$. Note that this is also the value of the vevs at the top. Another similarity is the relation $f_2=\omu e_2$, which can be deduced in the same way as for the trefoil.

Taking now $j$ odd, we can compute the braiding transformation
\begin{align}
& f_{-;1,3}^{(j)}=f_{-;1,3}^{(j+2)}-f_{1}^{(j+2)} f_{2} \cr
& f_{1}^{(j)}=f_{1}^{(j+2)}\left(1+e_{2} f_{2}\right)-f_{-;1,3}^{(j+2)} e_{2}\, .
\end{align}
This is a recursion suitable to a nice matrix form
\begin{equation}
    {f_{-;1,3}^{(1)}\choose f_{1}^{(1)}} = \left(
\begin{array}{cc}
 1 & -f_2  \\
 -e_2 & 1+e_2f_2 \\
\end{array}
\right)^l {f_{-;1,3}^{(2l+1)}\choose f_{1}^{(2l+1)}}\, .
\end{equation}
Braiding once more from $(0)$ to $(1)$, applying the cups and caps and using the relation between $e_2$ and $f_2$ to eliminate $f_2$, we get
\begin{equation}
    {\omu  \left(g^2 \left(e_2^2 \omu +1\right)-\olambda ^2\right)\choose e_2 g^2 \omu} = \left(
\begin{array}{cc}
 1 & -e_2 \omu  \\
 -e_2 & e_2^2 \omu +1 \\
\end{array}
\right)^l {e_2 g \omu \choose -g \left(\olambda ^2-1\right)}\, .
\end{equation}
The augmentation variety is obtained by eliminating $e_2$ from the system of equations. 

For the unknot $T(1,1)$ we recover
\begin{equation}
    \mathcal{A}_{T(1,1)}=g^2\omu-\olambda^2\omu+\olambda^4-\olambda^2 \, ,
\end{equation}
with
\begin{equation}
    e_2=\frac{g}{\olambda ^2}-\frac{1}{g}\, .
\end{equation}
For the trefoil $T(2,3)$
\begin{equation}
    \mathcal{A}_{3_1}=g^6 \omu ^2+2 g^4 \olambda ^4 \omu -g^4 \olambda ^2 \omu ^2-g^4 \olambda ^2 \omu +g^4 \omu +g^2 \olambda ^8-g^2 \olambda ^6 \omu -g^2 \olambda
   ^6-2 g^2 \olambda ^4 \omu +\olambda ^8 \omu\, ,
\end{equation}
for $T(2,5)$
\begin{align}
    \mathcal{A}_{5_1}=&\;g^{10} \omu ^3+3 g^8 \olambda ^4 \omu ^2-g^8 \olambda ^2 \omu ^3-g^8 \olambda ^2 \omu ^2+2 g^8 \omu ^2+3 g^6 \olambda ^8 \omu -2 g^6 \olambda
   ^6 \omu ^2-2 g^6 \olambda ^6 \omu \cr
    &-4 g^6 \olambda ^4 \omu ^2+2 g^6 \olambda ^4 \omu -g^6 \olambda ^2 \omu ^2-g^6 \olambda ^2 \omu +g^6 \omu
   +g^4 \olambda ^{12}-g^4 \olambda ^{10} \omu -g^4 \olambda ^{10}\cr
    &+2 g^4 \olambda ^8 \omu ^2-4 g^4 \olambda ^8 \omu +2 g^4 \olambda ^6 \omu
   ^2+2 g^4 \olambda ^6 \omu +g^4 \olambda ^4 \omu ^2-2 g^4 \olambda ^4 \omu +2 g^2 \olambda ^{12} \omu \cr
    &-g^2 \olambda ^{10} \omu ^2-g^2
   \olambda ^{10} \omu -2 g^2 \olambda ^8 \omu ^2+g^2 \olambda ^8 \omu +\olambda ^{12} \omu ^2\, ,
\end{align}
and for $T(2,7)$
\begin{align}
    \mathcal{A}_{7_1}=&\;g^{14} \omu ^4+4 g^{12} \olambda ^4 \omu ^3-g^{12} \olambda ^2 \omu ^4-g^{12} \olambda ^2 \omu ^3+3 g^{12} \omu ^3+6 g^{10} \olambda ^8 \omu
   ^2-3 g^{10} \olambda ^6 \omu ^3\cr
   &-3 g^{10} \olambda ^6 \omu ^2-6 g^{10} \olambda ^4 \omu ^3+6 g^{10} \olambda ^4 \omu ^2-2 g^{10} \olambda
   ^2 \omu ^3-2 g^{10} \olambda ^2 \omu ^2+3 g^{10} \omu ^2\cr
   &+4 g^8 \olambda ^{12} \omu -3 g^8 \olambda ^{10} \omu ^2-3 g^8 \olambda ^{10} \omu
   +3 g^8 \olambda ^8 \omu ^3-12 g^8 \olambda ^8 \omu ^2+3 g^8 \olambda ^8 \omu +4 g^8 \olambda ^6 \omu ^3\cr
   &+2 g^8 \olambda ^6 \omu ^2-2 g^8
   \olambda ^6 \omu +2 g^8 \olambda ^4 \omu ^3-8 g^8 \olambda ^4 \omu ^2+2 g^8 \olambda ^4 \omu -g^8 \olambda ^2 \omu ^2-g^8 \olambda ^2 \omu
   +g^8 \omu \cr
   &+g^6 \olambda ^{16}-g^6 \olambda ^{14} \omu -g^6 \olambda ^{14}+6 g^6 \olambda ^{12} \omu ^2-6 g^6 \olambda ^{12} \omu -2 g^6
   \olambda ^{10} \omu ^3+2 g^6 \olambda ^{10} \omu ^2\cr
   &+4 g^6 \olambda ^{10} \omu -4 g^6 \olambda ^8 \omu ^3+10 g^6 \olambda ^8 \omu ^2-4 g^6
   \olambda ^8 \omu -g^6 \olambda ^6 \omu ^3+g^6 \olambda ^6 \omu ^2+2 g^6 \olambda ^6 \omu \cr
   &+2 g^6 \olambda ^4 \omu ^2-2 g^6 \olambda ^4 \omu +3
   g^4 \olambda ^{16} \omu -2 g^4 \olambda ^{14} \omu ^2-2 g^4 \olambda ^{14} \omu +2 g^4 \olambda ^{12} \omu ^3-8 g^4 \olambda ^{12} \omu
   ^2\cr
   &+2 g^4 \olambda ^{12} \omu +2 g^4 \olambda ^{10} \omu ^3+g^4 \olambda ^{10} \omu ^2-g^4 \olambda ^{10} \omu +g^4 \olambda ^8 \omu ^3-4
   g^4 \olambda ^8 \omu ^2+g^4 \olambda ^8 \omu \cr
   &+3 g^2 \olambda ^{16} \omu ^2-g^2 \olambda ^{14} \omu ^3-g^2 \olambda ^{14} \omu ^2-2 g^2
   \olambda ^{12} \omu ^3+2 g^2 \olambda ^{12} \omu ^2+\olambda ^{16} \omu ^3\, .
\end{align}
In the $g\to 1$ limit, the tree-level augmentation variety of the torus knot $T(2,2l+1)$ is
\begin{equation} \label{eq:torus_tree}
    (\olambda^2-1)(1+\omu)^l(\omu-\olambda^{2(2l+1)})=0\, .
\end{equation}

\paragraph{Torus link $T(2,2l)$.} This link has a Schubert presentation in terms of 4 strands and the insertion of the braiding $B_2^{2l}$. In this case, the braiding does not keep $e_2$ and $f_2$ invariant, but denoting $e_2=e_2^{(0)}$ and $f_2=f_2^{(0)}$, it still leads to nice relations
\begin{align}
    f_2^{(2i)}&=\left(\frac{g}{\olambda_1\olambda_2}\right)^{2i}f_2 \cr
    e_2^{(2i)}&=\left(\frac{g}{\olambda_1\olambda_2}\right)^{-2i}e_2\, .
\end{align}
We can get relations between vevs in a similar way as for $T(2,2l+1)$. Keeping track of the new factors in various places, we obtain
\begin{align}
    \omu_1 {g^2 f_2\choose g^2-\olambda_1^2} &= M_1\cdots M_l\; {gf_2 \left(\frac{g}{\olambda_1\olambda_2}\right)^{2l} \choose -g \left(\olambda_1^2-1\right)}\cr
    \omu_2{g^2 f_2\choose g^2-\olambda_2^2} &= M_1\cdots M_l\; {gf_2 \left(\frac{g}{\olambda_1\olambda_2}\right)^{2l} \choose -g \left(\olambda_2^2-1\right)}\, ,
\end{align}
where the matrix $M_i$ is
\begin{equation}
    M_i = \left(\begin{array}{cc}
 1 & -f_2\left(\frac{g}{\olambda_1\olambda_2}\right)^{2i}  \\
 -e_2\left(\frac{g}{\olambda_1\olambda_2}\right)^{1-2i} & 1+\frac{g}{\olambda_1\olambda_2}e_2f_2 \\
\end{array}\right)\, .
\end{equation}

There is always a disconnected saddle in which the vev of $e_2f_2$ is zero, and this part of the variety takes the same form as for two unknots. The interesting saddles are in the connected part of the augmentation variety. 

For the Hopf link $T(2,2)$
\begin{equation}
    \omu_1=\frac{g}{\olambda_2^2}\qquad \qquad \omu_2=\frac{g}{\olambda_1^2}\, ,
\end{equation}
For $T(2,4)$
\begin{align}
    -g^4-g \olambda _2^2(\olambda_1^2-\olambda_2^2) \omu_1 +\olambda_1^2 \olambda_2^6 \omu_1^2&=0 \cr
    -g^4+g \olambda _1^2(\olambda_1^2-\olambda_2^2) \omu_2 +\olambda_1^6 \olambda_2^2 \omu_2^2&=0\, .
\end{align}
In the $g\to 1$ limit, the disconnected part of the tree-level augmentation variety of $T(2,4)$ reduces to two solutions for $(\omu_1,\omu_2)$:
\begin{equation}
    \omu_1 =\frac{1}{\olambda_2^4} \qquad \qquad \omu_2=\frac{1}{\olambda_1^4}\, ,
\end{equation}
and
\begin{equation}
    \omu_1 = \omu_2=-\frac{1}{\olambda_1^2\olambda_2^2}\, .
\end{equation}

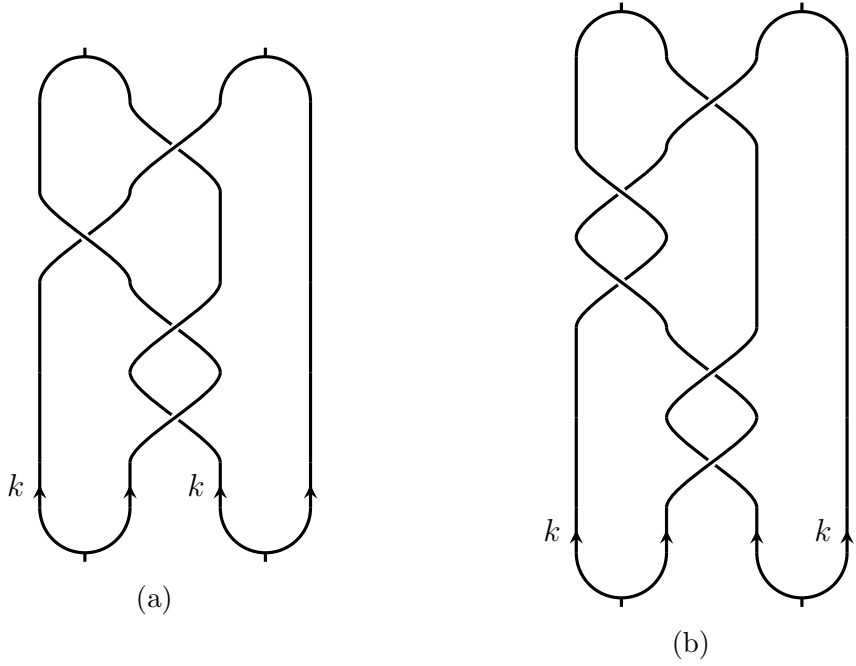
\begin{figure}[h]
\begin{subfigure}[t]{0.45\textwidth}
\centering
    \begin{tikzpicture}[baseline=(current bounding box.center), scale=0.6]
    \cupmodule{0}{2}
    \neutralmodule{0}{$k$,,$k$,}
    \braidmodule{1}{2}
    \braidmodule{3}{2}
    \braidmoduleI{5}{1}
    \braidmodule{7}{2}
    \capmodule{9}{2}
\end{tikzpicture}
\caption{}
\hfill
\end{subfigure}
\begin{subfigure}[t]{0.45\textwidth}
\centering
    \begin{tikzpicture}[baseline=(current bounding box.center), scale=0.6]
    \cupmodule{0}{2}
    \neutralmodule{0}{$k$,,,$k$}
    \braidmodule{1}{2}
    \braidmodule{3}{2}
    \braidmoduleI{5}{1}
    \braidmoduleI{7}{1}
    \braidmodule{9}{2}
    \capmodule{11}{2}
\end{tikzpicture}
\caption{}
\end{subfigure}
\caption{(a) The figure-eight knot ($4_1$) (b) The bowstring knot ($5_2$)}
\label{fig:8bow}
\end{figure}

\paragraph{Figure-eight knot ($4_1$).} This knot has a 4-strand Schubert presentation with braiding $B_2 B_1^{-1} B_2^2$, see Figure \ref{fig:8bow}. 
\begin{align}
\mathcal{A}_{4_1} = \, & -g^3 \olambda^4 + g^3 \olambda^6 + g^6 \omu 
- 2 g^4 \olambda^2 \omu + 2 g^2 \olambda^8 \omu - \olambda^{10} \omu  - g^5 \omu^2 \notag \\
&+ 2 g^5 \olambda^2 \omu^2 - 2 g \olambda^8 \omu^2 
+ g \olambda^{10} \omu^2  + g^4 \olambda^4 \omu^3 - g^2 \olambda^6 \omu^3\, .
\end{align}

\paragraph{The bowstring knot ($5_2$).} This knot has a 4-strand Schubert presentation with braiding \( B_2 B_1^{-2} B_2^2 \), see Figure \ref{fig:8bow}.
\begin{align}
\mathcal{A}_{5_2} = \; & g^2 \olambda^{26} - g^2 \olambda^{28} 
- 2 g^6 \olambda^{16} \omu + 4 g^6 \olambda^{18} \omu 
+ 3 g^4 \olambda^{20} \omu - 3 g^6 \olambda^{20} \omu  - 2 g^4 \olambda^{22} \omu - g^4 \olambda^{24} \omu 
\notag \\
&+ 2 g^2 \olambda^{26} \omu - \olambda^{28} \omu 
+ g^{10} \olambda^6 \omu^2 - 3 g^{10} \olambda^8 \omu^2  - 4 g^8 \olambda^{10} \omu^2 + 5 g^{10} \olambda^{10} \omu^2 
+ 3 g^8 \olambda^{12} \omu^2 \notag \\
&- 3 g^{10} \olambda^{12} \omu^2 
+ 6 g^6 \olambda^{14} \omu^2  - 4 g^8 \olambda^{14} \omu^2 + 3 g^6 \olambda^{16} \omu^2 
- 3 g^8 \olambda^{16} \omu^2 - 4 g^4 \olambda^{18} \omu^2 
\notag \\
&+ 5 g^6 \olambda^{18} \omu^2  - 3 g^4 \olambda^{20} \omu^2 + g^2 \olambda^{22} \omu^2 
- g^{14} \omu^3 + 2 g^{14} \olambda^2 \omu^3 
- g^{14} \olambda^4 \omu^3 - 2 g^{12} \olambda^6 \omu^3  \notag \\
&+ 3 g^{10} \olambda^8 \omu^3 - 3 g^{12} \olambda^8 \omu^3 
+ 4 g^{10} \olambda^{10} \omu^3 - 2 g^8 \olambda^{12} \omu^3 
- g^{16} \omu^4 + g^{14} \olambda^2 \omu^4\, .
\end{align}

\begin{figure}[h]
\begin{subfigure}[t]{0.45\textwidth}
\centering
    \begin{tikzpicture}[baseline=(current bounding box.center), scale=0.6] 
        \cupmodule{0}{2}
        \neutralmodule{0}{$k$,,,$k$}
        \braidmodule{1}{2}
        \braidmoduleI{3}{3}
        \braidmoduleI{5}{3}
        \braidmoduleI{7}{3}
        \braidmodule{9}{2}
        \braidmodule{11}{2}
        \capmodule{13}{2}
    \end{tikzpicture}
    \caption{}
    \label{fig:61a}
\end{subfigure}
\hfill
\begin{subfigure}[t]{0.45\textwidth}
\centering
    \begin{tikzpicture}[baseline=(current bounding box.center), scale=0.6] 
        \cupmodule{0}{2}
        \neutralmodule{0}{$k$,,,$k$}
        \braidmoduleI{1}{2}
        \braidmodule{3}{3}
        \braidmodule{5}{3}
        \braidmodule{7}{3}
        \braidmoduleI{9}{2}
        \braidmoduleI{11}{2}
        \capmodule{13}{2}
    \end{tikzpicture}
    \caption{}
    \label{fig:61b}
\end{subfigure}
\caption{Stevedore's knot (a) and its mirror (b).}
\label{fig:61}
\end{figure}
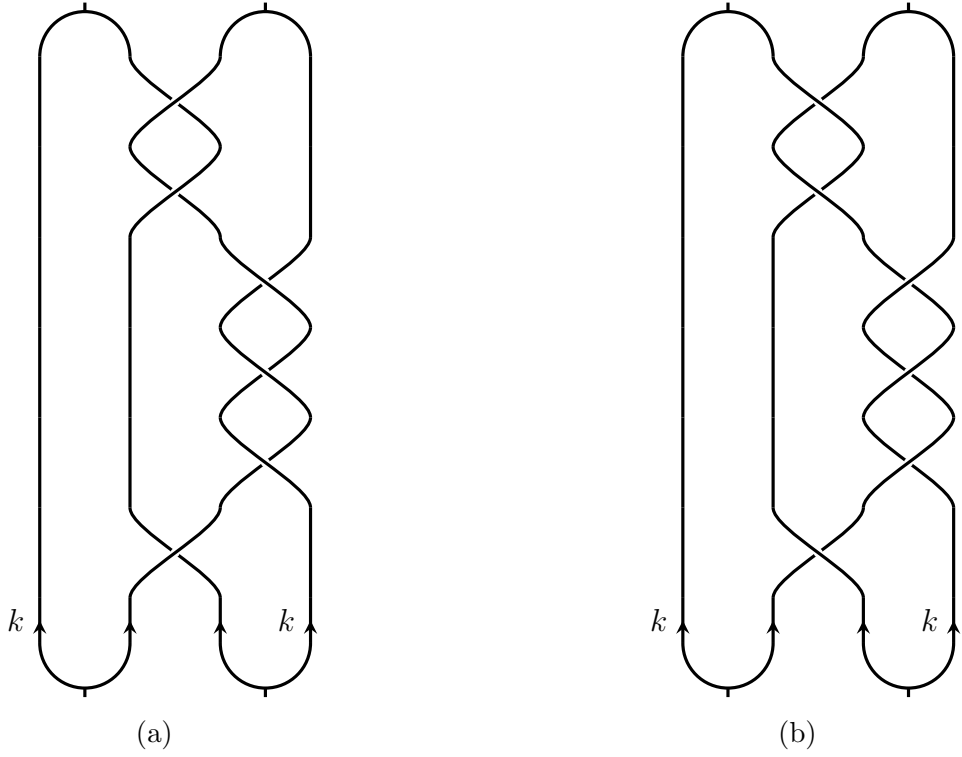

\paragraph{Stevedore's knot ($6_1$).} This knot has a 4-strand Schubert presentation with braiding \( B_2^2 B_3^{-3} B_2 \), see Figure \ref{fig:61a}.
\begin{align}
\mathcal{A}_{6_1} &= g^3 \olambda^{24} - g^3 \olambda^{26} - 2 g^8 \olambda^{16} \omu + 4 g^6 \olambda^{18} \omu + g^8 \olambda^{18} \omu - g^6 \olambda^{20} \omu - 2 g^6 \olambda^{22} \omu + g^4 \olambda^{26} \omu  \notag \\
&\quad - 2 g^2 \olambda^{28} \omu + \olambda^{30} \omu + g^{13} \olambda^8 \omu^2 - 4 g^{11} \olambda^{10} \omu^2 + 6 g^9 \olambda^{12} \omu^2 - 2 g^{11} \olambda^{12} \omu^2 \notag \\
&\quad - g^9 \olambda^{14} \omu^2 + 3 g^9 \olambda^{16} \omu^2 - 4 g^7 \olambda^{20} \omu^2 - 3 g^5 \olambda^{22} \omu^2 + 4 g^3 \olambda^{24} \omu^2 + 2 g^5 \olambda^{24} \omu^2 \notag \\
&\quad - 2 g^3 \olambda^{26} \omu^2 - 2 g^{14} \olambda^4 \omu^3 + 4 g^{12} \olambda^6 \omu^3 + 2 g^{14} \olambda^6 \omu^3 - 3 g^{12} \olambda^8 \omu^3 \notag \\
&\quad - 4 g^{12} \olambda^{10} \omu^3 + 3 g^{10} \olambda^{14} \omu^3 - g^8 \olambda^{16} \omu^3 + 6 g^6 \olambda^{18} \omu^3 - 2 g^8 \olambda^{18} \omu^3 + g^{15} \omu^4 \notag \\
&\quad - 2 g^{15} \olambda^2 \omu^4 + g^{15} \olambda^4 \omu^4 - 2 g^{13} \olambda^8 \omu^4 - g^{11} \olambda^{10} \omu^4 + 4 g^9 \olambda^{12} \omu^4 + g^{11} \olambda^{12} \omu^4 \notag \\
&\quad - 2 g^9 \olambda^{14} \omu^4 - g^{14} \olambda^4 \omu^5 + g^{12} \olambda^6 \omu^5\, .
\end{align}

\subsection{Additional considerations}

\paragraph{Mirrors.}
For every chiral knot, one can associate a mirror knot, which is obtained by replacing all its braid group elements by their inverses. For example, the mirror of the Stevedore knot is given by \( B_2^{-2} B_3^{3} B_{2}^{-2} \). See Figure \ref{fig:61b}. One can easily compute the augmentation of a mirror knot by performing the following transformation
\begin{align}
    \mathcal{A}_{\text{mirror}} (\omu, \olambda, g) &= \mathcal{A} \left( \frac{1}{\omu}, \frac{\olambda}{g}, \frac{1}{g} \right)\, ,
\end{align}
which is consistent with the fact that the mirror of the mirror is the original knot.

\paragraph{Symmetric vs antisymmetric powers.} 
The augmentation polynomial of symmetric powers of the fundamental representation can be obtained from the one of antisymmetric powers by replacing $N$ with $-N$, i.e. $\fq\to -\fq$, and vice versa. This means
\begin{align}
    \mathcal{A}_{S^{\bullet} \mathbb{C}^N} (\omu, \olambda, g) &= \mathcal{A}_{\Lambda^{\bullet} \mathbb{C}^N} \left( \omu, \frac{1}{\olambda}, \frac{1}{g} \right)\, .
\end{align}

\paragraph{Symmetry.}
The augmentation variety has a ${k}\to{N-k}$ symmetry
\begin{align}
    \mathcal{A}(\omu, \olambda, g) &= \mathcal{A}\left(\dfrac{1}{\omu}, \dfrac{g}{\olambda}, g\right)\, .
\end{align}

\begin{figure}[h]
\centering
\begin{subfigure}[t]{0.4\textwidth}
    \centering
    \begin{tikzpicture}[baseline=(current bounding box.center), scale=0.6]
        \neutralmodule{0}{$k$,$k$}
        \braidmoduletwo{1}{1}
    \end{tikzpicture}
    \caption{}
\end{subfigure}
\quad
\begin{subfigure}[t]{0.4\textwidth}
    \centering
    \begin{tikzpicture}[baseline=(current bounding box.center), scale=0.6]
        \neutralmodule{0}{$k$,$k$}
        \braidmoduletwoI{1}{1}
    \end{tikzpicture}
    \caption{}
\end{subfigure}
\caption{(a) Crossing number $+1$ (b) Crossing number $-1$}
\label{fig:cross}
\end{figure}
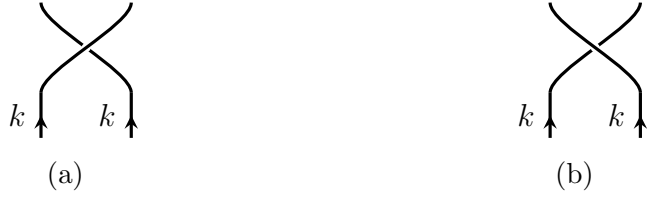

\paragraph{Framing.} All our calculations are done in vertical framing. To change to canonical framing, we have to add the twisting factor to $\omu$ 
\begin{align}
    \mathcal{A}_{\text{canonical}} (\omu, \olambda, g) &= \mathcal{A} \left( \omu \left(\frac{g}{\olambda^2} \right)^x, \olambda, g \right) \, 
\end{align}
where $x$ is the self-linking number of the given presentation of the knot in vertical framing, which can be computed by counting crossings with signs given by Figure \ref{fig:cross}.

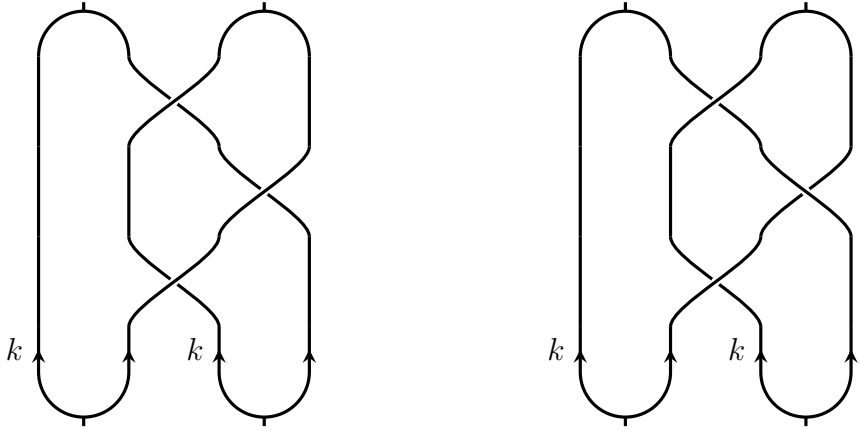
\begin{figure}
\begin{subfigure}[t]{0.45\textwidth}
\centering
    \begin{tikzpicture}[baseline=(current bounding box.center), scale=0.6]
    \cupmodule{0}{2}
    \neutralmodule{0}{$k$,,$k$,}
    \braidmodule{1}{2}
    \braidmodule{3}{3}
    \braidmodule{5}{2}
    \capmodule{7}{2}
\end{tikzpicture}
\hfill
\end{subfigure}
\begin{subfigure}[t]{0.45\textwidth}
\centering
    \begin{tikzpicture}[baseline=(current bounding box.center), scale=0.6]
    \cupmodule{0}{2}
    \neutralmodule{0}{$k$,,$k$,}
    \braidmodule{1}{2}
    \braidmoduleI{3}{3}
    \braidmodule{5}{2}
    \capmodule{7}{2}
\end{tikzpicture}
\end{subfigure}
\caption{The unknot (left) and left-handed trefoil (right).}
\label{fig:other_pres}
\end{figure}

\paragraph{Different presentations.}
Our construction is consistent with three-dimensional manipulations of the strands and rungs (up to framing). To check this, we can present known knots in different ways and see that we recover the same augmentation varieties. We will illustrate it with the unknot and left-handed trefoil of Figure \ref{fig:other_pres}.

For the unknot we can start with $f^u_1=-\omu (g\olambda^{-1}-g^{-1}\olambda)$, braid it to obtain $f^a_{+;1,4}=-f^a_2$, and braid it back to $-g\olambda^{-2}e^u_3=g\olambda^{-2}(\olambda-\olambda^{-1})$. This gives the augmentation variety of the unknot, accounting for the framing.

For the left-handed trefoil we can start with the rungs $f^u_1$, $f^u_{-;1,3}$ and $f^u_{-;1,4}$, and perform the usual manipulations to obtain relations
\begin{align}
    -\omu (g^2-\olambda^2)&=g\olambda f^a_2 (\olambda^2-2+ge^a_2f^a_2)\cr
   \omu\olambda e^a_2 &= \olambda^2-1+g e^a_2 f^a_2\cr
    g\omu e^a_2&=-\olambda^2f^a_2 \, ,
\end{align}
and obtain the augmentation variety
\begin{equation}
    0 = g^6 \omu -2 g^4 \olambda ^4 \omu -g^4 \olambda ^2 \omu -g^4 \olambda ^2+g^4+g^2 \olambda ^8 \omu -g^2 \olambda ^6 \omu ^2-g^2 \olambda ^6 \omu
   +2 g^2 \olambda ^4 \omu +\olambda ^8 \omu ^2\, ,
\end{equation}
which is mapped to the usual right-handed trefoil augmentation variety by the chirality reversal map $\omu\to \omu^{-1}$, $\olambda \to \olambda g^{-1}$, $g\to g^{-1}$.

\section{Augmentation variety from symmetric HOMFLY} \label{app:saddle}
In this section, we test the relationship between augmentation varieties and recursion relations for colored HOMFLY polynomials for several knots. As a source for the latter, we will employ \cite{Kawagoe:2021onh}. This computes the HOMFLY polynomials for symmetric powers of the fundamental representation for an infinite sequence of ``twist'' knots, which include the trefoil and figure eight:
\begin{align}
W_{3_1,S^n \bC^N} &= \sum_{i=0}^n 
(-1)^i a^{2i} q^{i(i-1)}
\left[ \begin{array}{@{\,}c@{\,}}n \\ i \end{array} \right]
\{n+i-1;a \}_i \{i-2;a \}_i \, \\
W_{4_1,S^n \bC^N}  &=  \sum_{i=0}^n 
\left[ \begin{array}{@{\,}c@{\,}}n \\ i \end{array} \right]
\{n+i-1;a \}_i \{i-2;a \}_i\, ,
\end{align}
with $a = \fq^N$ (Corollary $5.2$ in the reference). Although our main interest is anti-symmetric powers of the fundamental representation, we expect that the augmentation varieties for symmetric and anti-symmetric power knot invariants must be related by $N \to -N$. This will indeed be the case, i.e. the two varieties are related by $a = g^{-1}$.

The reference \cite{Kawagoe:2021onh} expresses the answer as sums over one or two variables of products of $\fq$-integers. In order to find the large $N$ saddles of such an expression, we do a saddle evaluation of the sums themselves. The sum is either dominated by an intermediate summand or by one of the endpoints. The first type of saddles we consider thus extremize the summand with respect to the summation variables, requiring the ratio of consecutive terms to be $1$. The second just specializes the summand to the endpoints of the sum. 

As an example, consider the trefoil invariant $W_{3_1,S^n \bC^N}$. We look for a value of $s=\fq^i$ such that the ratio 
\begin{equation}
    - a^2 s^2 \frac{\olambda/s-s/\olambda}{s-1/s}(a \olambda s - a^{-1} \olambda^{-1} s^{-1})(a s - a^{-1} s^{-1}) 
\end{equation}
of consecutive terms is $1$. Here we denote $\fq^n = \olambda$. The saddle for $i$ is thus
\begin{equation}
    (\olambda^2-s^2)(a^2 \olambda^2 s^2 - 1)(a^2 s^2 - 1) + \olambda^2(s^2-1)=0\, .
\end{equation}
We can expand it to 
\begin{equation}
    a^4 \olambda^4 s^4 - a^2 \olambda^4 s^2 - a^2 \olambda^2 s^2  + \olambda^2 s^2=(a^4 \olambda^2 s^4 - a^2 \olambda^2 s^2 - a^2 s^2 + 1)s^2\, ,
\end{equation}
and remove a factor of $s^2$:
\begin{equation}
        a^4 \olambda^2 s^4 -a^4 \olambda^4 s^2- a^2 \olambda^2 s^2 - a^2 s^2 + a^2 \olambda^4+ a^2 \olambda^2- \olambda^2+ 1=0 \, .
\end{equation}

After replacing the sum with the dominant term, we compute $\omu$ by taking the ratio of the dominant terms evaluated at $n+1$ and $n$:
\begin{equation}
    \frac{\olambda-\olambda^{-1}}{\olambda s^{-1} - \olambda^{-1} s}\frac{a \olambda s - a^{-1} \olambda^{-1} s^{-1}}{a \olambda - a^{-1} \olambda^{-1}}\, .
\end{equation}
We should include an extra factor of $\frac{a \olambda - a^{-1} \olambda^{-1}}{\olambda- \olambda^{-1}}$ to go from HOMFLY to the knot correlation function. We get 
\begin{equation}
    \omu = \frac{a \olambda s - a^{-1} \olambda^{-1} s^{-1}}{\olambda s^{-1} - \olambda^{-1} s} \, ,
\end{equation}
i.e. 
\begin{equation}
   a (\olambda^2  - s^2) \omu = a^2 \olambda^2 s^2 - 1 \, , 
\end{equation}
and 
\begin{equation}
  s^2 = a^{-1} \frac{a \olambda^2 \omu +1}{a \olambda^2+\omu}   \, ,
\end{equation}
leading to 
\begin{equation}
        a^2 \olambda^2 (a \olambda^2 \omu +1)^2 -a (a^2 \olambda^4 + \olambda^2 + 1)(a \olambda^2 \omu +1)(a \olambda^2+\omu) + (a^2 \olambda^4+ a^2 \olambda^2- \olambda^2+ 1)(a \olambda^2+\omu)^2=0 \, .
\end{equation}
Expanding out:
\begin{equation}
    (1- \olambda^2) \omu^2+ (- a^4 \olambda^8 + a^2 \olambda^6  + 2 (a^2-1)\olambda^4 + \olambda^2- 1) a \omu  + a^2 \olambda^6(a^2 \olambda^2-1)=0 \, .
\end{equation}
This matches \cite{Aganagic:2012jb} et al. under $Q \to a^2$, $\lambda_1 \to a/\omu$, $\mu_1 \to \olambda^2 a^2$. It also matches our answer under $g \to 1/a$, $\omu \to a^3 \olambda^6/\omu$. The former encodes $N \to -N$, the latter is due to different framing conventions. 

The ``endpoint'' saddles are a bit more confusing. The $i=0$ endpoint 
gives $\omu=1$. The $i=n$ endpoint gives $\omu = - a^2 \olambda^2 (\olambda^2 a - \olambda^{-2} a^{-1})^2 \frac{a \olambda - a^{-1} \olambda^{-1}}{\olambda- \olambda^{-1}}$. We do not understand the role of these saddles and we will disregard them from now on.

We can repeat the analysis for the figure eight knot answer. Now the intermediate saddle condition is 
\begin{equation}
    (\olambda^2 - s^2 )(a^2 \olambda^2 s^2 - 1) (a^2 s^2 - 1) + a^2 \olambda^2 s^2 (s^2-1) = 0 \, ,
\end{equation}
while 
\begin{equation}
    \omu = \frac{a \olambda s - a^{-1} \olambda^{-1} s^{-1}}{\olambda s^{-1} - \olambda^{-1} s} 
\end{equation}
gives again 
\begin{equation}
  s^2 = a^{-1} \frac{a \olambda^2 \omu +1}{a \olambda^2+\omu}   \, ,
\end{equation}
and thus the augmentation variety 
\begin{equation}
    a^3 \olambda^4 (\olambda^2-1) \omu^3 + \left(a^6 \olambda^{10}-4 a^4 \olambda^6+4 a^2 \olambda^4-1\right)\omu(\omu -a) - a^2 \olambda^4 (a^2 \olambda^2-1)\, .
\end{equation}

Finally, we can extend the analysis to the infinite family of twist knots $K_p$ from Theorem $5.1$ in the reference:
\begin{align*}
W_{K_p,S^n \bC^N} &=\sum_{i=0}^n   a^{i} q^{\frac{  i(i-1)  }{  2  }   } \{i\}! s_{i,p}
\left[ \begin{array}{@{\,}c@{\,}}n \\ i \end{array} \right] \{ n+i-1;a \}_i \{i-2;a \}_i \, , 
\end{align*}
where 
\begin{align*}
s_{i,p} 
& = \sum_{k=0}^i   (-1)^k   \frac{( a^k q^{k(k-1)})^{2p} }{ \{ k \}! \{ i-k \}!}.
\frac{\{ 2k-1 ; a \}}{ \{ i+k-1; a \}_{i+1} } \, .
\end{align*}
We need first to estimate $s_{i+1,p}/s_{i,p}$ by a saddle evaluation of the second sum. If $\fq^k = u$, 
\begin{equation}
    1 = - (a u^2)^{2p}\frac{s/u-u/s}{u-1/u}\frac{a u-1/a/u}{a u s-1/a/u/s} \, ,
\end{equation}
i.e. 
\begin{equation}
    (u^2-1)(a^2 u^2 s^2-1) = - (a u^2)^{2p}(s^2-u^2)(a^2 u^2-1) \, .
\end{equation}
Then the ratio is 
\begin{equation}
    \frac{s^2 u^2 a}{(s^2-u^2)(a^2 s^2 u^2-1)}\, .
\end{equation}
The saddle equation for the $i$ sum becomes 
\begin{equation}
    1 = a s (s-1/s)\frac{s^2 u^2 a}{(s^2-u^2)(a^2 s^2 u^2-1)}\frac{\olambda/s-s/\olambda}{s-1/s}(a \olambda s-1/a/\olambda/s) (a s-1/a/s)\, ,
\end{equation}
i.e. 
\begin{equation}
    \olambda^2 (s^2-u^2)(a^2 s^2 u^2-1)= u^2(\olambda^2-s^2)(a^2 \olambda^2 s^2-1) (a^2 s^2-1)\, .
\end{equation}
It reduces to 
\begin{equation}
    -a^4 l^2 s^4 u^2-a^2 l^4 u^2+a^2 l^2 u^4-a^2 l^2 u^2+s^2 \left(a^4 l^4 u^2+a^2
   u^2\right)+l^2-u^2=0\, .
\end{equation}
Finally, the final answer is estimated:
\begin{equation}
    \omu = (a\olambda s-1/a/\olambda/s)/(\olambda/s-s/\olambda)\, ,
\end{equation}
i.e. 
\begin{equation}
  s^2 = a^{-1} \frac{a \olambda^2 \omu +1}{a \olambda^2+\omu}   \, ,
\end{equation}
which we can plug back into the two previous equations to get two lengthy equations in $\omu$, $\olambda$ and $u$.

\bibliographystyle{JHEP}

\bibliography{mono}

\end{document}